\documentclass[twoside,11pt]{article}
\usepackage{jmlr2e}
\usepackage{caption}
\usepackage{amsmath}
\usepackage{subcaption}
\usepackage{multirow}
\usepackage{pdflscape}
\usepackage{makecell}

\allowdisplaybreaks

\newcommand{\sups}[2]{{#1}^{({#2})}}
\newcommand{\subs}[2]{{#1}_{({#2})}}
\newcommand{\la}{\langle}
\newcommand{\ra}{\rangle}
\newcommand{\mb}{\boldsymbol}
\newcommand{\N}{{\mathcal N}}
\newcommand{\TPB}{{\mathcal {TPB}}}
\newcommand{\TPBN}{{\mathcal {TPBN}}}
\newcommand{\tr}{\textrm{tr}}
\newcommand{\diag}{\textrm{diag}}
\newcommand{\GIG}{\mathcal {GIG}}

\DeclareMathOperator*{\argmax}{arg\,max}

\jmlrheading{}{}{}{}{}{}

\ShortHeadings{}{}
\firstpageno{1}

\begin{document}

\title{Bayesian group factor analysis with structured sparsity}

\author{\name Shiwen Zhao \email shiwen.zhao@duke.edu \\
       \addr Computational Biology and Bioinformatics Program\\
       \addr Department of Statistical Science \\
       Duke University\\
       Durham, NC 27705, USA
       \AND
       \name Chuan Gao \email chuan.gao@duke.edu \\
       \addr Department of Statistical Science\\
       Duke University\\
       Durham, NC 27705, USA
       \AND
       \name Sayan Mukherjee \email sayan@stat.duke.edu \\
       \addr Departments of Statistical Science, Computer Science, Mathematics \\
       Duke University\\
       Durham, NC 27705, USA
       \AND
       \name Barbara E Engelhardt \email bee@princeton.edu \\
       \addr Department of Computer Science \\
       Center for Statistics and Machine Learning \\
       Princeton University\\
       Princeton, NJ 08540, USA
       }
\editor{}

\maketitle

\begin{abstract}
  Latent factor models are the canonical statistical tool for
  exploratory analyses of low-dimensional linear structure for an
  observation matrix with $p$ features across $n$ samples. We develop
  a structured Bayesian group factor analysis model that extends the
  factor model to multiple coupled observation matrices; in the case
  of two observations, this reduces to a Bayesian model of canonical
  correlation analysis. The main contribution of this work is to
  carefully define a structured Bayesian prior that encourages both
  element-wise and column-wise shrinkage and leads to desirable
  behavior on high-dimensional data. In particular, our model puts a
  structured prior on the joint factor loading matrix, regularizing at
  three levels, which enables element-wise sparsity and unsupervised
  recovery of latent factors corresponding to structured variance
  across arbitrary subsets of the observations. In addition, our
  structured prior allows for both dense and sparse latent factors so
  that covariation among either all features or only a subset of
  features can both be recovered. We use fast parameter-expanded
  expectation-maximization for parameter estimation in this model.  We
  validate our method on both simulated data with substantial structure
  and real data, comparing against a number of state-of-the-art
  approaches. These results illustrate useful properties of our model,
  including i) recovering sparse signal in the presence of dense
  effects; ii) the ability to scale naturally to large numbers of
  observations; iii) flexible observation- and factor-specific
  regularization to recover factors with a wide variety of sparsity
  levels and percentage of variance explained; and iv) tractable
  inference that scales to modern genomic and document data sizes.
\end{abstract}


\section{Introduction}

Factor analysis models have attracted attention recently due to their
ability to perform exploratory analyses of the latent linear structure
in high dimensional data~\citep{west_bayesian_2003,
  carvalho_high-dimensional_2008, engelhardt_analysis_2010}. A latent
factor model finds a low dimensional representation $\mb x_i \in
\mathbb{R}^{k \times 1}$ of high-dimensional data with $p$ features,
$\mb y_i \in \mathbb{R}^{p \times 1}$ given $n$ samples.  A sample in
the low dimensional space is linearly projected to the original high
dimensional space through a \emph{loadings matrix} $\mb \Lambda \in
\mathbb{R}^{p\times k}$ with Gaussian noise $\mb \epsilon_i \in
\mathbb{R}^{p \times 1}$:
\begin{align}
  \mb y_i = \mb \Lambda \mb x_i + \mb \epsilon_i,
 \label{eq:FAModel}
\end{align} 
for $i = 1,\cdots,n$.  It is often assumed $\mb x_i$ follows a
$\N_k(\mb 0, \mb I_k)$ distribution, where $\mb I_k$ is the identity
matrix of dimension $k$, and $\mb \epsilon_i \sim \N_p(\mb 0, \mb
\Sigma)$, where $\mb \Sigma$ is a $p \times p$ diagonal covariance
matrix with $\sigma^2_j$ for $j = 1,\cdots,p$ on the diagonal.  In
many applications of factor analysis, the number of latent factors $k$
is much smaller than the number of features $p$ and the number of
samples $n$. Integrating over factor $\mb x_i$, this factor analysis
model produces a low-rank estimation of the feature covariance
matrix. In particular, the covariance of $\mb y_i$, $\mb \Omega \in
\mathbb{R}^{p \times p}$, is estimated as
\begin{align}
  \mb \Omega = \mb \Lambda \mb \Lambda^T + \mb \Sigma = \sum_{h=1}^k
  \mb \lambda_{\cdot h} \mb \lambda^T_{\cdot h} + \mb \Sigma, 
  \label{eq:FACov}
\end{align}
where $\mb \lambda_{\cdot h}$ is the $h^{th}$ column of $\mb
\Lambda$. This factorization suggests that each factor separately
contributes to the covariance of the sample through its
corresponding loading. Traditional exploratory data analysis methods
including principal component analysis
(PCA)~\citep{hotelling_analysis_1933}, independent component analysis
(ICA)~\citep{comon_independent_1994}, and canonical correlation
analysis (CCA)~\citep{hotelling_relations_1936} all have
interpretations as latent factor models. Indeed, the field of latent
variable models is extremely broad, and robust unifying frameworks are
desirable~\citep{Cunningham2014}.

Considering latent factor models (Equation~\ref{eq:FAModel}) as
capturing a low-rank estimate of the feature covariance matrix, we can
characterize canonical correlation analysis (CCA) as modeling paired
observations $\sups{\mb y}{1}_i \in \mathbb{R}^{p_1 \times 1}$ and
$\sups{\mb y}{2}_i \in \mathbb{R}^{p_2 \times 1}$ across $n$ samples
to identify a linear latent space for which the correlations between
the two observations are
maximized~\citep{hotelling_relations_1936,bach_probabilistic_2005}. The
Bayesian CCA (BCCA) model extends this covariance representation to
two observations: the combined loading matrix jointly models
covariance structure shared across both observations and covariance
local to each observation~\citep{klami_bayesian_2013}. Group factor
analysis (GFA) models further extend this representation to $m$
coupled observations for the same sample, modeling, in its fullest
generality, the covariance associated with every subset of
observations~\citep{virtanen_bayesian_2012, klami_group_2014}. GFA
becomes intractable when $m$ is large due to exponential explosion of
covariance matrices to estimate.

In a latent factor model, the loading matrix $\mb \Lambda$ plays an
important role in the subspace mapping. In applications where there
are fewer samples than features---the $n \ll p$
scenario~\citep{west_bayesian_2003}---it is essential to include
strong regularization on the loading matrix because the optimization
problem is under-constrained and has many equivalent solutions that
optimize the data likelihood. In the machine learning and statistics
literature, priors or penalties are used to regularize the elements of
the loading matrix, occasionally by inducing sparsity. Element-wise
sparsity corresponds to \emph{feature selection}. This has the effect
that a latent factor contributes to variation in only a subset of the
observed features, generating interpretable results
\citep{west_bayesian_2003, carvalho_high-dimensional_2008,
  knowles_nonparametric_2011}. For example, in gene expression
analysis, sparse factor loadings are interpreted as non-disjoint
clusters of co-regulated genes~\citep{pournara_factor_2007,
  lucas_latent_2010, gao_latent_2013}.
 
Imposing element-wise sparsity in latent factor models has been
studied through regularization via $\ell_1$ type
penalties~\citep{zou_sparse_2006, witten_penalized_2009,
  salzmann_factorized_2010}. More recently, Bayesian shrinkage methods
using sparsity-inducing priors have been introduced for latent factor
models~\citep{archambeau_sparse_2008, carvalho_high-dimensional_2008,
  virtanen_bayesian_2012, bhattacharya_sparse_2011,
  klami_bayesian_2013}.  The spike-and-slab
prior~\citep{mitchell_bayesian_1988}, the classic two-groups Bayesian
sparsity-inducing prior, has been used for sparse Bayesian latent
factor models~\citep{carvalho_high-dimensional_2008}. A
computationally tractable one-group prior, the automatic relevance
determination (ARD) prior \citep{neal_bayesian_1995,
  tipping_sparse_2001}, has also been used to induce sparsity in
latent factor models~\citep{engelhardt_analysis_2010,
  pruteanu-malinici_automatic_2011}.  More sophisticated structured
regularization approaches for linear models have been studied in
classical statistics~\citep{zou_regularization_2005,
  kowalski_structured_2009, jenatton_structured_2011,
  huang_learning_2011}.

Global structured regularization of the loading matrix, in fact, has
been used to extend latent factor models to multiple observations.
The BCCA model \citep{klami_bayesian_2013} assumes a latent factor
model for each observation through a shared latent vector $\mb x_i \in
\mathbb{R}^{k \times 1}$. This BCCA model may be written as a latent
factor model by vertical concatenation of observations, loading
matrices, and Gaussian residual errors. By inducing group-wise
sparsity---explicit blocks of zeros---in the combined loading matrix,
the covariance shared across the two observations and the covariance
local to each observation are
estimated~\citep{klami_probabilistic_2008, klami_bayesian_2013}.
Extensions of this approach to multiple coupled observations
$\sups{\mb y}{1}_i \in \mathbb{R}^{p_1 \times 1}, \cdots, \sups{\mb
  y}{m}_i \in \mathbb{R}^{p_m \times 1}$ have resulted in group factor
analysis models (GFA)~ \citep{archambeau_sparse_2008,
  salzmann_factorized_2010, jia_factorized_2010,
  virtanen_bayesian_2012}.

In addition to the linear factor models mentioned above, flexible
non-linear latent factor models have been developed. The Gaussian
process latent variable model (GPLVM)
\citep{lawrence_probabilistic_2005} extends Equation
(\ref{eq:FAModel}) to non-linear mappings with a Gaussian process
prior on latent variables. Extensions of GPLVM include models that
allow multiple observations~\citep{shon_learning_2005,
  ek_ambiguity_2008, salzmann_factorized_2010,
  damianou_manifold_2012}. Although our focus will be on linear maps,
we will keep the non-linear possibility open for model extensions, and
we will include the GPLVM model in our model comparisons on simulated
data.


The primary contribution of this study is that we develop a GFA model
using Bayesian shrinkage with hierarchical structure that encourages
both element-wise and column-wise sparsity; the resulting flexible
Bayesian GFA model is called BASS (Bayesian group factor Analysis with
Structured Sparsity). The structured sparsity in our model is achieved
with multi-scale application of a hierarchical sparsity-inducing prior
that has a computationally tractable representation as a scale mixture
of normals, the three parameter Beta prior
($\mathcal{TPB}$)~\citep{armagan_generalized_2011, gao_latent_2013}.
Our BASS model i) shrinks the loading matrix globally, removing
factors that are not supported in the data; ii) shrinks loading
columns to decouple latent spaces from arbitrary subsets of observations;
iii) allows factor loadings to have either an element-wise sparse or a
non-sparse prior, combining interpretability with
dimension reduction. In addition, we developed a parameter expanded
expectation maximization (PX-EM) method based on rotation augmentation
to tractably find \emph{maximum a posteriori} estimates of the model
parameters~\citep{rockova2014fast}. PX-EM has the same computational
complexity as the standard EM algorithm, but produces more robust
solutions by enabling fast searching over posterior modes.

In Section $2$ we review current work in sparse latent factor models
and describe our BASS model. In Sections $3$ and $4$, we briefly
review Bayesian shrinkage priors and introduce the structured
hierarchical prior in BASS.  In Section $5$, we introduce our PX-EM
algorithms for parameter estimation. In Section $6$, we show the
behavior of our model for recovering simulated sparse signals among
$m$ observation matrices and compare these results with
state-of-the-art methods.  In Section $7$, we present results that
illustrate the performance of BASS on three real data sets. We first
show that the estimates of shared factors from BASS can be used to
perform multi-label learning and prediction in the Mulan Library data
and the 20 Newsgroups data. Then we demonstrate that BASS can be used
to find biologically meaningful structure and construct
condition-specific co-regulated gene networks using the sparse factors
specific to observations.  We conclude by considering possible
extensions to this model in Section $8$.

\section{Bayesian Group Factor Model}
\label{sec:Model}

\subsection{Latent Factor Models}
\begin{figure}
  \centering
  \includegraphics[width=0.8\textwidth]{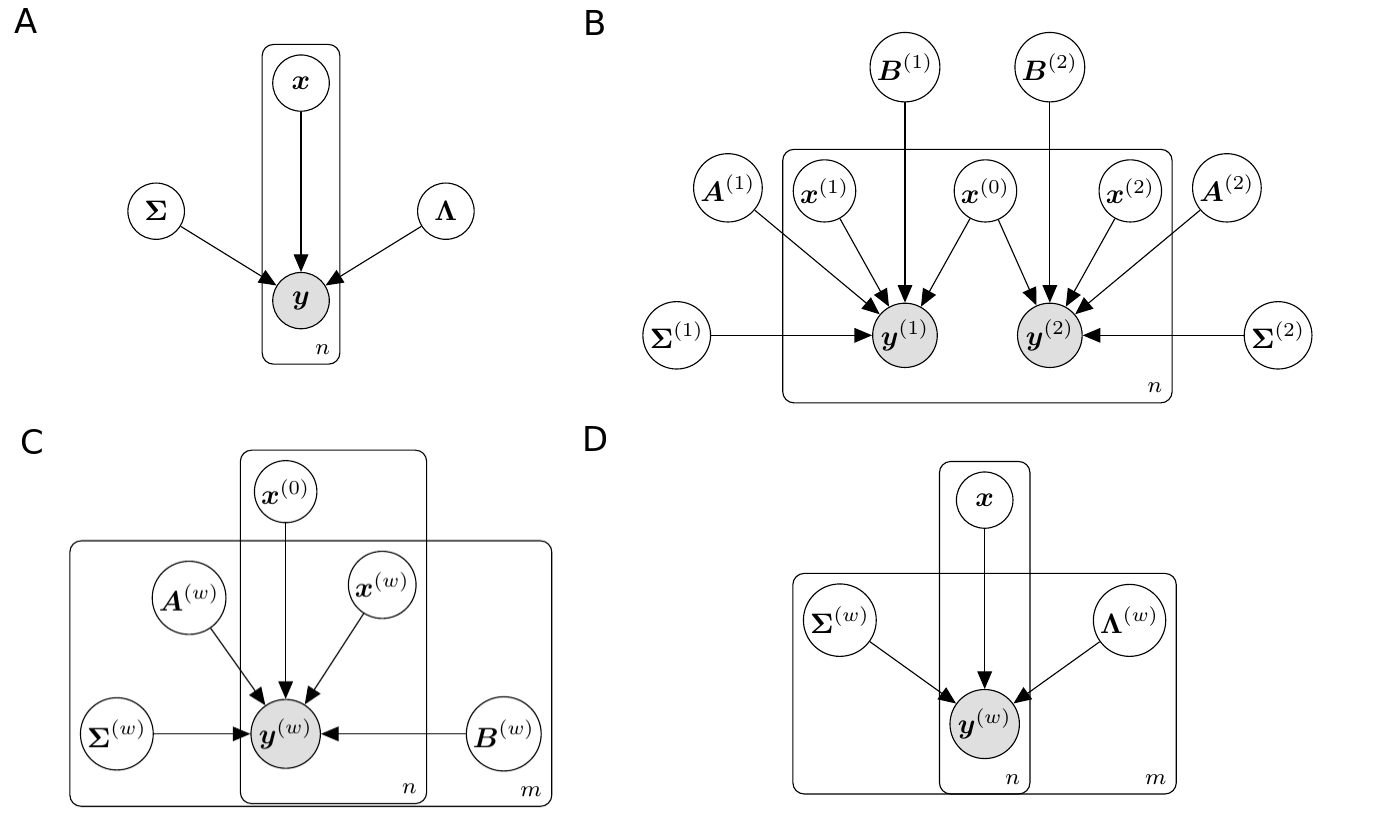}
  \caption{\textbf{Graphical representation of different latent
      factor models.} Panel A: Factor analysis model. Panel B:
    Bayesian canonical correlation analysis model (BCCA). Panel C:
    An extension of BCCA model to multiple observations. Panel D:
    Bayesian group factor analysis model developed in this work.}
  \label{fig:Graph}
\end{figure}

Factor analysis has been extensively used for dimension reduction and
low dimensional covariance matrix estimation. For concreteness, we
re-write the basic factor analysis model here as
\begin{align}
  \mb y_i = \mb \Lambda \mb x_i + \mb \epsilon_i, \notag
\end{align}
where $\mb y_i \in \mathbb{R}^{p \times 1}$ is modeled as a linear
transformation of a latent vector $\mb x_i \in \mathbb{R}^{k \times
  1}$ through loading matrix $\mb \Lambda \in \mathbb{R}^{p\times k}$
(Figure~\ref{fig:Graph}A). Here, $\mb x_i$ is assumed to follow a
$\N_k(\mb 0, \mb I_k)$ distribution, where $\mb I_k$ is the
$k$-dimensional identity matrix, and $\mb \epsilon_i \sim \N_p(\mb 0,
\mb \Sigma)$, where $\mb \Sigma$ is a $p \times p$ diagonal
matrix. With an isotropic noise assumption, $\mb \Sigma = \mb I
\sigma^2$, this model has a probabilistic principal components
analysis interpretation~\citep{roweis_em_1998,
  tipping_probabilistic_1999}. For factor analysis, and in this work,
it is assumed that $\mb \Sigma = \mbox{diag}(\sigma^2_1,\cdots, \sigma^2_p)$
representing independent idiosyncratic
noise~\citep{tipping_mixtures_1999}.

Integrating over the factors $\mb x_i$, we see that the covariance of
$\mb y_i$ is estimated with a low-rank matrix factorization: $\mb
\Lambda \mb \Lambda^T + \mb \Sigma$. We further let $\mb Y = [\mb y_1
, \cdots \mb y_n]$ be the collection of $n$ observations of $\mb y_i$, and
similarly let $\mb X = [\mb x_1, \cdots, \mb x_n]$ and $\mb E = [\mb
\epsilon_1, \cdots, \mb \epsilon_n]$. Then the factor analysis model
for the sample $\mb Y$ is written as
\begin{align}
  \mb Y = \mb \Lambda \mb X + \mb E. 
\end{align}

\subsection{Probabilistic Canonical Correlation Analysis}

In the context of two paired observations $\sups{\mb y}{1}_i \in
\mathbb{R}^{p_1 \times 1}$ and $\sups{\mb y}{2}_i \in \mathbb{R}^{p_2
  \times 1}$ on the same $n$ samples, canonical correlation analysis
(CCA) seeks to find linear projections (canonical directions) such
that the sample correlations in the projected space are mutually
maximized~\citep{hotelling_relations_1936}. The work of interpreting
CCA as a probabilistic model can be traced back to classical
descriptions by~\cite{bach_probabilistic_2005}. With a common latent
factor, $\mb x_i \in \mathbb{R}^{k\times 1}$, $\sups{\mb y}{1}_i$ and
$\sups{\mb y}{2}_i$ are modeled as
\begin{align}
  \sups{\mb y}{1}_i &= \sups{\mb \Lambda}{1} \mb x_i + \sups{\mb e}{1}_i, \notag \\
  \sups{\mb y}{2}_i &= \sups{\mb \Lambda}{2} \mb x_i + \sups{\mb
    e}{2}_i. \label{eq:ProbCCA}
\end{align}
In this model, the errors are distributed as $\sups{\mb e}{1}_i \sim
\N_{p_1} (\mb 0,\sups{\mb \Psi}{1})$ and $\sups{\mb e}{2}_i \sim
\N_{p_2} (\mb 0, \sups{\mb \Psi}{2})$, where $\sups{\mb \Psi}{1}$ and
$\sups{\mb \Psi}{2}$ are positive semi-definite matrices, and not
necessarily diagonal, allowing dependencies among the residual errors
within an observation. The maximum likelihood estimates of the loading
matrices in the classical CCA framework, $\sups{\mb \Lambda}{1}$ and
$\sups{\mb \Lambda}{2}$, are the first $k$ canonical directions up to
orthogonal transformations~\citep{bach_probabilistic_2005}.

\subsection{Bayesian CCA with Group-wise Sparsity}

Building on the probabilistic CCA model, a Bayesian CCA (BCCA) model
has the following form~\citep{klami_bayesian_2013}
\begin{align}
  \sups{\mb y}{1}_i &= \sups{\mb A}{1} \sups{\mb x}{0}_i + \sups{\mb
    B}{1} \sups{\mb x}{1}_i +
  \sups{\mb \epsilon}{1}_i, \notag \\
  \sups{\mb y}{2}_i &= \sups{\mb A}{2} \sups{\mb x}{0}_i + \sups{\mb
    B}{2} \sups{\mb x}{2}_i + \sups{\mb
    \epsilon}{2}_i, \label{eq:BCCA}
\end{align}
with $\sups{\mb x}{0}_i \in \mathbb{R}^{k_0 \times 1}$, $\sups{\mb
  x}{1}_i \in \mathbb{R}^{k_1 \times 1}$ and $\sups{\mb x}{2}_i \in
\mathbb{R}^{k_2 \times 1}$ (Figure~\ref{fig:Graph}B). The latent
vector $\sups{\mb x}{0}_i$ is shared by both $\sups{\mb y}{1}_i$ and
$\sups{\mb y}{2}_i$, and captures their common variation through
loading matrices $\sups{\mb A}{1}$ and $\sups{\mb A}{2}$. Two
additional latent vectors, $\sups{\mb x}{1}_i$ and $\sups{\mb
  x}{2}_i$, are specific to each observation; they are multiplied by
observation-specific loading matrices $\sups{\mb B}{1}$ and $\sups{\mb
  B}{2}$. The two residual error terms are $\sups{\mb \epsilon}{1}_i
\sim \N_{p_1}(\mb 0, \sups{\mb \Sigma}{1})$ and $\sups{\mb
  \epsilon}{2}_i \sim \N_{p_2}(\mb 0, \sups{\mb \Sigma}{2})$, where
$\sups{\mb \Sigma}{1}$ and $\sups{\mb \Sigma}{2}$ are diagonal
matrices. This model was originally called inter-battery factor
analysis (IBFA) \citep{browne_maximum-likelihood_1979} and recently
has been studied under a full Bayesian inference framework
\citep{klami_bayesian_2013}. It may be interpreted as the
probabilistic CCA model (Equation~\ref{eq:ProbCCA}) with an additional
low-rank factorization of the observation-specific error covariance
matrices.  In particular, we re-write the residual error term specific
to observation $w$ ($w = 1, 2$) from the probabilistic CCA model
(Equation~\ref{eq:ProbCCA}) as $\sups{\mb e}{w}_i = \sups{\mb B}{w}
\sups{\mb x}{w}_i + \sups{\mb \epsilon}{w}_i$; then marginally
$\sups{\mb e}{w}_i \sim \N_{p_w}(\mb 0, \sups{\mb \Psi}{w})$ where
$\sups{\mb \Psi}{w} = \sups{\mb B}{w} (\sups{\mb B}{w})^T + \sups{\mb
  \Sigma}{w}$.

\cite{klami_bayesian_2013} re-wrote the BCCA model as a factor
analysis model with group-wise sparsity in the loading matrix. Let
$\mb y_i \in \mathbb{R}^{p \times 1}$ (where $p = p_1 + p_2$) be the
vertical concatenation of $\sups{\mb y}{1}_i$ and $\sups{\mb y}{2}_i$;
let $\mb x_i \in \mathbb{R}^{k \times 1}$ (where $k = k_0 + k_1 + k_2$)
be the vertical concatenation of $\sups{\mb x}{0}_i$, $\sups{\mb
  x}{1}_i$ and $\sups{\mb x}{2}_i$; and let $\mb \epsilon_i \in
\mathbb{R}^{p \times 1}$ be the vertical concatenation of the two
residual errors. Then, the BCCA model (Equation~\ref{eq:BCCA}) may be
written as a factor analysis model
\begin{align}
  \mb y_i = \mb \Lambda \mb x_i + \mb \epsilon_i, \notag
\end{align}
with $\mb \epsilon_i \sim N_p(\mb 0, \mb \Sigma)$, where
\begin{align}
  \mb \Lambda = \begin{bmatrix}
    \sups{\mb A}{1} & \sups{\mb B}{1} & \mb 0 \\
    \sups{\mb A}{2} & \mb 0 & \sups{\mb B}{2}
  \end{bmatrix}, \quad \mb \Sigma = \begin{bmatrix}
    \sups{\mb \Sigma}{1} & \mb 0 \\
    \mb 0 & \sups{\mb \Sigma}{2}
  \end{bmatrix}. \label{eq:GWSCCA}
\end{align}
The structure in the loading matrix $\mb \Lambda$ has a specific
meaning: the non-zero columns (those in $\sups{\mb A}{1}$ and
$\sups{\mb A}{2}$) project the shared latent factors (i.e., the first
$k_0$ elements of $\mb x_i$) to $\sups{\mb y}{1}_i$ and $\sups{\mb
  y}{2}_i$, respectively; these latent factors represent the
covariance shared across the observations. The columns with zero
blocks (those in $[\sups{\mb B}{1}; \mb 0]$ or $[\mb 0; \sups{\mb
    B}{2}]$) relate specific factors to only one of the two
observations; they model covariance specific to that
observation. Under this model, the sparsity structure of $\mb \Lambda$
is estimated via factor-wise shrinkage.

\subsection{Extensions to Multiple Observations}

Classical and Bayesian extensions of the CCA model to allow multiple
observations ($m>2$) have been proposed~\citep{mcdonald_three_1970,
  browne_factor_1980, archambeau_sparse_2008,qu_sparse_2011,
  ray_bayesian_2014}. Generally, these approaches partition the latent
variables into those that are shared and those that are
observation-specific as follows:
\begin{align}
  \sups{\mb y}{w}_i = \sups{\mb A}{w} \sups{\mb x}{0}_i + \sups{\mb
    B}{w} \sups{\mb x}{w}_i + \sups{\mb \epsilon}{w}_i \quad
  \mbox{for} \quad w = 1,\cdots,m. \label{eq:GFA}
\end{align}
By vertical concatenation of $\sups{\mb y}{w}_i$, $\sups{\mb x}{w}_i$
and $\sups{\mb \epsilon}{w}_i$, this model can be viewed as a latent
factor model (Equation~\ref{eq:FAModel}) with the joint loading matrix
$\mb \Lambda$ having a similar group-wise sparsity pattern as the
BCCA model
\begin{equation}
  \mb \Lambda = \begin{bmatrix}
    \sups{\mb A}{1} & \sups{\mb B}{1} & \cdots & \mb 0 \\
    \sups{\mb A}{2} & \mb 0 & \cdots & \mb 0 \\
    \vdots & \vdots & \ddots & \vdots \\
    \sups{\mb A}{m} & \mb 0 & \cdots & \sups{\mb B}{m} 
  \end{bmatrix}. \label{eq:GFA_lambda}
\end{equation}
Here, the first block column ($\sups{\mb A}{w}$) is a non-zero loading
matrix across the features of all observations, and the remaining
columns have a block diagonal structure with observation-specific
loading matrices ($\sups{\mb B}{w}$) on the diagonal. However, those
extensions are limited by the strict diagonal structure of the loading
matrix. Structuring the loading matrix in this way prevents this model
from capturing covariance structure among arbitrary subsets of
observations. On the other hand, there are an exponential number of
possible subsets of observations, making estimation of covariance
structure among all observation subsets intractable for large $m$.

The structure on $\mb \Lambda$ in Equation (\ref{eq:GFA_lambda}) has
been relaxed to model covariance among subsets of the observations
~\citep{jia_factorized_2010, virtanen_bayesian_2012,
  klami_group_2014}. In the relaxed formulation, each observation
$\sups{\mb y}{w}_i$ is modeled by its own loading matrix $\sups{\mb
  \Lambda}{w}$ and a shared latent vector $\mb x_i$
(Figure~\ref{fig:Graph}D):
\begin{align}
  \sups{\mb y}{w}_i = \sups{\mb \Lambda}{w} \mb x_i + \sups{\mb
    \epsilon}{w}_i \quad \mbox{for} \quad w =
  1,\dots,m. \label{eq:BGFA}
\end{align}
By allowing columns in $\sups{\mb \Lambda} {w}$ to be zero, the model
decouples certain latent factors from certain observations.  The
covariance structure of an arbitrary subset of observations is modeled
by factors with non-zero loading columns corresponding to the
observations in that subset. Factors that correspond to non-zero
entries for only one observation capture covariance specific to that
observation.  Two different approaches have been proposed to achieve
column-wise shrinkage in this framework: Bayesian
shrinkage~\citep{virtanen_bayesian_2012, klami_group_2014} and
explicit penalties~\citep{jia_factorized_2010}. The group factor
analysis (GFA) model puts an ARD
prior~\citep{tipping_sparse_2001} on the loading column for each
observation to allow column-wise
shrinkage~\citep{virtanen_bayesian_2012, klami_group_2014}:
\begin{align}
  \sups{\lambda}{w}_{j h} &\sim \N \left( 0, \left(\sups{\alpha}{w}_h\right)^{-1}
  \right) \quad \mbox{for} \quad j = 1,\dots, p_w \notag, \\ 
  \quad \sups{\alpha}{w}_h &\sim Ga(a_0,b_0) \notag,
\end{align}
for observation $w = 1, \dots, m$ and $h = 1, \dots, k$. This prior
assumes that each element of each observation-specific loading
$\sups{\lambda}{w}_{\cdot h}$ is jointly regularized. 
This prior encourages the posterior probability of
$\sups{\alpha}{w}_h$ to take on large values and also values near
zero, pushing all elements of $\sups{\lambda}{w}_{\cdot h}$ toward
zero or imposing minimal shrinkage, enabling observation-specific,
column-wise sparsity.

Other work puts alternative structured regularizations on $\sups{\mb
  \Lambda}{w}$~\citep{jia_factorized_2010}.  To induce
observation-specific, column-wise sparsity they used mixed norms: an
$\ell_1$ norm penalizes each observation-specific column, and either
$\ell_2$ or $\ell_\infty$ norms penalize the elements in an
observation-specific column:
\begin{align}
  \phi(\sups{\mb \Lambda}{w}) = \sum_{h=1}^k || \sups{\mb
    \lambda}{w}_{\cdot h} ||_2 \qquad \mbox{or} \qquad  \phi(\sups{\mb \Lambda}{w}) = \sum_{h=1}^k ||
  \sups{ \mb \lambda}{w}_{\cdot h} ||_\infty. \notag
\end{align}
The $\ell_1$ norm penalty achieves observation-specific column-wise
shrinkage. Both of these mixed norm penalties create a bi-convex
problem in $\mb \Lambda$ and $\mb X$.

These two approaches of adaptive structured regularization in GFA
models capture covariance uniquely shared among arbitrary subsets of
the observations and avoid modeling shared covariance in non-maximal
subsets. But neither the ARD approach nor the mixed norm penalties
encourages element-wise sparsity within loading columns. Adding
element-wise sparsity is important because it results in interpretable
latent factors, where features with non-zero loadings in a specific
factor have an interpretation as a cluster~\citep{west_bayesian_2003,
  carvalho_high-dimensional_2008}.  To induce element-wise sparsity,
one can either use Bayesian shrinkage on individual
loadings~\citep{carvalho_horseshoe_2010} or a mixed norm with $\ell_1$
type penalties on each element (i.e., $\sum_{h=1}^k \sum_{j=1}^p
|\sups{\lambda}{w}_{jh}|$). 

A more recent GFA model is a step toward both column-wise and
element-wise sparsity~\citep{khan_identification_2014}. In this model,
element-wise sparsity is achieved by using independent ARD priors on
each loading element and column-wise sparsity is achieved by a
spike-and-slab type prior on loading columns. However, independent ARD
priors do not allow the model to adjust shrinkage levels within each
factor, and this approach does not separate sparse and dense factors.
One contribution of this work is to define a carefully structured
Bayesian shrinkage prior on the loading matrix of a GFA model that
encourages both element-wise and column-wise shrinkage, and that
models sparse and dense factors jointly.

%

\section{Bayesian Structured Sparsity}

The column-wise sparse structure of $\mb \Lambda$ in GFA models
belongs to a general class of structured sparsity methods that has
drawn attention recently~\citep{zou_regularization_2005,
  yuan_model_2006, jenatton_structured_2011, jenatton_structured_2009,
  kowalski_sparse_2009, kowalski_structured_2009, zhao_composite_2009,
  huang_learning_2011, jia_factorized_2010}. For example, in
structured sparse PCA, the loading matrix is constrained to have
specific patterns~\citep{jenatton_structured_2009}.  Later
\cite{jenatton_structured_2011} and \cite{huang_learning_2011}
discussed more general structured variable selection methods in a
regression framework. However, there has been little work in using
Bayesian structured sparsity, with some
exceptions~\citep{kyung_penalized_2010, engelhardt_bayesian_2014}).
Starting from Bayesian sparse priors, we propose a structured
hierarchical sparse prior that includes three levels of shrinkage,
which is conceptually similar to tree structured
shrinkage~\citep{romberg_bayesian_2001}, or global-local priors in the
regression framework~\citep{polson_shrink_2010}.

\subsection{Bayesian Sparsity-Inducing Priors}

Bayesian shrinkage priors have been widely used in latent factor
models due to their flexible and interpretable
solutions~\citep{west_bayesian_2003, carvalho_high-dimensional_2008,
  polson_shrink_2010, knowles_nonparametric_2011,
  bhattacharya_sparse_2011}. In Bayesian statistics, a regularizing
term, $\phi(\mb \Lambda)$, may be viewed as a marginal prior
proportional to $\exp(-\phi(\mb \Lambda))$; the regularized optimum
then becomes the maximum a posteriori (MAP)
solution~\citep{polson_shrink_2010}. For example, the well known
$\ell_2$ penalty for coefficients in linear regression models
corresponds to Gaussian priors, also known as ridge regression or
Tikhonov regularization~\citep{hoerl_ridge_1970}. In contrast, an
$\ell_1$ penalty corresponds to double exponential or Laplace
priors, also known as the Bayesian
Lasso~\citep{tibshirani_regression_1996, park_bayesian_2008,
  Hans2009}.

When the goal of regularization is to induce sparsity, the prior
distribution should be chosen so that it has substantial probability
mass around zero, which draws small effects toward zero, and heavy
tails, which allows large signals to escape from substantial
shrinkage~\citep{o1979outlier, carvalho_horseshoe_2010,
  armagan_generalized_2011}. The canonical Bayesian sparsity-inducing
prior is the spike-and-slab prior, which is a mixture of a point mass
at zero and a flat distribution across the space of real values, often
modeled as a Gaussian with a large variance term
\citep{mitchell_bayesian_1988, west_bayesian_2003}. The spike-and-slab
prior has elegant interpretability by estimating the probability that
certain loadings are excluded, modeled by the `spike' distribution, or
included, modeled by the `slab' distribution
\citep{carvalho_high-dimensional_2008}. This interpretability comes at
the cost of having exponentially many possible configurations of model
inclusion parameters in the loading matrix.

Recently, scale mixtures of normal priors have been proposed as a
computationally efficient alternative to the two component
spike-and-slab prior~\citep{west_scale_1987, carvalho_horseshoe_2010,
  polson_shrink_2010, armagan_generalized_Pareto_2011,
  armagan_generalized_2011, bhattacharya_bayesian_2012}: such priors
generally assume normal distributions with a mixed variance term. The
mixing distribution of the variance allows strong shrinkage near zero
but weak regularization away from zero.  For example, the
inverse-gamma distribution on the variance term results in an ARD
prior~\citep{tipping_sparse_2001}, and an exponential distribution on
the variance term results in a Laplace
prior~\citep{park_bayesian_2008}. The horseshoe prior, with a half
Cauchy distribution on the standard deviation as the mixing density,
has become popular due to its strong shrinkage and heavy tails
~\citep{carvalho_horseshoe_2010}. 

A more general class of beta mixtures of normals is the three
parameter beta distribution~\citep{armagan_generalized_2011}. Although
these continuous shrinkage priors do not directly model the
probability of feature inclusion, it has been shown in the regression
framework that two layers of regularization---global
regularization, across all coefficients, and local regularization,
specific to each coefficient~\citep{polson_shrink_2010})---has behavior
that is similar to the spike-and-slab prior in effectively modeling
signal and noise separately, but with computational
tractability~\citep{carvalho_handling_2009}. In this study, we extend
and structure the beta mixture of normals prior to three levels of
hierarchy to induce desirable behavior in the context of GFA models.

\subsection{Three Parameter Beta Prior}

\begin{figure}
  \centering
  \includegraphics[width=0.7\textwidth]{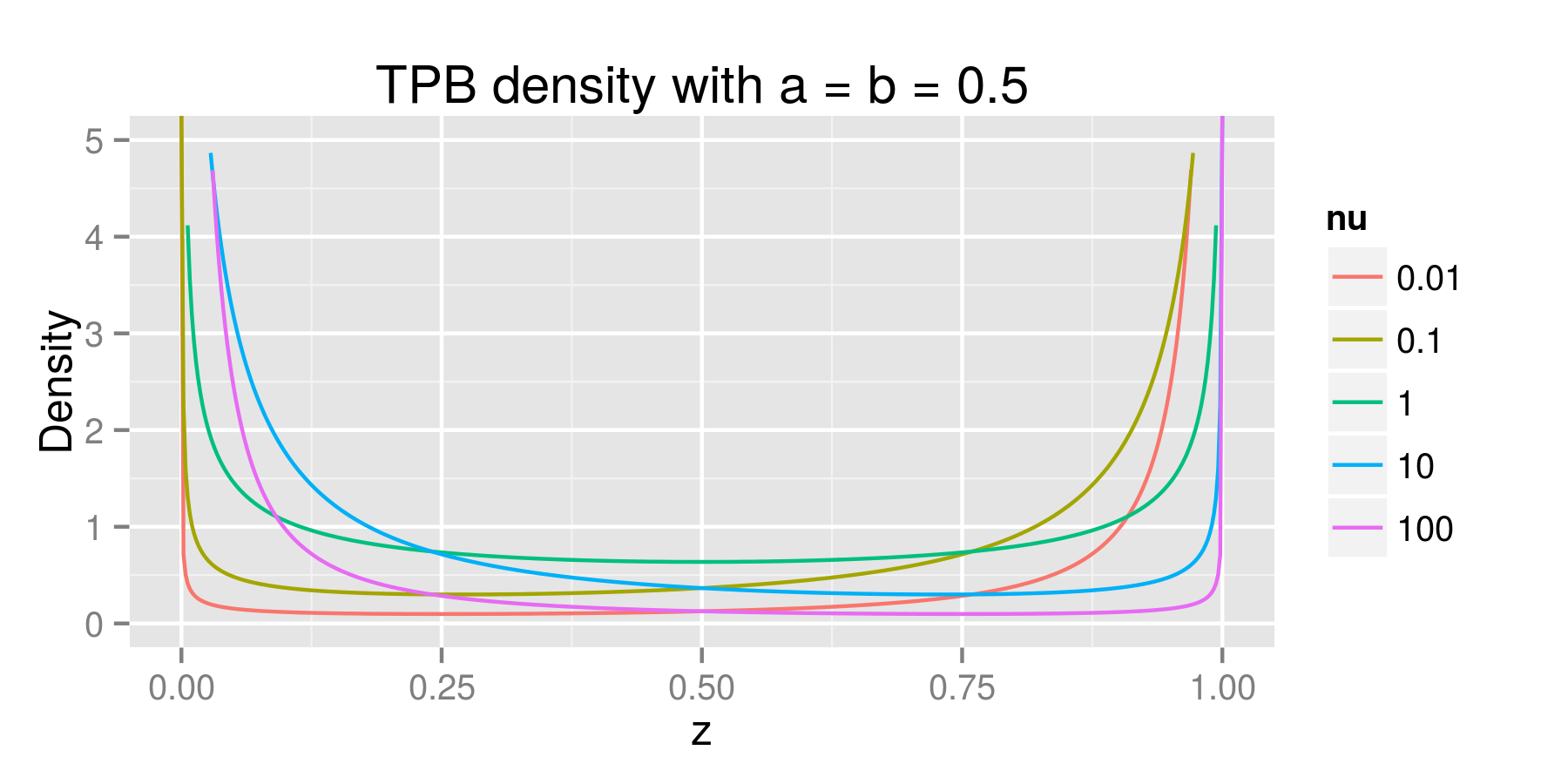}
  \caption{{\bf Density of three parameter beta ($\TPB$) distribution
      with different values of $\nu$.}}
  \label{fig:TPBpdf}
\end{figure}

The three parameter beta ($\TPB$) distribution for a random variable
$Z \in (0,1)$ has the following density \citep{armagan_generalized_2011}:
\begin{align}
  f(z; a,b,\nu) = \frac{\Gamma(a+b)}{\Gamma{(a)} \Gamma{(b)}} \nu^b
  z^{b-1} (1-z)^{a-1} \{1 + (\nu - 1)z\}^{-(a+b)}, \label{TPBdist}
\end{align}
where $a, b, \phi>0$. We denote this distribution as
$\mathcal{TPB}(a,b,\nu)$.  When $0<a<1$ and $0<b<1$, the distribution
is bimodal, with modes at $0$ and $1$ (Figure \ref{fig:TPBpdf}). The
\emph{variance parameter} $\nu$ gives the distribution freedom: with
fixed $a$ and $b$, smaller values of $\nu$ put greater probability on
$z=1$, while larger values of $\nu$ move the probability mass towards
$z=0$ \citep{armagan_generalized_2011}.  With $\nu=1$, this
distribution is identical to a beta distribution, $Be(b,a)$.

Let $\lambda$ denote the parameter to which we are applying
sparsity-inducing regularization. We assign the following
$\mathcal{TPB}$ normal scale mixture distribution, $\mathcal{TPBN}$,
to $\lambda$:
\begin{align}
  \lambda | \varphi \sim \N\left(0,\frac 1{\varphi} - 1\right), \qquad \mbox{with} \qquad
  \varphi \sim \mathcal{TPB}(a,b,\nu), \notag
\end{align}
where the \emph{shrinkage parameter} $\varphi$ follows a $\TPB$
distribution.  With $a = b = 1/2$ and $\nu = 1$, this prior becomes
the horseshoe prior~\citep{carvalho_horseshoe_2010,
  armagan_generalized_2011, gao_latent_2013}. The bimodal property of
$\varphi$ induces two distinct shrinkage behaviors: the mode near one
encourages $\frac 1{\varphi} - 1$ towards zero and induces strong
shrinkage on $\lambda$; the mode near zero encourages $\frac
1{\varphi} - 1$ large, creating a diffuse prior on $\lambda$. Further
decreasing the variance parameter $\nu$ puts more support on stronger
shrinkage \citep{armagan_generalized_2011, gao_latent_2013}.  If we
let $\theta = \frac 1{\varphi} - 1$, then this mixture has the
following hierarchical representation:
\begin{align}
  \lambda \sim \N(0,\theta), \quad \theta \sim Ga(a,\delta), \quad
  \delta \sim Ga(b,\nu). \notag
\end{align}
Note the difference between the ARD prior and the $\mathcal{TPB}$: the
ARD prior induces sparsity using an inverse gamma prior on $\theta$,
whereas the $\mathcal{TPB}$ induces sparsity by using a gamma prior on
the $\theta$ variable and then regularizing the rate parameter
$\delta$ using a second gamma prior. These differences lead to
different behavior of ARD and the $\mathcal{TPB}$
in theory~\citep{polson_shrink_2010} and in practice, as we show
below.

\subsection{Global-Factor-Local Shrinkage}

The flexible representation of the $\TPB$ prior makes it an ideal
choice for latent factor models.  \cite{gao_latent_2013} extended the
$\TPB$ prior to three levels of regularization on a
loading matrix:
\begin{alignat}{3}
  \varrho  &\sim \TPB(e,f,\nu), && \qquad  \textrm{Global}  \notag \\
  \zeta_h &\sim \TPB\left(c,d,\frac 1{\varrho} - 1\right), && \qquad \textrm{Factor-specific}  \notag \\
  \varphi_{jh} &\sim \TPB\left(a,b,\frac 1{\zeta_h} -1\right), && \qquad \textrm{Local}  \notag \\
  \lambda_{jh} &\sim \N\left(0, \frac 1{\varphi_{jh}} -
  1\right). && \label{eq:3LevelTPB1}
\end{alignat}
At each of the three levels, a $\TPB$ distribution is used to induce
sparsity via its estimated variance parameter ($\nu$ in Equation
\ref{TPBdist}), which in turn is regularized using a $\TPB$
distribution. Specifically, the global shrinkage parameter $\varrho$
applies strong shrinkage across the $k$ columns of the loading matrix
and jointly adjusts the support of column-specific parameter
$\zeta_h$, $h \in \{1,\dots,k\}$ at either zero or one. This can be
interpreted as inducing sufficient shrinkage across loading columns to
recover the number of factors supported by the observed data. In
particular, when $\zeta_h$ is close to one, all elements of column $h$
are close to zero, effectively removing the $h^{th}$ component. When
near zero, the factor-specific regularization parameter $\zeta_h$
adjusts the shrinkage applied to each element of the $h^{th}$ loading
column, estimating the column-wise shrinkage by borrowing strength
across all elements in that column. The local shrinkage parameter,
$\varphi_{jh}$, creates element-wise sparsity in the loading matrix
through a $\TPBN$. Three levels of shrinkage allow us to model both
column-wise and element-wise shrinkage simultaneously, and give the
model nonparametric behavior in the number of factors via model
selection.

Equivalently, this global-factor-local shrinkage prior can be written
as:
\begin{alignat}{3}
  \textrm{Global} \quad &
  \begin{cases}
    \gamma \sim Ga(f,\nu), \notag \\
    \eta \sim Ga(e,\gamma)), \notag 
  \end{cases} \notag \\
  \textrm{Factor-specific} \quad &
  \begin{cases}
    \tau_h \sim Ga(d,\eta), \notag \\
    \phi_h \sim Ga(c,\tau_h), \notag 
  \end{cases} \notag \\
  \textrm{Local} \quad &
  \begin{cases}
    \delta_{jh} \sim Ga(b,\phi_h), \notag \\
    \theta_{jh} \sim Ga(a,\delta_{jh}), \notag
  \end{cases} \notag \\
  \lambda_{jh} &\sim \N(0,\theta_{jh}). \label{eq:3LevelTPB2}
\end{alignat}

We further extend our prior to jointly model sparse and dense
components by assigning to the local shrinkage parameter a
two-component mixture distribution~\citep{gao_latent_2013}:
\begin{align}
  \theta_{jh} &\sim \pi Ga(a,\delta_{jh}) + (1-\pi)
  \delta_{\phi_h}(\cdot), \quad
  \label{eq:3LevelTPB2z}
\end{align}
where $\delta_{\phi_h}(\cdot)$ is the Dirac delta function centered at
$\phi_h$. The motivation for this two component mixture is that, in
real applications such as the analysis of gene expression data, it has
been shown that much of the variation in the observation is due to
technical (e.g., batch) or biological effects (e.g., sex, ethnicity),
which impact a large number of features~\citep{Leek2010}. Therefore,
loadings corresponding to these effects will often not be sparse. A
two-component mixture (Equation~\ref{eq:3LevelTPB2z}) allows the prior
on the loading (Equation~\ref{eq:3LevelTPB1}) to select between
element-wise sparsity or column-wise sparsity. Element-wise sparsity
is encouraged via the $\TPBN$ prior. Column-wise sparsity jointly
regularizes each element of the column with a shared variance term:
$\lambda_{jh} \sim \N\left(0,\frac 1{\zeta_h} - 1\right)$. Modeling
each element in a column using a shared regularized variance term has
two possible behaviors: i) $\zeta_h$ in Equation (\ref{eq:3LevelTPB1})
is close to $1$ and the entire column is shrunk towards zero,
effectively removing this factor; ii) $\zeta_h$ is close to zero, and
all elements of the column have a shared Gaussian distribution,
inducing only non-zero elements in that loading. We call included
factors that have only non-zero elements \emph{dense factors}.

Jointly modeling sparse and dense factors effectively combines
low-rank covariance factorization with interpretability
\citep{zou_sparse_2006, parkhomenko_sparse_2009}. The dense factors
capture the broad effects of observation confounders, model a low-rank
approximation of the covariance matrix, and usually account for a
large proportion of variance
explained~\citep{chandrasekaran_rank-sparsity_2011}. The sparse
factors, on the other hand, capture the small groups of interacting
features in a (possibly) high dimensional sparse space, and usually
account for a small proportion of the variance explained.

We introduce indicator variables $z_h$, $h = 1, \dots, k$, to indicate
which mixture component each $\theta_{jh}$ is generated from Equation
~(\ref{eq:3LevelTPB2z}), where $z_h = 1$ means $\theta_{jh} \sim Ga(a,
\delta_{jh})$ and $z_h = 0$ means $\theta_{jh} \sim
\delta_{\phi_h}(\cdot)$. Thus, a component is a sparse factor when
$z_h=1$ and either a dense factor or eliminated when $z_h=0$. We let
$\mb z = [z_1,\dots,z_k]$ and put a Bernoulli distribution with
parameter $\pi$ on $z_h$. We further let $\pi$ have a flat beta
distribution $Be(1,1)$. This construct allows us to quantify the
posterior probability that each factor $h$ is generated from each
mixture component type via $z_h$.

\section{Bayesian Group Factor Analysis with Structured Sparsity}

In this work, we use global-factor-local $\TPB$ priors in the GFA
model to enable both element-wise and column-wise
shrinkage. Specifically, we put a $\TPB$ prior independently on each
loading matrix corresponding to the $w^{th}$ observation, $\sups{\mb
  \Lambda}{w}$.  Let $\mb Z = [\sups{\mb z}{1}; \dots ; \sups{\mb
    z}{m}] \in \mathbb{R}^{m \times k}$. The indicator variable
$\sups{z}{w}_h$ is associated with the $h^{th}$ factor and specific to
observation $w$.  When $\sups{z}{w}_h = 1$, the $h^{th}$ factor has a
sparse loading for observation $w$; when $\sups{z}{w}_h = 0$, then
either the $h^{th}$ factor has a dense loading column for observation
$w$, or observation $w$ is not represented in that loading column. A
zero loading column for observation $w$ effectively decouples the
factor from that observation, leading to the column-wise sparse
behavior in previous GFA models~\citep{virtanen_bayesian_2012,
  klami_group_2014}. In our model, factors that include no
observations in the associated loading column are removed from the
model. We refer to this model as Bayesian group factor Analysis with
Structured Sparsity (\emph{BASS}).

We summarize BASS as follows. The generative model for $m$ coupled
observations $\sups{\mb y}{w}_i$ with $w = 1,\dots,m$ and $i =
1,\cdots, n$ is
\begin{align}
  \sups{\mb y}{w}_i = \sups{\mb \Lambda}{w} \mb x_i + \sups{\mb
    \epsilon}{w}_i, \quad \textrm{for} \quad w = 1,\dots,m. \notag
\end{align}
This model is written as a latent factor model by concatenating the $m$
feature vectors into vector $\mb y_i$
\begin{align}
  \mb y_i &= \mb \Lambda \mb x_i + \mb \epsilon_i, \notag \\
  \mb x_i &\sim \N_k(0,\mb I_k), \notag \\
  \mb \epsilon_i &\sim \N_p(0, \mb \Sigma), \label{eq:GFAfull}
\end{align}
where $\mb \Sigma = \mbox{diag}(\sigma^2_1,\cdots,\sigma^2_p)$. We put
independent global-factor-local $\TPB$ priors (Equation~\ref{eq:3LevelTPB2}) on $\sups{\mb \Lambda}{w}$:
\begin{alignat}{3}
  \textrm{Global} \quad &
  \begin{cases}
    \sups{\gamma}{w} \sim Ga(f,\nu), \notag \\
    \sups{\eta}{w} \sim Ga(e,\sups{\gamma}{w})), \notag 
  \end{cases} \notag \\
  \textrm{Factor-specific} \quad &
  \begin{cases}
    \sups{\tau}{w}_h \sim Ga(d,\sups{\eta}{w}), \notag \\
    \sups{\phi}{w}_h \sim Ga(c,\sups{\tau}{w}_h), \notag 
  \end{cases} \notag \\
  \textrm{Local} \quad &
  \begin{cases}
    \sups{\delta}{w}_{jh} \sim Ga(b,\sups{\phi}{w}_h), \notag \\
    \sups{\theta}{w}_{jh} \sim Ga(a,\sups{\delta}{w}_{jh}), \notag
  \end{cases} \notag \\
  \sups{\lambda}{w}_{jh} &\sim
  \N(0,\sups{\theta}{w}_{jh}). \notag
\end{alignat}
We allow local shrinkage to follow a two-component mixture
\begin{align}
  \sups{\theta}{w}_{jh} &\sim \sups{\pi}{w}
  Ga(a,\sups{\delta}{w}_{jh}) + (1- \sups{\pi}{w})
  \delta_{\sups{\phi}{w}_h}(\cdot), \notag
\end{align}
where the mixture proportion has a Beta distribution
\begin{align}
  \sups{\pi}{w} \sim Be(1,1). \notag
\end{align}
We put a conjugate inverse gamma distribution on the residual variance parameters
\begin{align}
  \sigma^{-2}_j & \sim Ga(a_\sigma,b_\sigma). \notag
\end{align}

In our application of this model, we set the hyperparameters of the
global-factor-local $\TPB$ prior to $a = b = c = d = e = f = 0.5$,
which recapitulates the horseshoe prior at all three levels of the
hierarchy. The hyperparameters for the error variances, $a_{\sigma}$
and $b_{\sigma}$, were set to $1$ and $0.3$ respectively to allow a
relatively wide support of variances~\citep{bhattacharya_sparse_2011}.
When there are only two coupled observations, the BASS framework is
identical to a Bayesian CCA model (Equation~\ref{eq:BCCA}) due to its
column-wise shrinkage.

\section{Parameter Estimation}

Given our setup, the full joint distribution of the BASS model
factorizes as
\begin{align}
  p(\mb Y, \mb X, &\mb \Lambda, \mb \Theta, \mb \Delta, \mb \Phi, \mb
  T, \mb \eta, \mb \gamma, \mb Z, \mb \Sigma, \mb \pi) \notag \\
  &= p(\mb Y|\mb \Lambda, \mb X, \mb \Sigma) p(\mb X)  \notag \\
  & \qquad \times p(\mb \Lambda | \mb \Theta) p(\mb \Theta | \mb
  \Delta, \mb Z, \mb \Phi) p(\mb \Delta | \mb \Phi) p(\mb \Phi | \mb
  T) p(\mb T| \mb \eta) p(\mb \eta | \mb \gamma) \notag \\
  & \qquad \times p(\mb \Sigma) p(\mb Z|\mb \pi) p(\mb
  \pi), \label{eq:jointFull}
\end{align}
where $\mb \Theta = \{\sups{\theta}{w}_{jh}\}$, $\mb \Delta =
\{\sups{\delta}{w}_{jh}\}$, $\mb \Phi = \{\sups{\phi}{w}_h\}$, $\mb T
= \{\sups{\tau}{w}_h \}$, $\mb \eta = \{\sups{\eta}{w}\}$ and $\mb
\gamma = \{\sups{\gamma}{w}\}$ are the collections of the
global-factor-local $\TPB$ prior parameters. The posterior
distributions of model parameters may be either simulated through
Markov chain Monte Carlo (MCMC) methods or approximated using
variational Bayes approaches. We derive an MCMC algorithm based on a
Gibbs sampler (Appendix~\ref{app:MCMC}). The MCMC algorithm updates
the joint loading matrix row by row using block updates, enabling
relatively fast mixing~\citep{bhattacharya_sparse_2011}. 

In many applications, we are interested in a single point estimate of
the parameters instead of the complete posterior estimate; thus, often
an expectation maximization (EM) algorithm is used to find a
\emph{maximum a posteriori} (MAP) estimate of model parameters using
conjungate gradient optimization~\citep{dempster_maximum_1977}. In EM,
the latent factors $\mb X$ and the indicator variables $\mb Z$ are
treated as missing data and their expectations estimated in the E-step
conditioned on the current values of the parameters; then the model
parameters are optimized in the M-step conditioning on the current
expectations of the latent variables. Let $\mb \Xi = \{\mb \Lambda,
\mb \Theta, \mb \Delta, \mb \Phi, \mb T, \mb \eta, \mb \gamma, \mb
\pi, \mb \Sigma \}$ be the collection of the parameters optimized in
the M-step. The expected complete log likelihood, denoted $Q(\cdot)$,
may be written as
\begin{align}
  Q(\mb \Xi | \subs{\mb \Xi}{s}) = \mathbb{E}_{\mb X, \mb Z| \subs{\mb
      \Xi}{s}, \mb Y} \left[\log \left(p(\mb \Xi, \mb X, \mb Z| \mb
  Y)\right)\right]. \label{eq:EMQfun}
\end{align}
Since $\mb X$ and $\mb Z$ are conditionally independent given $\mb
\Xi$, the expectation may be calculated using the full conditional
distributions of $\mb X$ and $\mb Z$ derived for the MCMC
algorithm. The derivation of the EM algorithm for BASS is
then straightforward (Appendix~\ref{app:EM}); note that, when estimating
$\mb \Lambda$, the loading columns specific to each observation are
estimated jointly.

\subsection{Identifiability}

The latent factor model (Equation~\ref{eq:FAModel}) is unidentifiable
up to orthonormal rotations: for any orthogonal matrix $\mb P$ with
$\mb P^T \mb P = \mb I$, letting $\mb \Lambda' = \mb \Lambda \mb P^T$
and $\mb x' = \mb P \mb x$ produces the same estimate of the data
covariance matrix and has an identical likelihood. When using factor
analysis for prediction or covariance estimation, rotational
invariance is irrelevant. However, for all applications that interpret
the factors or use individual factors or loadings for downstream
analysis, this rotational invariance cannot be ignored.  One
traditional solution is to restrict the loading matrix to be lower
triangular \citep{west_bayesian_2003,
  carvalho_high-dimensional_2008}. This solution gives a special role
to the first $k-1$ features in $\mb y$--namely, that the $h^{th}$
feature does not contribute to the $k-h^{th}$ through the $k^{th}$
factor. For this reason, the lower triangular approach does not
generalize easily and requires domain knowledge that may not be
available~\citep{carvalho_high-dimensional_2008}.

In the BASS model, we have rotational invariance when we right
multiply the joint loading matrix by $\mb P^T$ and left multiply $\mb
x$ by $\mb P$, producing an identical covariance matrix and
likelihood. This rotation invariance is addressed in BASS because the
non-sparse rotations of the loading matrix violates the prior
structure induced by the observation-wise and element-wise sparsity.


Scale invariance is a second identifiability problem inherent in
latent factor models. In particular, scale invariance means that a
loading can be multiplied by a non-zero constant and the corresponding
factor by the inverse of that constant, and this will result in the
same data likelihood. This problem we and others have addressed
satisfactorily by using posterior probabilities as optimization
objectives instead of likelihoods and by including regularizing priors
on the factors that restrict the magnitude of the constant. We make an
effort to not interpret the relative or absolute scale of the factors
or loadings including sign beyond setting a reasonable threshold for
zero.

Finally, factor analysis is non-identifiable up to \emph{label
  switching}, or jointly shuffling the $h=1,\dots,k$ indices of the
loadings and factors--assuming we do not take the lower-triangular
approach. Other approaches put distributions on the loading sparsity
or proportion of variance explained in order to address this
problem~\citep{bhattacharya_sparse_2011}. We do not explicitly order
or interpret the order of the factors, so we do not address this
non-identifiability in the model. Label switching is handled here and
elsewhere by a post-processing step, such as ordering factors
according to proportion of variance explained. In our simulation
studies, we interpret results with this non-identifiability in mind.

\subsection{Sparse rotations via PX-EM}

Another general problem with latent factor models that BASS is not
immune to is the convergence to local optima and sensitivity to
parameter initializations. Once the model parameters are initialized,
the EM algorithm may be stuck in locally optimal but globally
supoptimal regions with undesirable factor orientations. 
To address this problem, we take advantage of the rotational
invariance of the factor analysis framework. Parameter expansion (PX)
has been shown to reduce the initialization dependence by introducing
auxiliary variables that rotate the current loading matrix estimate to
best respect our prior while keeping the likelihood
stable~\citep{liu_parameter_1998, dyk_art_2001}.

We extend our model (Equation~\ref{eq:GFAfull}) using parameter
expansion $\mb R$, a positive definite $k \times k$ matrix, as
\begin{align}
  \mb y_i &= \mb \Lambda \mb R^{-1}_L \mb x_i + \mb \epsilon_i, \notag \\
  \mb x_i &\sim \N_k(\mb 0, \mb R), \notag \\
  \mb \epsilon_i &\sim \N_k(\mb 0, \mb \Sigma), \label{eq:PXBGFA}
\end{align}
where $\mb R_L$ is the lower triangular matrix of the Cholesky
decomposition of $\mb R$. The covariance of $\mb y_i$ is invariant
under this expansion, and, correspondingly, the likelihood is
stable. Note $\mb R^{-1}_L$ is not an orthogonal matrix; however,
because it is full rank, it can be transformed into an orthogonal
matrix times a rotation matrix via a polar
decomposition~\citep{rockova2014fast}. We let $\mb \Lambda^\star = \mb
\Lambda \mb R^{-1}_L $ and assign our BASS $\TPBN$ prior on this
\emph{rotated} loading matrix.

We let $\mb \Xi^\star = \{\mb \Lambda^\star, \mb \Theta, \mb \Delta,
\mb \Phi, \mb T, \mb \eta, \mb \gamma, \mb \pi, \mb \Sigma \}$, and
the parameters of our expanded model are $\{ \mb \Xi^\star \cup \mb R
\}$. The EM algorithm in this expanded parameter space generates a
sequence of parameter estimates $\{ \subs{\mb \Xi^\star}{1} \cup
\subs{\mb R}{1}, \subs{\mb \Xi^\star}{2} \cup \subs{\mb R}{2}, \cdots
\}$, which corresponds to a sequence of parameter estimates in the
original space $\{\subs{\mb \Xi}{1}, \subs{\mb \Xi}{2}, \cdots\}$,
where $\mb \Lambda$ is recovered via $\mb \Lambda^\star \mb R_L$
\citep{rockova2014fast}. We initialize $\subs{\mb R}{0} = \mb
I_k$. The expected complete log likelihood of this PX BASS model is
written as
\begin{align}
  Q(\mb \Xi^\star, \mb R | \subs{\mb \Xi}{s}) = \mathbb{E}_{\mb X, \mb
    Z| \subs{\mb \Xi}{s}, \mb Y, \mb R_0} \log\big(p(\mb \Xi^\star,
  \mb R, \mb X, \mb Z| \mb Y)\big). \label{eq:PXEMQfun}
\end{align}

In our parameter expanded EM (PX-EM) for BASS, the conditional
distributions of $\mb X$ and $\mb Z$ still factorize in the
expectation. However, the distribution of $\mb x_i$ depends on
expansion parameter $\mb R$. The full joint distribution
(Equation~\ref{eq:GFAfull}) has a single change in $p(\mb X)$, with
$\mb \Lambda^\star$ in the place of $\mb \Lambda$. In the M-step, the
$\mb R$ that maximizes Equation (\ref{eq:PXEMQfun}) is
\begin{align}
  \subs{\mb R}{s} = \argmax_{\mb R} Q(\mb \Xi^\star, \mb R | \subs{\mb
    \Xi}{s}) = \argmax_{\mb R} \bigg(\textrm{const}- \frac{n}{2} \log
  |\mb R| - \frac{1}{2} \tr \big(\mb R^{-1} \mb S^{XX}\big)\bigg),
  \notag
\end{align}
where $\mb S^{XX} = \sum_{i=1}^n \la \mb x_{\cdot i} \mb x_{\cdot i}^T
\ra$. The solution is $\subs{\mb R}{s} = \frac{1}{n} \mb S^{XX}$. For
the E-step, $\mb \Lambda$ is first calculated and the expectation is
taken in the original space (details in Appendix \ref{app:PX-EM}).

Note that the proposed PX-EM for the BASS model keeps the likelihood
invariant but does not keep the prior invariant after transformation
of $\mb \Lambda$. This is different from the PX-EM studied by
\cite{liu_parameter_1998}, as discussed in
\cite{rockova2014fast}. Because the resulting posterior is not
invariant, we run PX-EM only for a few iterations and then switch to
the EM algorithm. The effect is that the BASS model is substantially
less sensitive to initialization (see Results).  By
introducing expansion parameter $\mb R$, the posterior modes in the
original space are intersected with equal likelihood curves indexed by
$\mb R$ in expanded space. Those curves facilitate traversal between
posterior modes in the original space and encourage initial parameter
estimates with appropriate sparse structure in the loading matrix
\citep{rockova2014fast}. 

\subsection{Computational Complexity}

The computational complexity of the block Gibbs sampler for the BASS
model is demanding. Updating each loading row requires the inversion
of a $k \times k$ matrix with $O(k^3)$ complexity and then calculating
means with $O(k^2n)$ complexity. The complexity of updating the full
loading matrix repeats this calculation $p$ times. Other updates are
of lower order relative to updating the loading. Our Gibbs sampler has
$O(k^3 p + k^2 p n)$ complexity per iteration, which makes MCMC
difficult to apply when $p$ is large.

In the BASS EM algorithm, the E-step has complexity $O(k^3)$ for a
matrix inversion, complexity $O(k^2p + kpn)$ for calculating the first
moment, and complexity $O(k^2 n)$ for calculating the second
moment. Calculations in the M-step are all of a lower order. Thus, the
EM algorithm has complexity $O(k^3 + k^2 p + k^2 n + kpn)$ per
iteration.

Our PX-EM algorithm for the BASS model requires an additional Cholesky
decomposition with complexity $O(k^3)$ and a matrix multiplication
with complexity $O(k^2 p)$ above the EM algorithm. The total
complexity is therefore the same as the original EM algorithm,
although in practice we note that the constants have a noticable
negative impact on the running time. 

\section{Simulations and Comparisons}

We demonstrate the performance of our model on simulated data in three
settings: paired observations, four observations, and ten
observations.

\subsection{Simulations}
\subsubsection{Simulations with Paired Observations (CCA)}

We simulated two data sets with $p_1 = 100$, $p_2 = 120$ in order to
compare results from our method to results from state-of-the-art CCA
methods. The number of samples in these simulations was $n=\{20, 30,
40, 50\}$, chosen to be smaller than both $p_1$ and $p_2$ to reflect
the large $p$, small $n$ regime~\citep{west_bayesian_2003} that
motivated our structured approach. We first simulated observations
with only sparse latent factors (\emph{Sim1}). In particular, we set
$k = 6$, where two sparse factors are shared by both observations
(factors $1$ and $2$; in Table \ref{tab:LoadingZv2}), two sparse
factors are specific to $\sups{\mb y}{1}$ (factors $3$ and $4$; Table
\ref{tab:LoadingZv2}), and two sparse factors are specific to
$\sups{\mb y}{2}$ (factors $5$ and $6$; Table
\ref{tab:LoadingZv2}). The elements in the sparse loading matrix were
randomly generated from a $\N(0, 4)$ Gaussian distribution, and
sparsity was induced by setting $90\%$ of the elements in each loading
column to zero at random (Figure~\ref{fig:ResultsSimV2}A). We zeroed
values of the sparse loadings for which the absolute values were less
than $0.5$. Latent factors $\mb x$ were generated from $\N_6(0,\mb
I_6)$. Residual error was generated by first generating the $p = p_1 +
p_2$ diagonals on the residual covariance matrix $\mb \Sigma$ from a
uniform distribution on $(0.5,1.5)$, and then generating each column
of the error matrix from $\N_p(\mb 0, \mb \Sigma)$.

We performed a second simulation that included both sparse and dense
latent factors (\emph{Sim2}). In particular, we extended \emph{Sim1}
to $k=8$ latent factors, where one of the shared sparse factors is now
dense, and two dense factors, each specific to one observation, were
added. For all dense factors, each loading was generated according to
a $\N(0, 4)$ Gaussian distribution (Table~\ref{tab:LoadingZv2};
Figure~\ref{fig:ResultsSimV2}B).

\begin{table}[h!]
  \centering
  \begin{tabular}{l*{15}{c}}
    \hline
    & \multicolumn{6}{c}{\emph{Sim1}} & & \multicolumn{8}{c}{\emph{Sim2}} \\
    \cline{2-7}  \cline{9-16}
    Factors & 1 & 2 & 3 & 4 & 5 & 6 & & 1 & 2 & 3 & 4 & 5 & 6 & 7 & 8 \\
    \hline
    $\sups{\mb Y}{1}$ & S & S & S & S & - & - & & S & D & S & S & D & - & - & - \\
    $\sups{\mb Y}{2}$ & S & S & - & - & S & S & & S & D & - & - & - & S & S & D \\
    \hline
  \end{tabular}
  \caption{{\bf Latent factors in \emph{Sim1} and \emph{Sim2} with two
      observation matrices.} S represents a sparse vector; D
    represents a dense vector; - represents no contribution to that
    observation from the factor.}
  \label{tab:LoadingZv2}
\end{table}

\subsubsection{Simulations with Four Observations (GFA)}

We performed two simulations (\emph{Sim3} and \emph{Sim4}) including
four observations with $p_1 = 70$, $p_2 = 60$, $p_3 = 50$ and $p_4 =
40$. The number of samples, as above, was set to
$n=\{20,30,40,50\}$. In \emph{Sim3}, we let $k=6$ and only simulated
sparse factors: the first three factors are specific to $\sups{\mb
  y}{1}$, $\sups{\mb y}{2}$ and $\sups{\mb y}{3}$, respectively, and
the last three correspond to different subsets of the observations
(Table~\ref{tab:LoadingZv4}). In \emph{Sim4} we let $k=8$, and, as
with \emph{Sim2}, included both sparse and dense factors
(Table~\ref{tab:LoadingZv4}). Samples in these two simulations were
generated following the same method as in the simulations with two
observations.

\begin{table}
  \centering
  \begin{tabular}{l*{15}{c}}
    \hline
    & \multicolumn{6}{c}{\emph{Sim3}} & & \multicolumn{8}{c}{\emph{Sim4}} \\
    \cline{2-7}   \cline{9-16}
    Factors & 1 & 2 & 3 & 4 & 5 & 6 & & 1 & 2 & 3 & 4 & 5 & 6 & 7 & 8 \\
    \hline
    $\sups{\mb Y}{1}$ & S & - & - & S & - & - & & S & - & - & - & D & - & - & - \\
    $\sups{\mb Y}{2}$ & - & S & - & S & S & S & & - & S & - & S & - & D & - & - \\
    $\sups{\mb Y}{3}$ & - & - & S & - & S & S & & - & - & S & S & - & - & D & - \\
    $\sups{\mb Y}{4}$ & - & - & - & - & - & S & & - & - & S & - & - & - & - & D \\
    \hline
  \end{tabular}
  \caption{{\bf Latent factors in \emph{Sim3} and \emph{Sim4} with
      four observation matrices.} S represents a sparse vector; D
    represents a dense vector; - represents no contribution to that
    observation from the factor. }
  \label{tab:LoadingZv4}
\end{table}

\subsubsection{Simulations with Ten Observations (GFA)}
To further evaluate BASS on multiple observations, We performed two
additional simulations (\emph{Sim5} and \emph{Sim6}) on ten coupled
observations with $p_w = 50$ for $w = 1,\cdots,10$. The number of
samples was set to $n=\{20,30,40,50\}$. In \emph{Sim5}, we let $k=8$
and only simulated sparse factors (Table~\ref{tab:LoadingZv10}).  In
\emph{Sim6} we let $k=10$ and simulated both sparse and dense factors
(Table~\ref{tab:LoadingZv10}). Samples in these two simulations were
generated following the same method as in the simulations with two
observations.

\begin{table}
\small
  \centering
  \begin{tabular}{l*{19}{c}}
    \hline
    & \multicolumn{8}{c}{\emph{Sim5}} & & \multicolumn{10}{c}{\emph{Sim6}} \\
    \cline{2-9}   \cline{11-20}
    Factors & 1 & 2 & 3 & 4 & 5 & 6 & 7 & 8 & & 1 & 2 & 3 & 4 & 5 & 6 & 7 & 8 & 9 & 10\\
    \hline
    $\sups{\mb Y}{1}$ & S & - & - & - & - & - & - & - & & S & - & - & - & - & - & D & - & - & -\\
    $\sups{\mb Y}{2}$ & S & - & - & S & - & - & - & - & & S & - & - & S & - & - & D & - & - & -\\
    $\sups{\mb Y}{3}$ & S & - & - & S & S & - & - & - & & - & - & - & S & - & - & D & D & - & -\\
    $\sups{\mb Y}{4}$ & S & S & - & S & S & - & S & - & & - & S & - & S & - & - & D & D & - & -\\
    $\sups{\mb Y}{5}$ & - & S & - & S & S & - & S & - & & - & S & - & S & S & - & - & D & D & -\\
    $\sups{\mb Y}{6}$ & - & S & - & - & - & - & S & S & & - & S & - & - & S & - & - & D & D & -\\
    $\sups{\mb Y}{7}$ & - & - & S & - & - & - & S & S & & - & S & S & - & S & - & - & - & D & D\\
    $\sups{\mb Y}{8}$ & - & - & S & - & - & - & S & S & & - & - & S & - & S & - & - & - & D & D\\
    $\sups{\mb Y}{9}$ & - & - & S & - & - & - & - & S & & - & - & S & - & - & - & - & - & - & D\\
    $\sups{\mb Y}{10}$ & - & - & S & - & - & S & - & - & & - & - & S & - & - & S & - & - & - & D\\
    \hline
  \end{tabular}
  \caption{{\bf Latent factors in \emph{Sim5} and \emph{Sim6} with
      four observation matrices.} S represents a sparse vector; D
    represents a dense vector; - represents no contribution to that
    observation from the factor. }
  \label{tab:LoadingZv10}
\end{table}

\subsection{Methods for Comparison}

We compared BASS to five available linear models that accept multiple
observations: the Bayesian group factor analysis model with an ARD
prior (GFA)~\citep{klami_bayesian_2013}, an extension of GFA that
allows element-wise sparsity with independent ARD priors (sGFA)
\citep{khan_identification_2014, suvitaival_cross-organism_2014}, a
regularized version of CCA (RCCA)~\citep{gonzalez_cca:_2008}, sparse
CCA (SCCA)~\citep{witten_extensions_2009}, and Bayesian joint factor
analysis (JFA)~\citep{ray_bayesian_2014}. We also included the linear
version of a flexible non-linear model, manifold relevance
determination (MRD) \citep{damianou_manifold_2012}. To evaluate the
sensitivity of BASS to initialization, we compare three different
initialization methods: random initialization (EM), 
$50$ iterations of MCMC (EM-MCMC), and $20$ iterations of PX-EM
(PX-EM); each of these were followed with EM until convergence,
reached when both the number of non-zero loadings do not change for
$t$ iterations and the log likelihood changes $< 1\times 10^{-5}$
within $t$ iterations.  We performed $20$ runs for each version of
inference in BASS: EM, MCMC-EM, and PX-EM. In \emph{Sim1} and
\emph{Sim3}, we set the initial number of factors to $k=10$. In
\emph{Sim2}, \emph{Sim4}, \emph{Sim5}, and \emph{Sim6}, we set the
initial number of factors to $15$.

The GFA model~\citep{klami_bayesian_2013} uses an ARD prior to
encourage column-wise shrinkage of the loading matrix, but not
sparsity within the loadings. The computational complexity of this GFA
model with variational updates is $O(k^3 m + k^2p + k^2n+kpn)$ per
iteration, which is nearly identical to BASS but includes an
additional factor $m$, the number of observations, scaling the $k^3$
term. In our simulations, we ran the GFA model with the factor number
set to the correct value.

The sGFA model~\citep{khan_identification_2014} encourages
element-wise sparsity using independent ARD priors on loading
elements. Loading columns are modeled with a spike-and-slab type
mixture to encourage column-wise sparsity. Inference is performed with
a Gibbs sampler without using block updates.  Its complexity is $O(k^3
+ k^2pn)$ per iteration, which, when $k$ is large, will dominate the
per-iteration complexity of BASS; furthermore, Gibbs samplers
typically require greater numbers of iterations than EM-based
methods. We ran the sGFA model with the correct number of factors in
our six simulations.

We ran the regularized version of classical CCA (RCCA) for comparison
in \emph{Sim1} and \emph{Sim2}~\citep{gonzalez_cca:_2008}.  Classical
CCA tries to find $k$ canonical projection directions $\mb u_h$ and
$\mb v_h$ ($h = 1,\dots,k$) for $\sups{\mb Y}{1}$ and $\sups{\mb
  Y}{2}$ respectively such that i) the correlation between $\mb u^T_h
\sups{\mb Y}{1}$ and $\mb v^T_h \sups{\mb Y}{2}$ is maximized for $h =
1,\dots, k$; and ii) $\mb u^T_{h'} \sups{\mb Y}{1}$ is orthogonal to
$\mb u^T_h \sups{\mb Y}{1}$ with $h' \ne h$, and similarly for $\mb
v_h$ and $\sups{\mb Y}{2}$. Let these two projection matrices be
denoted $\mb U = [\mb u_1,\dots,\mb u_k] \in \mathbb{R}^{p_1 \times
  k}$ and $\mb V = [\mb v_1,\dots,\mb v_k] \in \mathbb{R}^{p_2 \times
  k}$. These matrices are the maximum likelihood estimates of the
shared loading matrices in the Bayesian CCA model up to orthogonal
transformations~\citep{bach_probabilistic_2005}. However, classical
CCA requires the observation covariance matrices to be non-singular
and thus is not applicable in the current simulations where $n < p_1,
p_2$. Therefore, we used a regularized version of CCA
(RCCA)~\citep{gonzalez_cca:_2008}, which regularizes CCA using an
$\ell_2$-type penalty by adding $\lambda_1 \mb I_{p_1}$ and $\lambda_2
\mb I_{p_2}$ to the two sample covariance matrices. The effect of this
penalty is not to induce sparsity but instead to allow application to
$p\gg n$ data sets. The two regularization parameters $\lambda_1$ and
$\lambda_2$ were chosen according to leave-one-out cross-validation
with the search space defined on a $11 \times 11$ grid from $0.0001$
to $0.01$. The projection directions $\mb U$ and $\mb V$ were
estimated using the best regularization parameters. We let $\mb P' =
[\mb U ; \mb V]$; this matrix was comparable to the simulated loading
matrix up to orthogonal transformations.  We calculated the matrix
$\mb P$ such that the Frobenius norm between $\mb \Lambda' \mb P^T$
and simulated $\mb \Lambda$ was minimized, with the constraint that
$\mb P^T \mb P = \mb I$. This was done by the constraint preserving
updates of the objective function \citep{wen_feasible_2013}. After
finding the optimal orthogonal transformation matrix, we recovered
$\mb \Lambda' \mb P^T$ as the estimated loading matrix.  We chose $6$
and $8$ regularized projections for comparison in \emph{Sim1} and
\emph{Sim2} respectively, representing the true number of latent
linear factors. RCCA does not apply to multiple coupled observations,
therefore was not included in further simulations.

The sparse CCA (SCCA) method~\citep{witten_extensions_2009} maximizes
correlation between two observations after projecting the original
space with a sparsity-inducing penalty on the projection directions,
producing sparse matrices $\mb U$ and $\mb V$. This method is encoded
in the R package {\tt PMA} \citep{RPackage_PMA}. For \emph{Sim1} and
\emph{Sim2}, as with RCCA, we found an optimal orthogonal
transformation matrix $\mb P$ such that the Frobenius norm between
$\mb \Lambda_{S} \mb P^T$ and simulated $\mb \Lambda$ was minimized,
where $\mb \Lambda_{S}$ was the vertical concatenation of the
recovered sparse $\mb U$ and $\mb V$. We chose $6$ and $8$ sparse
projections in \emph{Sim1} and \emph{Sim2} for comparison,
respectively, representing the true number of latent linear
factors. Because both RCCA and SCCA are both deterministic and greedy,
the results for $k < 6$ are all implicitly available by subsetting the
factors in the $k=6$ results.

An extension of SCCA allows for multiple observations
\citep{witten_extensions_2009}. For \emph{Sim3} and \emph{Sim4}, we
recovered four sparse projection matrices $\sups{\mb U}{1}, \sups{\mb
  U}{2}, \sups{\mb U}{3}, \sups{\mb U}{4}$, and for \emph{Sim5} and
\emph{Sim6}, we recovered ten projection matrices. $\mb \Lambda_{S}$
was calculated with the concatenation of those projection matrices.
Then the orthogonal transformation matrix $\mb P$ was calculated
similarly by minimizing the Frobenius norm between $\mb \Lambda_{S}
\mb P^T$ and the true loading matrix $\mb \Lambda$. The number of
canonical projections was set to $6$ in \emph{Sim3}, $8$ in
\emph{Sim4} and \emph{Sim5}, and $10$ in \emph{Sim6}, again
corresponding to the true number of latent factors.

The Bayesian joint factor analysis model
(JFA)~\citep{ray_bayesian_2014} puts an Indian buffet process (IBP)
prior~\citep{griffiths_indian_2011} on the factors, inducing
element-wise sparsity, and an ARD prior on the variance of the
loadings. The idea of putting an IBP on a latent factor model, which
gives desirable non-parametric behavior in the number of latent
factors and also produces element-wise sparsity in the loading matrix,
was described for the Nonparametric Sparse Factor Analysis (NSFA)
model~\citep{knowles_nonparametric_2011}. Similarly, in JFA,
element-wise sparsity is thus encouraged both in the factors and in
the loadings. JFA partitions latent factors into a fixed number of
observation-specific factors and factors shared by all observations,
and does not include column-wise sparsity. Its complexity is $O(k^3 +
k^2pn)$ per iteration of the Gibbs sampler. We ran JFA on our
simulations with the number of factors set to the correct
values. Because the JFA model uses a sparsity-inducing prior instead
of an independent Gaussian prior on the latent factors, the resulting
model does not have a closed form posterior predictive distribution
(Equation~\ref{eq:Prediction}); therefore, we excluded the JFA model
from prediction results.


The non-linear manifold relevance determination (MRD) model
\citep{damianou_manifold_2012} extends the notable Gaussian process
latent variable (GPLVM) model \citep{lawrence_probabilistic_2005} to
include multiple observations. A GPLVM puts a Gaussian process prior
on the latent variable space. GPLVM has an interpretation of a dual
probabilistic PCA model that marginalizes loading columns using
Gaussian priors. MRD extends GPLVM by putting multiple weight vectors
on the latent variables using a Gaussian process kernel. Each of the
weight vectors corresponds to one observation, therefore they
determine a soft partition of latent variable space. The complexity of
MRD is quadratic in the number of samples $n$ per iteration using a
sparse Gaussian process.  Posterior inference and prediction using the
MRD model was performed with Matlab package {\tt vargplvm}
\citep{damianou_manifold_2012}.  We used the linear kernel with
feature selection (i.e., {\tt Linard2} kernel), meaning that we used
the linear version of this model for accurate comparison. We ran the
MRD model on our simulated data with the correct number of factors.

We summarize the parameter choices for all methods here:
\begin{itemize}
 \item[sGFA] We used the \verb$getDefaulOpts$ function in the sGFA
   package to set the default parameters. In particular, the ARD prior
   was set to $Ga(10^{-3},10^{-3})$. The prior on inclusion
   probability is set to $Beta(1,1)$. Total MCMC iterations were set
   to $10^5$ with {\tt sampling iterations} set to $1000$ and {\tt
     thinning steps} set to $5$.
 \item[GFA] We used the \verb$getDefaultOpts()$ function in the GFA package to set the default parameters. In particular, the ARD prior for both loading and error variance was set to $Ga(10^{-14}, 10^{-14})$. The {\tt maximum iteration} parameter was set to $10^5$, and the ``L-BFGS'' optimization method was used. 
 \item[RCCA] The regularization parameter was chosen using leave-one-out cross-validation on an $11 \times 11$ grid from $0.0001$ to $0.01$ using the function \verb$estim.regul$ in the \verb$CCA$ package.
 \item[SCCA] We used the \verb$PMA$ package with Lasso penalty (the {\tt typex} and {\tt typez} parameters in the function \verb$CCA$ were set to ``standard''). This corresponds to setting the $\ell_1$ bound of the projection vector to $0.3\sqrt{p_w}$ for $w = 1,2$.
 \item[JFA] The ARD prior for both the loading and factor scores were set to $Ga(10^{-5}, 10^{-5})$. The parameters of the beta process prior were set to $\alpha = 0.1$ and $c = 10^4$. The MCMC iterations were set to $1000$ with $200$ iterations of burn-in. As is the default, we did not thin the chain.
 \item[MRD] We used the \verb$svargplvm_init$ function in the \verb$GPLVM$ package to initialize parameters. In particular, the \verb$linar2$ kernel was chosen for all observations. Latent variables were initialized by concatenating the observation matrices first (the `concatenated' option) and then performing PCA. Other parameters were set by \verb$svargplvm_init$ with the default options.
\end{itemize}

\subsection{Metrics for Comparison}

To quantitatively compare the results of BASS with the alternative
methods, we used the sparse and dense stability
indices~\citep{gao_latent_2013} to compare the simulated loadings with
the recovered loadings. The sparse stability index (SSI) measures the
similarity between columns of sparse matrices. SSI is invariant to
column scale and label switching, but it penalizes factor splitting
and matrix rotation; larger values of SSI indicate better
recovery. Let $\mb C \in \mathbb{R}^{k_1\times k_2}$ be the absolute
correlation matrix of columns of two sparse loading matrices. Then SSI
is calculated by
\begin{align}
  SSI &= \frac{1}{2 k_1} \sum_{h_1 = 1}^{k_1} \bigg( \max(\mb
  c_{h_1,\cdot}) - \frac{\sum_{h_2 = 1}^{k_2} I(c_{h_1,h_2} >
    \overline{\mb c_{h_1,\cdot}}) c_{h_1,h_2}}{ k_2 - 1} \bigg) \notag \\
  &+ \frac{1}{2 k_2} \sum_{h_2 = 1}^{k_2} \bigg( \max(\mb
  c_{\cdot,h_2}) - \frac{\sum_{h_1 = 1}^{k_1} I(c_{h_1,h_2} >
    \overline{\mb c_{\cdot,h_2}}) c_{h_1,h_2}}{ k_1 - 1}
  \bigg). \label{eq:SSI}
\end{align}

The dense stability index (DSI) quantifies the difference between
dense matrix columns, and is invariant to orthogonal matrix rotation,
factor switching, and scale; DSI values closer to zero indicate better
recovery. Let $\mb M_1$ and $\mb M_2$ be the dense matrices. DSI is
calculated by
\begin{align}
  DSI &= \frac{1}{p^2} tr(\mb M_1 \mb M_1^T - \mb M_2 \mb
  M_2^T). \label{eq:DSI}
\end{align}

We extended the stability indices to allow multiple coupled
observations as in our simulations. In \emph{Sim1}, \emph{Sim3}, and
\emph{Sim5}, all factors are sparse, and SSIs were calculated between
the true sparse loading matrices and recovered sparse loading
matrices. In \emph{Sim2}, \emph{Sim4}, and \emph{Sim6}, because none
of the methods other than BASS explicitly distinguishes sparse and
dense factors, we categorized each recovered factor as follows. We
first selected a global sparsity threshold on the elements of the
combined loading matrix; here we set that value to $0.15$. Elements
below this threshold were set to zero in the loading matrix.  Then we
chose the first five loading columns with the fewest non-zero elements
as the sparse loadings in \emph{Sim2}, first four such loadings as the
sparse loadings in \emph{Sim4}, and first six such loadings as sparse
in \emph{Sim6}. The remaining loading columns were considered dense
loadings and were not zeroed according to the global sparsity
threshold. We found that varying the sparsity threshold did not affect
the separation of sparse and dense loadings significantly across
methods. SSIs were then calculated for the true sparse loading matrix
and the recovered sparse loadings across methods.

To calculate DSIs, we treated the loading matrices $\sups{\mb
  \Lambda}{w}$ for each observation separately, and calculated the DSI
for the recovered dense components of each observation. The final DSI
for each method was the sum of the $m$ separate DSIs. Because the
loading matrix is marginalized out in the MRD
model~\citep{lawrence_probabilistic_2005}, we excluded MRD from this
DSI comparison.

We further evaluated the prediction performance of BASS and other
methods. In the BASS model (Equation~\ref{eq:BGFA}), the joint
distribution of any one observation $\sups{\mb y}{w}_i$ and all other
observations $\sups{\mb y}{-w}_i$ can be written as
\begin{align}
  \begin{pmatrix}
    \sups{\mb y}{w}_i \\
    \sups{\mb y}{-w}_i
  \end{pmatrix} \sim
  \N \begin{bmatrix}
    \begin{pmatrix}
      \mb 0 \\
      \mb 0 
    \end{pmatrix} ,
    \begin{pmatrix}
      \sups{\mb \Lambda}{w} (\sups{\mb \Lambda}{w})^T + \sups{\mb
        \Sigma}{w} &
      \sups{\mb \Lambda}{w} (\sups{\mb \Lambda}{-w})^T \\
      \sups{\mb \Lambda}{-w} (\sups{\mb \Lambda}{w})^T & \sups{\mb
        \Lambda}{-w} (\sups{\mb \Lambda}{-w})^T + \sups{\mb
        \Sigma}{-w}
    \end{pmatrix}
  \end{bmatrix}, \notag
\end{align}
where $\sups{\mb \Lambda}{-w}$ and $\sups{\mb \Sigma}{-w}$ are the
loading matrix and residual covariance excluding the $w^{th}$
observation. Therefore, the conditional distribution of $\sups{\mb
  y}{w}_i$ is a multivariate response in a multivariate linear
regression model, where $\sups{\mb y}{-w}_i$ are the predictors; the
mean term takes the form
\begin{align}
  \mathbb{E} (\sups{\mb y}{w}_i | \sups{\mb y}{-w}_i) &= \sups{\mb
    \Lambda}{w} (\sups{\mb \Lambda}{-w})^T \big(\sups{\mb \Lambda}{-w}
  (\sups{\mb \Lambda}{-w})^T + \sups{\mb \Sigma}{-w}\big)^{-1}
  \sups{\mb y}{-w}_i \notag \\
  &= \sum_{h=1}^k \sups{\mb \lambda}{w}_{\cdot h} (\sups{\mb
    \lambda}{-w}_{\cdot h})^T \big(\sups{\mb \Lambda}{-w} (\sups{\mb
    \Lambda}{-w})^T + \sups{\mb \Sigma}{-w}\big)^{-1} \sups{\mb
    y}{-w}_i. \label{eq:Prediction}
\end{align}
We used this conditional distribution to predict specific observations
given others. For the six simulations, we used the simulated data as
training data. Then, we generated 200 samples as test data using the
true model parameters. For each simulation study, we chose at least
one observation in the test data as the response and used the other
observations and model parameters estimated from the training data to
perform prediction. Mean squared error (MSE) is used to evaluate the
prediction performance. For \emph{Sim1} and \emph{Sim2}, $\sups{\mb
  y}{2}_i$ was the response; for \emph{Sim3} and \emph{Sim4},
$\sups{\mb y}{3}_i$ was the response; and for \emph{Sim5} and
\emph{Sim6}, $\sups{\mb y}{8}_i$, $\sups{\mb y}{9}_i$ and $\sups{\mb
  y}{10}_i$ were the responses.

\subsection{Results of the Simulation Comparison}

\begin{figure}
  \centering
  \includegraphics[width=1\textwidth]{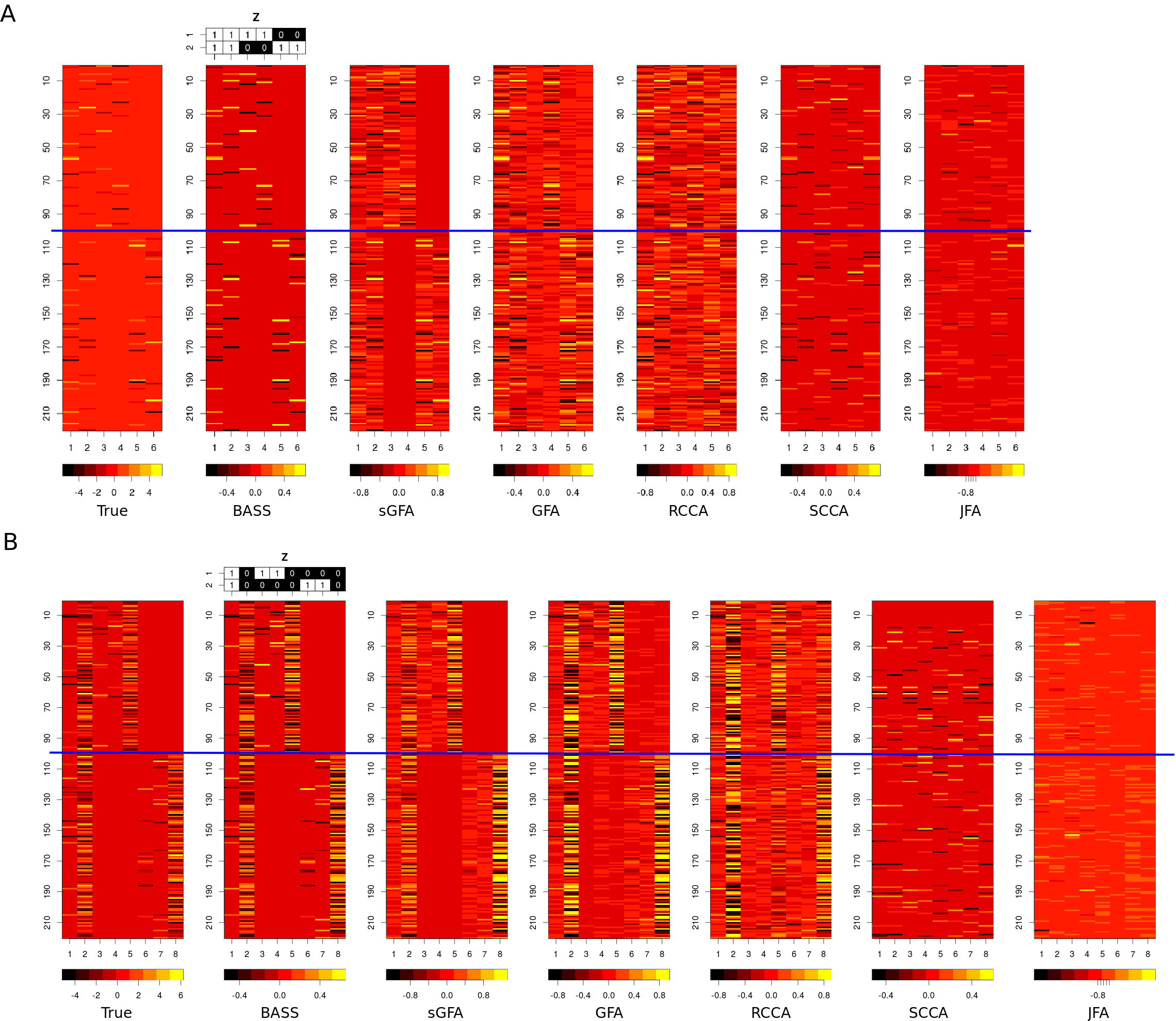}
  \caption{{\bf Simulation results with two paired observations.}  We
    reordered the columns of the recovered matrices and, where
    necessary, multiplied columns by $-1$ for easier visual
    comparisons. Horizontal lines separate the two observations. Panel
    A: Comparison of the recovered loading matrices using different
    models on \emph{Sim1}.  Panel B: Comparison of the recovered
    loading matrices using different models on \emph{Sim2}. }
  \label{fig:ResultsSimV2}
\end{figure}

We first evaluated the performance of BASS and the other methods in
terms of recovering the correct number of sparse and dense factors in
the six simulations (Figures S3-S8). We calculated the percentage of
correctly identified factors across $20$ runs in the simulations with
$n=200$ (Table~\ref{tab:Percentage}).  Qualitatively, BASS recovered
the closest matches to the simulated loading matrices across all
methods (Figures \ref{fig:ResultsSimV2}, S1, S2). The correctly
estimated loading matrices by the three different BASS initialization
approaches produced similar results; we only plot matrices from the
PX-EM method.

\begin{table}
  \centering
  \begin{tabular}{l*{3}{c}}
    \hline
    & EM & MCMC-EM & PX-EM \\
    \hline
    \emph{Sim1} & 79.17\% & 99.17\% & 91.67\% \\
    \emph{Sim2} & 61.25\% & 93.75\% & 85.62\% \\
    \emph{Sim3} & 50.00\% & 78.57\% & 73.57\% \\
    \emph{Sim4} & 62.78\% & 86.11\% & 82.78\% \\
    \emph{Sim5} & 17.22\% & 86.67\% & 66.67\% \\
    \emph{Sim6} & 13.64\% & 60.45\% & 62.73\% \\
    \hline
  \end{tabular}
  \caption{{\bf Percentage of latent factors correctly identified
      across $20$ runs with $n=40$.} The columns represent the runs of
    EM, EM initialized with MCMC, and EM initialized with PX-EM.}
  \label{tab:Percentage}
\end{table}

\subsubsection{Results on Simulations with Two Observations (CCA)}

\begin{figure}
  \centering
  \includegraphics[width=0.8\textwidth]{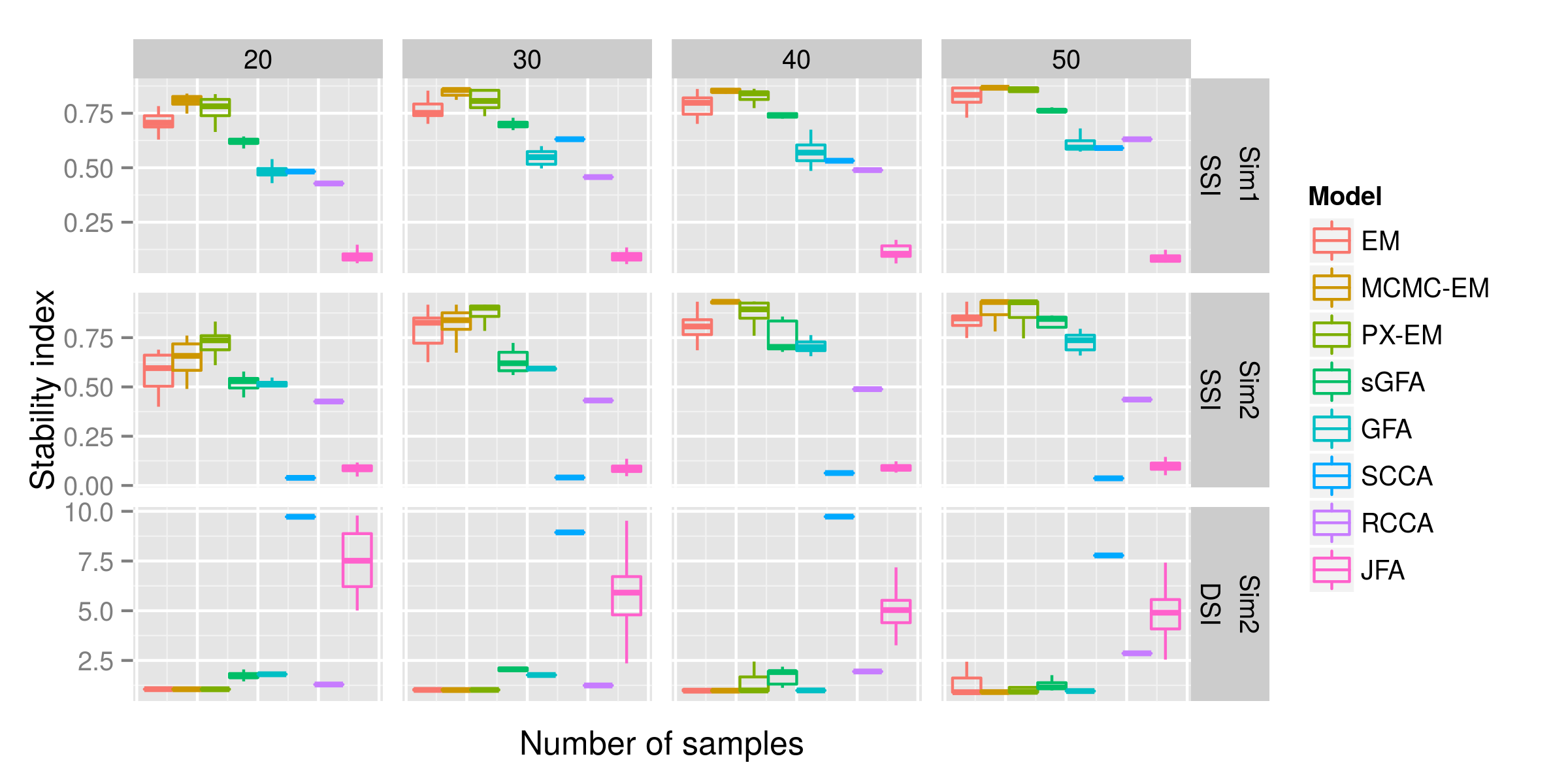}
  \caption{{\bf Comparison of stability indices on recovered loading
      matrices with two observations.} For SSI, a larger value
    indicates better recovery; for DSI, a smaller value indicates
    better recovery. The boundaries of the box are the first and third
    quartiles. The line extends to the highest and lowest observations
    that are within $1.5$ times the distance of the first and third
    quartiles beyond the box boundaries.}
  \label{fig:StabilityV2}
\end{figure}

Comparing results with two observations (\emph{Sim1} and \emph{Sim2}),
our model produced the best SSIs and DSIs among all methods across all
sample sizes (Figures \ref{fig:StabilityV2}).  The column-wise
spike-and-slab prior in sGFA recovered the appropriate number of
columns (Figure \ref{fig:ResultsSimV2}).  However, sGFA's performance
was limited for these simulations because the ARD prior does not
produce sufficient element-wise sparsity, resulting in low SSIs
(Figure \ref{fig:StabilityV2}). As a consequence of not matching
sparse loadings well, sGFA had difficulty recovering dense loadings,
especially with small sample sizes (Figure \ref{fig:StabilityV2}).

The GFA model had difficulty recovering sparse loadings because of
column-wise ARD priors with the same limitation (Figure
\ref{fig:ResultsSimV2}, Figure \ref{fig:StabilityV2}). Its dense
loadings were indirectly affected by the lack of sufficient sparsity
for small sample sizes (Figure \ref{fig:StabilityV2}). RCCA also had
difficulty in the two simulations because the recovered loadings were
not sufficiently sparse using the $\ell_2$-type penalty (Figure
\ref{fig:ResultsSimV2}). SCCA recovered shared sparse loadings well in
\emph{Sim1} (Figure \ref{fig:ResultsSimV2}). However SCCA does not
model local covariance structure, and therefore was unable to recover
the sparse loadings specific to either of the observations in
\emph{Sim1} (Figure \ref{fig:ResultsSimV2}A) resulting again in poor
SSIs (Figure \ref{fig:StabilityV2}). Adding dense loadings
deteriorated the performance of SCCA (Figures \ref{fig:ResultsSimV2}B,
\ref{fig:StabilityV2}). The JFA model did not recover the true
loadings matrix well because of insufficient sparsity in the loadings
and additional sparsity in the factors (Figure
\ref{fig:ResultsSimV2}). The SSIs and DSIs for JFA reflect this
data-model mismatch (Figure \ref{fig:StabilityV2}).

We next evaluated the predictive performance of these methods for two
observations. In \emph{Sim1}, SCCA achieved the best prediction
accuracy in three training sample sizes (Table \ref{tab:MSEV2}). We
attribute this to SCCA recovering well the shared sparse loadings
(Figure \ref{fig:ResultsSimV2}) because the prediction accuracy is a
function of only the shared loadings. Note
(Equation~\ref{eq:Prediction}) that zero columns in either $\sups{\mb
  \Lambda}{w}$ or $\sups{\mb \Lambda}{-w}$ decouple the contribution
of the corresponding factors to the prediction of $\sups{\mb
  y}{w}_i$. In \emph{Sim2}, shared sparse and dense factors contribute
to the prediction performance, and BASS achieved the best prediction
accuracy (Table \ref{tab:MSEV2}).

\begin{table}
  \centering
\tiny
    \begin{tabular}{ll*{14}{c}}
    \hline
    & & \multicolumn{6}{c}{BASS}  & &     &      &      &       \\
    \cline{3-8} 
    & \multirow{2}{*}{$n_t$} & \multicolumn{2}{c}{EM} & \multicolumn{2}{c}{MCMC-EM} & \multicolumn{2}{c}{PX-EM} & \multicolumn{2}{c}{sGFA} & \multicolumn{2}{c}{GFA} & \multicolumn{1}{c}{SCCA} & \multicolumn{1}{c}{RCCA} & \multicolumn{2}{c}{MRD-lin} \\
    & & Err & SE & Err & SE & Err & SE & Err & SE & Err & SE & Err & Err & Err & SE \\
    \hline
    \multirow{5}{*}{\emph{Sim1}} 
    & 10 & 1.00 & 0.024 & 1.03 & 0.024 & 1.02 & 0.028 & 1.00 & $<$1e-3 & 0.98 & 0.002 & \bf 0.88 & 1.01 & 1.08 & 0.024 \\
    & 30 & 0.90 & 0.022 & \bf 0.88 & 0.001 & 0.88 & 0.003 & 0.92 & 0.005 & 0.93 & 0.002 & 0.88 & 0.97  & 1.00 & 0.016\\
    & 50 & 0.88 & 0.011 & \bf 0.87 & 0.003 & 0.88 & 0.014 & 0.90 & 0.004 & 0.92 & 0.002 & 0.88 & 0.92 & 0.98 & 0.028 \\
    & 100 & 0.88 & 0.010 & \bf 0.87 & 0.001 & 0.87 & 0.005 & 0.89 & 0.003 & 0.89 & $<$1e-3 & 0.87 & 0.91 & 0.97 & 0.016\\
    & 200 & 0.88 & 0.007 & \bf 0.87 & 0.004 & 0.87 & 0.005 & 0.88 & 0.001 & 0.88 & $<$1e-3 & 0.87 & 0.95 & 1.16 & 0.202\\
    \hline
    \multirow{5}{*}{\emph{Sim2}} 
    & 10 & 0.80 & 0.161 & 0.82 & 0.162 & \bf 0.68 & 0.003 & 0.74 & 0.043 & 0.89 & 0.023 & 0.86 & 0.72 & 1.14 & 0.002 \\
    & 30 & 0.72 & 0.092 & 0.72 & 0.097 & 0.67 & 0.016 & 0.67 & 0.014 & \bf 0.66 & 0.006 & 0.86 & 0.70 & 1.15 & 0.034 \\
    & 50 & 0.71 & 0.155 & 0.70 & 0.155 & 0.65 & 0.105 & \bf 0.63 & 0.009 & 0.67 & $<$1e-3 & 0.85 & 0.72 & 1.17 & 0.009 \\
    & 100 & 0.63 & 0.066 & \bf 0.61 & 0.013 & 0.62 & 0.013 & 0.62 & 0.005 & 0.61 & 0.001 & 0.85 & 0.75 & 1.13 & 0.013 \\
    & 200 & 0.65 & 0.099 & \bf 0.61 & 0.012 & 0.63 & 0.020 & 0.62 & 0.007 & 0.61 & 0.002 & 0.85 & 0.81 & 1.55 & 0.591 \\
    \hline
  \end{tabular}
  \caption{{\bf Prediction accuracy with two observations on $n_s=200$
      test samples.}  Test samples $\sups{\mb y}{2}_i$ are treated as
    the response, and training samples $\sups{\mb y}{1}_i$ are used to
    estimate parameters in order to predict the response. Prediction
    accuracy is measured by mean squared error (MSE) between simulated
    $\sups{\mb y}{1}_i$ and $\mathbb{E}( \sups{\mb y}{1}_i | \sups{\mb
      y}{2}_i)$. Values presented are the mean MSE and standard
    deviation for $20$ runs of each model. If standard error (SE) is
    missing for a method, the method was deterministic and SE$=0$.}
  \label{tab:MSEV2}
\end{table}


\subsubsection{Results on Simulations with Four Observations (GFA)}

\begin{figure}
  \centering
  \includegraphics[width=0.8\textwidth]{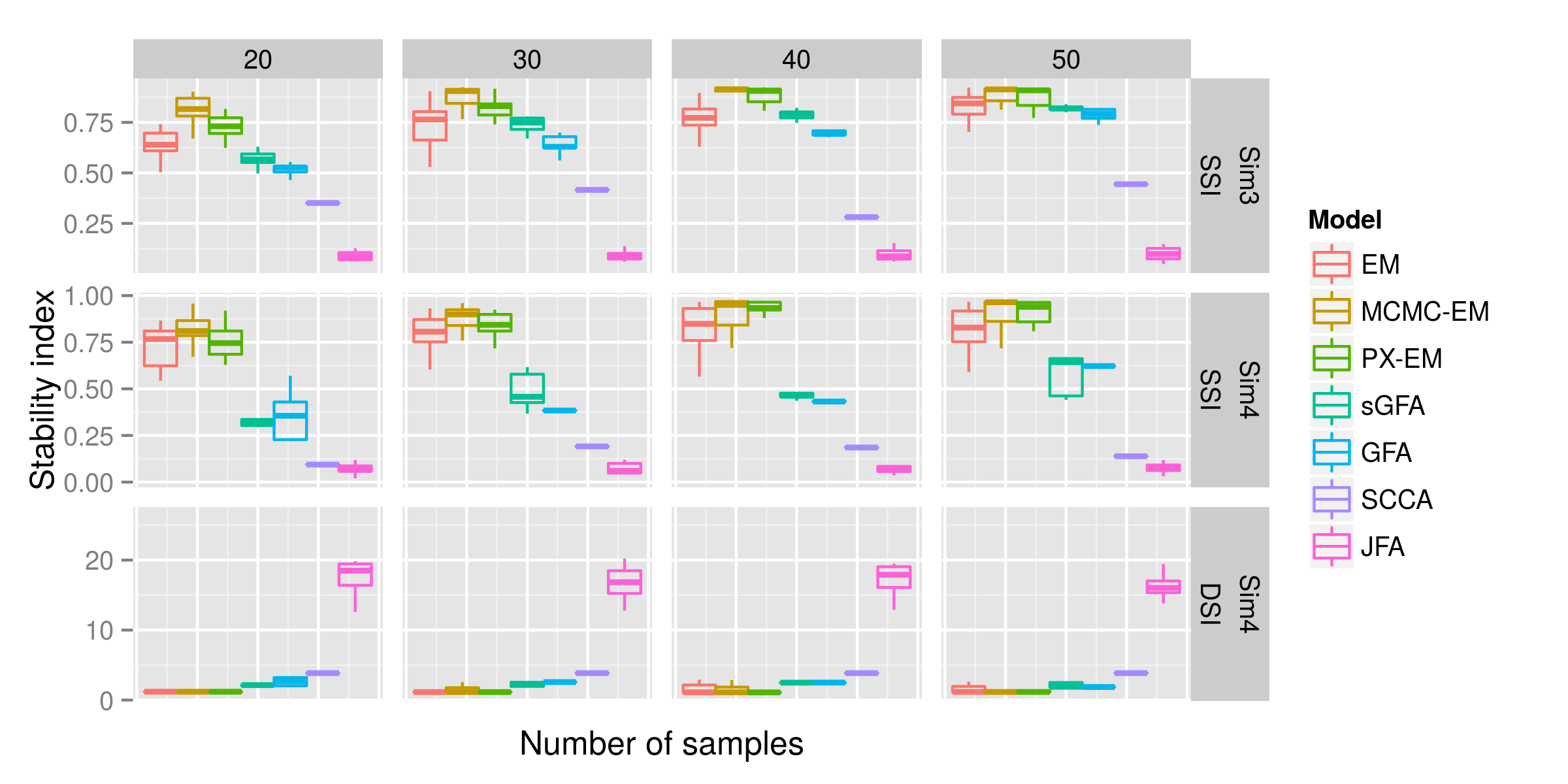}
  \caption{{\bf Comparison of stability indices on recovered loading
      matrices with four observations.} For SSI, a larger value
    indicates better recovery; for DSI, a smaller value indicates
    better recovery. The boundaries of the box are the first and third
    quartiles. The line extends to the highest and lowest values that
    are within 1.5 times the distance of the first and third
    quartiles beyond the box boundaries.}
  \label{fig:StabilityV4}
\end{figure}

For simulations with four observations (\emph{Sim3} and \emph{Sim4}),
BASS correctly recovered sparse and dense factors and their active
observations (Figure S1). sGFA achieved
column-wise sparsity for two observations; however, its sparsity level
within factors was insufficient to match the simulations. GFA results
produced insufficient column-wise sparsity: columns with zero values
were not effectively removed (Figure S1B).
Element-wise shrinkage in GFA was less effective than either BASS or
sGFA (Figure S1).  The results of SCCA and JFA did
not match the true loading matrices for the same reasons as discussed
in \emph{Sim1} and \emph{Sim2} (Figure S1). The
results using stability indices showed that BASS produced the best
SSIs and DSIs across models and almost all sample sizes (Figure
\ref{fig:StabilityV4}). sGFA achieved similar SSI values in
\emph{Sim3} with $n=40$ compared to BASS EM, but showed worse
performance for BASS MCMC-EM and PX-EM. The advantage of BASS relative
to the other methods is apparent in these SSI comparisons, which
specifically highlight interpretability and robust recovery of this
common type of latent structure (Figure \ref{fig:StabilityV4}).

\begin{table}
\tiny
  \centering
    \begin{tabular}{ll*{15}{c}}
    \hline
    & & \multicolumn{6}{c}{BASS}  & &     &      &      &       \\
    \cline{3-8} 
    & \multirow{2}{*}{$n_t$} & \multicolumn{2}{c}{EM} & \multicolumn{2}{c}{MCMC-EM} & \multicolumn{2}{c}{PX-EM} & \multicolumn{2}{c}{sGFA} & \multicolumn{2}{c}{GFA} & \multicolumn{1}{c}{SCCA} & \multicolumn{2}{c}{MRD-lin} \\
    & & Err & SE & Err & SE & Err & SE & Err & SE & Err & SE & Err & Err & SE \\
    \hline
    \multirow{5}{*}{\emph{Sim3}} 
    & 10 & 1.03 & 0.044 & 1.02 & 0.019 & 1.01 & 0.010 & 1.00 & $<$1e-3 & \bf 0.97 & 0.001 & 1.00 & 1.00 & $<$1e-3 \\
    & 30 & 0.91 & 0.049 & \bf 0.87 & 0.016 & 0.88 & 0.007 & 0.90 & 0.007 & 0.93 & 0.003 & 1.00 & 0.99 & 0.021 \\
    & 50 & \bf 0.85 & 0.019 & 0.85 & $<$1e-3 & 0.87 & 0.038 & 0.87 & 0.005 & 0.88 & 0.002 & 1.01 & 1.04 & 0.095 \\
    & 100 & 0.85 & 0.019 & \bf 0.84 & 0.002 & 0.84 & 0.003 & 0.86 & 0.004 & 0.87 & 0.001 & 1.11 & 0.92 & 0.014 \\
    & 200 & 0.84 & 0.001 & 0.84 & $<$1e-3 & 0.84 & 0.004 & 0.84 & 0.001 & \bf 0.83 & 0.001 & 1.13 & 1.16 & 0.140 \\
    \hline
    \multirow{5}{*}{\emph{Sim4}} 
    & 10 & 1.05 & 0.095 & 1.03 & 0.094 & 1.10 & 0.138 & \bf 1.00 & $<$1e-3 & 1.32 & 0.029 & 1.35 & 1.98 & 0.067 \\
    & 30 & 0.97 & 0.020 & \bf 0.95 & 0.015 & 0.96 & 0.013 & 0.97 & 0.007 & 1.03 & 0.003 & 1.40 & 1.50 & 0.090 \\
    & 50 & 0.94 & 0.013 & \bf 0.93 & 0.005 & 0.94 & 0.012 & 0.95 & 0.005 & 1.02 & 0.017 & 1.40 & 1.50 & 0.084 \\
    & 100 & \bf 0.93 & 0.015 & 0.93 & 0.007 & 0.93 & 0.010 & 0.94 & 0.003 & 0.96 & $<$1e-3 & 1.51 & 1.47 & 0.088 \\
    & 200 & 0.91 & 0.029 & 0.92 & 0.022 & \bf 0.89 & 0.047 & 0.93 & 0.001 & 0.89 & 0.001 & 1.77 & 1.58 & 0.132 \\
    \hline
  \end{tabular}
  \caption{{\bf Prediction accuracy with four observations on
      $n_s=200$ test samples.}  Test samples $\sups{\mb y}{3}_i$ are
    treated as the response, and training samples $\sups{\mb y}{1}_i$,
    $\sups{\mb y}{2}_i$, and $\sups{\mb y}{4}_i$ are used to estimate
    parameters in order to predict the response. Prediction accuracy
    is measured by mean squared error (MSE) between simulated
    $\sups{\mb y}{3}_i$ and $\mathbb{E}( \sups{\mb y}{3}_i | \sups{\mb
      y}{1}_i, \sups{\mb y}{2}_i, \sups{\mb y}{4}_i)$. Values
    presented are the mean MSE and standard deviation for $20$ runs of
    each model. Standard error (SE) is missing for SCCA because the
    method is deterministic and SE$=0$.}
  \label{tab:MSEV4}
\end{table}

In the context of prediction using four observation matrices, BASS
achieved the best prediction performance with $\sups{\mb y}{3}_i$ as
the response and the remaining observations as predictors (Table
\ref{tab:MSEV4}). In particular, the MCMC-initialized EM approach had
the best overall prediction performance across methods for these two
simulations.

\subsubsection{Results on Simulations with Ten Observations (GFA)}

\begin{figure}
  \centering
  \includegraphics[width=0.8\textwidth]{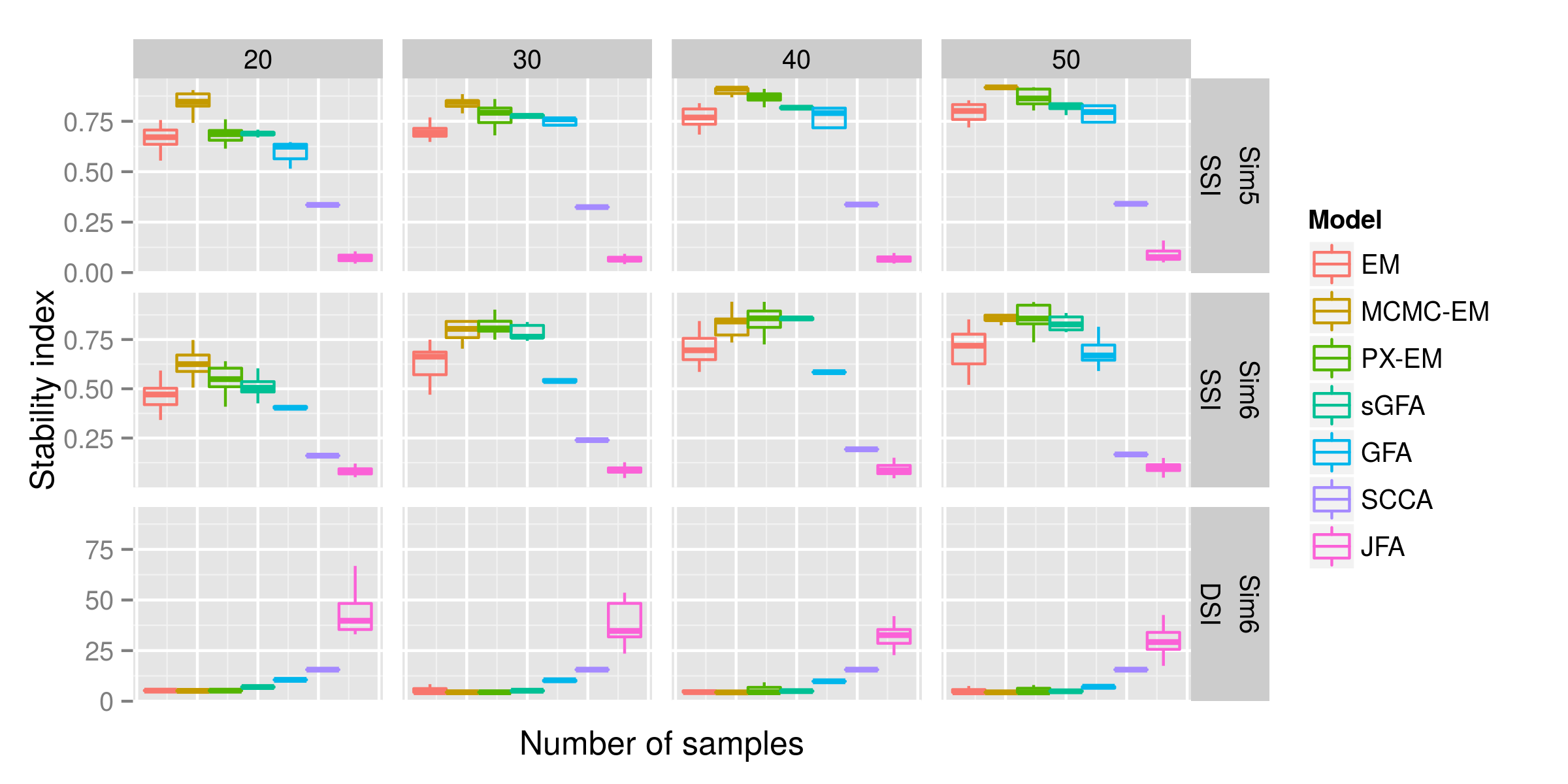}
  \caption{{\bf Comparison of stability indices on recovered loading
      matrices with ten observations.} For SSI, a larger value
    indicates better recovery; for DSI, a smaller value indicates
    better recovery. The boundaries of the box are the first and third
    quartiles. The line extends to the highest and lowest value
    within 1.5 times the distance of the first and third
    quartiles beyond the box boundaries.}
  \label{fig:StabilityV10}
\end{figure}

When we increased the number of observations to ten (\emph{Sim5} and
\emph{Sim6}), BASS still correctly recovered the sparse and dense
factors and their active observations (Figure S2). sGFA effectively performed column-wise
selection although element-wise sparsity remained inadequate (Figure
S2). GFA did not recover sufficient column-wise
or element-wise sparsity (Figure S2). SCCA and
JFA both failed to recover the true loading matrices (Figure
S2). For the stability indices, BASS with MCMC-EM
and PX-EM produced the best SSIs in \emph{Sim5} across all methods and
for almost all sample sizes (Figures \ref{fig:StabilityV10}). Here
sGFA achieved equal or better SSIs than BASS EM, highlighting the
sensitivity of BASS EM to initializations. GFA had equivalent or worse
SSIs than BASS EM. In this pair of simulations, the advantages of BASS
for flexible and robust column-wise and element-wise shrinkage are
apparent (Figures \ref{fig:StabilityV10}). BASS also achieved the best
prediction performance in \emph{Sim5} and
\emph{Sim6} with ten observations (Table~\ref{tab:MSEV4}).

\begin{table}
\tiny
  \centering
    \begin{tabular}{ll*{15}{c}}
    \hline
    & & \multicolumn{6}{c}{BASS}  & &     &      &      &       \\
    \cline{3-8} 
    & \multirow{2}{*}{$n_t$} & \multicolumn{2}{c}{EM} & \multicolumn{2}{c}{MCMC-EM} & \multicolumn{2}{c}{PX-EM} & \multicolumn{2}{c}{sGFA} & \multicolumn{2}{c}{GFA} & \multicolumn{1}{c}{SCCA} & \multicolumn{2}{c}{MRD-lin} \\
    & & Err & SE & Err & SE & Err & SE & Err & SE & Err & SE & Err & Err & SE \\
    \hline
    \multirow{5}{*}{\emph{Sim5}} 
    & 10 & 1.01 & 0.020 & 1.00 & 0.011 & 1.00 & 0.007 & \bf 0.99 & 0.008 & 1.00 & 0.002 & 0.99 & 1.49 & 0.001 \\
    & 30 & 0.88 & 0.031 & \bf 0.86 & 0.018 & 0.87 & 0.028 & 0.89 & 0.005 & 0.90 & 0.002 & 0.99 & 1.01 & 0.035 \\
    & 50 & 0.86 & 0.023 & \bf 0.85 & $<$1e-3 & 0.86 & 0.022 & 0.87 & 0.003 & 0.88 & 0.001 & 0.99 & 0.97 & 0.020 \\
    & 100 & \bf 0.85 & 0.007 & 0.85 & $<$1e-3 & 0.85 & 0.002 & 0.86 & 0.003 & 0.87 & 0.001 & 1.01 & 0.92 & 0.039 \\
    & 200 & 0.85 & 0.006 & 0.84 & $<$1e-3 & 0.84 & $<$1e-3 & 0.84 & 0.001 & \bf 0.83 & 0.001 & 0.96 & 1.06 & 0.105 \\
    \hline
    \multirow{5}{*}{\emph{Sim6}} 
    & 10 & 0.61 & 0.164 & 0.57 & 0.116 & \bf 0.51 & 0.031 & 0.58 & 0.012 & 0.75 & 0.011 & 0.97 & 1.00 & $<$1e-3 \\
    & 30 & 0.49 & 0.160 & 0.40 & 0.093 & \bf 0.38 & 0.007 & 0.43 & 0.006 & 0.40 & 0.005 & 0.98 & 0.46 & 0.006 \\
    & 50 & 0.44 & 0.099 & \bf 0.39 & 0.011 & 0.39 & 0.004 & 0.41 & 0.002 & 0.40 & 0.001 & 1.01 & 0.42 & 0.009 \\
    & 100 & \bf 0.39 & 0.033 & 0.39 & 0.004 & 0.39 & 0.011 & 0.39 & 0.002 & 0.39 & 0.001 & 0.97 & 0.52 & 0.249 \\
    & 200 & \bf 0.38 & 0.003 & 0.38 & 0.001 & 0.38 & 0.001 & 0.39 & 0.001 & 0.39 & 0.001 & 1.01 & 0.40 & 0.020 \\
    \hline
  \end{tabular}
  \caption{{\bf Prediction mean squared error with ten observations on
      $n_s=200$ test samples.}  Test samples $\sups{\mb y}{8}_i,
    \sups{\mb y}{9}_i $ and $\sups{\mb y}{10}_i$ are treated as the
    response and the rest of the observations are used as the training
    data to estimate parameters used to predict the
    response. Prediction accuracy is measured by mean squared error
    (MSE) between simulated responses and predicted responses. Values
    presented are the mean MSE and standard deviation for $20$ runs of
    each model. Standard error (SE) is missing for SCCA because the
    method is deterministic and SE$=0$.}
  \label{tab:MSEV10}
\end{table}

Across the three BASS methods, MCMC-EM had the most accurate
performance across nearly all simulation settings. However, this
performance boost comes with the price of running a small number of
Gibbs sampling iterations with complexity of $O(k^3 p + k^2 p n)$ per
iteration. When $p$ is large, even a few iterations are
computationally infeasible. PX-EM, on the other hand, has the same
complexity as EM, and showed robust and accurate simulation results
relative to EM. In the following real applications, we used BASS EM
initialized with a small number of iterations of PX-EM.

\section{Real Data Applications}

In this section we considered three real data applications of BASS. In
the first application, we evaluated the prediction performance for
multiple correlated response variables in the Mulan Library
\citep{mulan}. In the second application, we applied BASS to gene
expression data from the Cholesterol and Pharmacogenomic (CAP)
study. The data consist of expression measurements for about ten
thousands genes in 480 lymphoblastoid cell lines (LCLs) under two
experimental conditions~\citep{Mangravite2013,
  brown_integrative_2013}. BASS was used to detect sparse covariance
structures specific to each experimental condition. In the third
application, we applied BASS to approximately $20,000$ newsgroup
documents divided into $20$ newsgroups~\citep{newsgroup_1997} in order
to perform multiclass classification.
%
%

\subsection{Multivariate Response Prediction: The Mulan Library}

The Mulan Library consists of multiple data sets collected for the
purpose of evaluating multi-label predictions \citep{mulan}. This
library was used to test the Bayesian CCA model (GFA in our
simulations) for multi-label prediction vectors converted to multiple
binary label vectors (one-hot encoding)
\citep{klami_bayesian_2013}. There are two observations ($m=2$): the
matrix of labels were treated as one observation ($\sups{\mb Y}{1}$)
and the features were treated as another ($\sups{\mb Y}{2}$).
Recently Mulan added multiple regression data sets with continuous
variables. We chose ten benchmark data sets from the Mulan
Library. Four of them ({\tt bibtex, delicious, mediamill, scene}) have
binary labels as responses and were also studied in
\cite{klami_bayesian_2013}. Another six data sets ({\tt rf1, rf2,
  scm1d, scm20d, atp1d, atp7d}) have continuous responses (Table
\ref{tab:Mulan}). For all data sets, we removed features with
identical values for all samples in the training set as
uninformative. For the continuous response data sets, for each value,
we subtracted the mean and divided by the standard deviation of each
feature.

We ran BASS, sGFA, GFA, and MRD-lin on these ten data sets, and
compared the methods using prediction accuracy. For data sets with
binary labels, we quantified prediction error using the Hamming loss
between the predicted labels and true labels. The predicted labels on
the test samples were calculated using the same thresholding rules as
in~\cite{klami_bayesian_2013}. The value of the threshold was chosen
so that the Hamming loss between estimated labels and true labels in
training set was miminized. We used the R package
\verb;PresenceAbsence; and Matlab function \verb;perfcurve; to find
the thresholds to produce binary classifications from continuous
prediction estimates. In particular, the R package
\verb;PresenceAbsence; selects the threshold by maximizing the percent
correctly classified, which corresponds to minimizing the Hamming
loss.  For continuous target variables, mean squared error (MSE) was
used to evaluate prediction accuracy. We initialized BASS with $500$
factors and $50$ PX-EM iterations. The other models were set to the
default parameters with the number of factors set to
$\min(p_1,p_2,50)$ (see Simulations for details). All of the methods
were run $20$ times, and minimum error values were reported (Tables S1-S11).

BASS achieved the best prediction accuracy in five of the ten data
sets (Table~\ref{tab:Mulan}). For the data sets with a binary
response, sGFA produced the overall best performance compared with
other models, achieving the smallest MSE in all four data sets. GFA
had the most stable results in terms of SE in the four data sets. For
the continuous response, BASS outperformed the other models in four
out of six data sets. GFA again had the most stable MSE compared with
other methods. The good performance of BASS on the data sets with
continuous response variables may be attributed to the structured
sparsity on the loading matrix, achieving the intended gains in
generalization error from flexible regularization. Although the ARD
prior used in GFA did not produce consistently sparse loadings, this
model generated the most stable results.

\begin{table}
\footnotesize
  \centering
  \begin{tabular}{l*{12}{c}}
    \hline
    \multirow{2}{*}{Data Set} & \multirow{2}{*}{$p_1$} & \multirow{2}{*}{$p_2$}  
    & \multirow{2}{*}{$n_{t}$} & \multirow{2}{*}{$n_{s}$}  & \multicolumn{2}{c}{BASS}  
    & \multicolumn{2}{c}{sGFA} & \multicolumn{2}{c}{GFA} & \multicolumn{2}{c}{MRD-lin} \\
    & & & & & Err & SE & Err & SE & Err & SE & Err & SE  \\
    \hline
    bibtex & 1836 & 159 & 4880 & 2515  & 0.014 & 0.001  & 0.014 & 0.001 & \bf 0.014 & $<$1e-3 & 0.014 & 0.001 \\
    delicious & 983 & 500 & 12920 & 3185  & 0.016 & 0.001 & \bf 0.016 & $<$1e-3 & 0.017 & $<$1e-3 & 0.020 & $<$1e-3 \\
    mediamill & 120 & 101 & 30993 & 12914  & \bf 0.032 & 0.001 & 0.032 & 0.005 & 0.034 & $<$1e-3 & 0.043 & $<$1e-3 \\
    scene & 294 & 6 & 1211 & 1196 & 0.131 & 0.016 & \bf 0.123 & 0.029 & 0.130 & 0.002 & 0.138 & 0.026 \\
    \hline
    rf1 & 64 & 8 & 4108 & 5017  & \bf 0.292 & 0.050 & 0.390 & 0.008 & 0.309 & $<$1e-3 & 0.370 & 0.146 \\
    rf2 & 576 & 8 & 4108 & 5017  & \bf 0.271 & 0.027 & 0.478 & 0.004 & 0.427 & 0.001 & 0.438 & 0.160 \\
    scm1d & 280 & 16 & 8145 & 1658  & \bf 0.211 & 0.005 & 0.225 & 0.028 & 0.213 & $<$1e-3 & 0.212 & 0.163 \\
    scm20d & 61 & 16 & 7463 & 1503  & 0.650 & 0.015 & \bf 0.538 & 0.006 & 0.720 & 0.002 & 0.608 & 0.033 \\
    atp1d & 370 & 6 & 237 & 100  & \bf 0.176 & 0.032 & 0.208 & 0.006 & 0.201 & 0.001 & 0.219 & 0.113 \\
    atp7d & 370 & 6 & 196 & 100  & 0.597 & 0.063 & 0.537 & 0.015 & \bf 0.537 & 0.003 & 0.545 & 0.049 \\
    \hline
  \end{tabular}
  \caption{{\bf Multi-variate response prediction from Mulan library.}
    $p_1$: the number of features; $p_2$: the number of responses;
    $n_t$: the number of training samples; $n_s$: the number of test
    samples.  The first four data sets have binary responses, and the
    final six are continuous responses.  For binary responses, error
    (Err) is evaluated using Hamming loss between predicted labels and
    test labels in test samples.  For continuous responses, mean
    squared error (MSE) is used to quantify error.  Values shown
    are the minimum Hamming loss or MSE across $20$ runs of each
    method, and the standard error (SE), are quantified across
    those 20 runs.}
  \label{tab:Mulan}
\end{table}

\subsection{Gene Expression Data Analysis}

We applied our BASS model to gene expression data from the Cholesterol
and Pharmacogenomic (CAP) study, consisting of expression measurements
for $10,195$ genes in $480$ lymphoblastoid cell lines (LCLs) after
24-hour exposure to either a control buffer ($\sups{\mb Y}{1}$) or
$2\mu M$ simvastatin acid ($\sups{\mb
  Y}{2}$)~\citep{Mangravite2013,brown_integrative_2013}. In this
example, the number of observations $m=2$, representing gene
expression levels on the same samples and genes after the two
different exposures. The expression levels were prepossessed to adjust
for experimental traits (batch effects and cell growth rate) and
clinical traits of donors (age, BMI, smoking status and gender).  We
projected the adjusted expression levels to the quantiles of a
standard normal within gene to control for outlier effects and applied
BASS with the initial number of factors set to $k=2,000$. We performed
parameter estimation $100$ times on these data with $100$ iterations
of PX-EM to initialize EM. Across these $100$ runs, the estimated
number of recovered factors was approximately $870$ (Table S$2$), recovering only a few dense factors
(Table~S12) likely due to the adjustments made in the
prepossessing step. The total percent of variance explained (PVE) by
the recovered latent structure was $14.73\%$, leaving $85.27\%$ of the
total variance to be captured in the residual error.

We computed the PVE of the sparse factors alone
(Figure~S9A). The PVE for the $h^{th}$ factor was
calculated as the variance explained by the $h^{th}$ factor divided by
the total variance: $tr(\mb \lambda_{\cdot h} \mb \lambda^T_{\cdot h})
/ tr( \mb \Lambda \mb \Lambda^T + \mb \Sigma)$. Shared sparse factors
explained more variance than observation-specific sparse factors,
suggesting that variation in expression levels across genes was driven
by structure shared across the exposures to a greater degree than by
exposure-specific structure.  Moreover, $87.5\%$ of the
observation-specific sparse factors contained fewer than $100$ genes,
whereas only $0.7\%$ of those factors had greater than $500$
genes. The shared sparse factors had, on average, more genes than the
observation-specific factors: $72\%$ shared sparse factors had
fewer than $100$ genes, whereas $4.5\%$ had greater than
$500$ genes. (Figure~S9B).

The sparse factors specific to each observation characterize the local
sparse covariance estimates.  As we pursue more carefully
elsewhere~\citep{Gao2014}, we used observation-specific sparse factors
to a construct a gene co-expression network that is uniquely found in
the samples from that exposure while explicitly controlling for shared
covariance across exposures~\citep{zou2013}. The problem of
constructing condition specific co-expression networks has been
studied by both machine learning and computational biology communities
\citep{li_genome-wide_2002, ma_cosine_2011}.  Our BASS model provides
an alternative approach to solve this problem.  We denote $\sups{\mb
  B}{w}_s$ as the sparse loadings in $\sups{\mb B}{w}$ ($w \in
\{1,2\}$) and $\sups{\mb X}{w}_s$ as the factors corresponding to the
sparse loadings for observation $w$. Then, $\sups{\mb \Omega}{w}_s =
\sups{\mb B}{w}_s Var(\sups{\mb X}{w}_s) (\sups{\mb B}{w}_s)^T +
\sups{\mb \Sigma}{w}$ represents the regularized estimate of the
covariance matrix specific to each observation after controlling for
the contributions of the dense factors.

In our model, $Var(\sups{\mb X}{w}_s) = \mb I$, and so the covariance
 matrix becomes $\sups{\mb \Omega}{w}_s = \sups{\mb B}{w}_s (\sups{\mb
 B}{w}_s)^T + \sups{\mb \Sigma}{w}$. We inverted this positive
 definite covariance matrix to get a precision matrix $\sups{\mb R}{w}
 = (\sups{\mb \Omega}{w}_s)^{-1}$. The partial correlation between
 gene $j_1$ and $j_2$, representing the correlation between the two
 features conditioned on the remaining features, is then calculated by
 normalizing each entry in the precision
 matrix~\citep{edwards_introduction_2000, schafer_empirical_2005}:
\begin{align}
  \sups{\rho}{w}_{j_1 j_2} = - \frac{\sups{r}{w}_{j_1
      j_2}}{\sqrt{\sups{r}{w}_{j_1 j_1} \sups{r}{w}_{j_2
        j_2}}}. \notag
\end{align}
A partial correlation that is (near) zero for two genes suggests that
they are conditionally independent; non-zero partial correlation
implies a direct relationship between two genes, and a network edge is
added between the pair of genes. The resulting undirected network is
an instance of a Gaussian Markov random field, also known as a
Gaussian graphical model~\citep{edwards_introduction_2000,
  koller_probabilistic_2009}. We note that BASS was the only method
that allows for construction of the condition specific network: sGFA could
not be applied to data of this magnitude, GFA did not shrink the
column selection sufficiently to recover sparsity in the condition
specific covariance matrix, and SCCA only recovers shared sparse
projections.

We used the following method to combine the results of $100$ runs to
construct a single observation-specific gene co-expression network for
each observation. For each run, we first constructed a network by
connecting genes with partial correlation greater than a threshold
($0.01$). Then we combined the $100$ run-specific networks to
construct a single network by removing all network edges that appear
in fewer than than $50$ ($50\%$) of the networks. The two
observation-specific gene co-expression networks contain a total of
$160$ genes and $1,244$ edges (buffer treated, Figure
{\ref{fig:CAPNetwork}A), and $154$ genes and $1,030$ edges
  (statin-treated, Figure \ref{fig:CAPNetwork}B), respectively.

\begin{figure}
  \centering
  \includegraphics[width=0.9\textwidth]{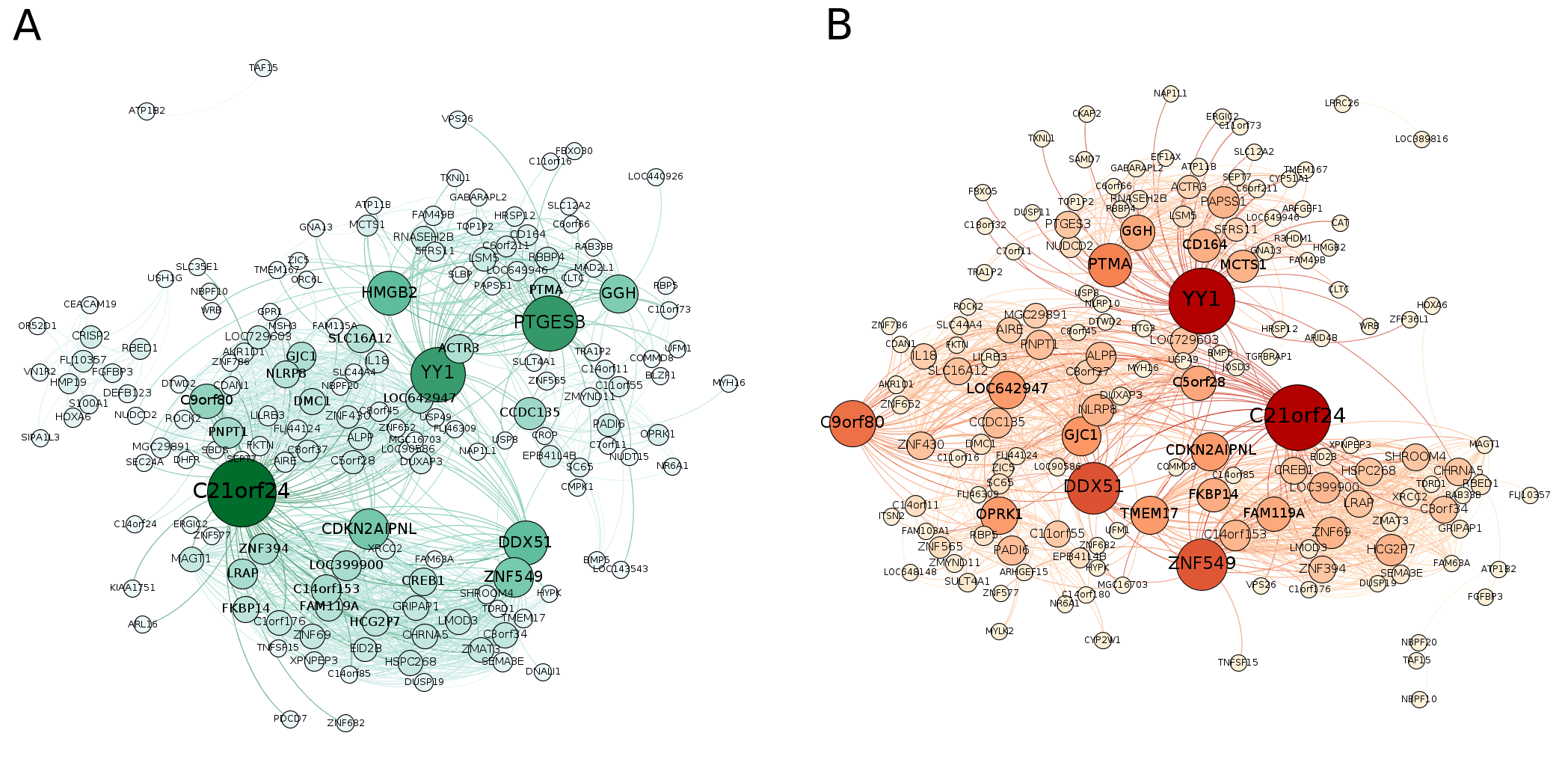}
  \caption{{\bf Observation-specific gene co-expression networks from
      the CAP data.} The two networks represent the co-expressed genes
    specific to buffer-treated samples (Panel A) and statin-treated
    samples (Panel B).  The node size is scaled according to
    the number of shortest paths from all vertices to all others that
    pass through that node (\emph{betweenness centrality}).}
  \label{fig:CAPNetwork}
\end{figure}

\subsection{$20$ Newgroups Analysis}

In this application, we appied BASS and related models to the problem
of multiclass label prediction in the $20$ Newsgroups data
\citep{newsgroup_1997}. The documents were processed so that
duplicates and headers were removed, resulting $18,846$ documents. The
data were downloaded using the \verb$scikit-learn$ Python package
\citep{scikit-learn}. We converted the raw data into TF-IDF feature
vectors and selected $319$ words using SVM feature selection from
\verb$scikit-learn$. One document had a zero vector across the subset
of vocabulary words and was removed. We further held out $100$ documents
at random from each newsgroup as test data (Table S14). 

We applied BASS to the transposed data matrices with the $20$
newsgroups as $20$ observations. We set the initial number of factors
to $k=1000$ and ran EM $100$ times from random starting points, each
with $100$ initial PX-EM iterations for initialization. There were
on average $820$ factors recovered in each iteration.

We analyzed the newsgroup-specific words in following way. We
calculated the Pearson correlation of each estimated loading and
newsgroup indicator vectors consisting of ones for all of the
documents in one newsgroup and zeros for documents in the other
groups. Then, for each newsgroup, the loadings with the ten largest
absolute value correlation coefficients were used to find the ten
words with the largest absolute value factor scores.  The results from
one run include, for example, the \verb;rec.autos; newsgroup with
`car', `dealer' and `oil' as top words, and the
\verb;rec.sport.baseball; newsgroup with `baseball', `braves' and
`runs' as top words (Table \ref{tab:top10words}).

We further partitioned the newsgroups into six classes according to
subject matter to analyze the shared words across newsgroups subgroups
(Table \ref{tab:shared10words}). As above, we calculated the Pearson
correlation with the binary indicator vectors for documents in
newsgroup subgroups, and we analyzed the top ten words in the ten
factors with largest absolute value correlation coefficients with these
subsets of newsgroups (Table \ref{tab:shared10words}). We found, for
example, that the newsgroups \verb;talk.religion.misc;,
\verb;alt.atheism; and \verb;soc.religion.christian; had `god',
`bible' and `christian' as top shared words. Examining one of the
selected shared loadings for this newsgroup subgroup (Figure
\ref{fig:newsgroup}A), we noticed that documents outside of these three
newsgroups, for the most part, have negligible loadings. 
This analysis highlights the ability of BASS to recover meaningful
shared structure among 20 observations.

\begin{landscape}
\begin{table}
  \centering
  \begin{tabular}{c*{19}{c}}
    \hline
    \multicolumn{2}{c}{\texttt{alt.atheism}} & \multicolumn{2}{c}{\texttt{comp.graphics}} & \multicolumn{2}{c}{\texttt{comp.os.ms-windows.misc}} & 
    \multicolumn{2}{c}{\texttt{comp.sys.ibm.pc.hardware}} & \multicolumn{2}{c}{\texttt{comp.sys.mac.hardware}} \\
    \hline
    islam & atheism & graphics & polygon & windows & file & ide & drive & mac & powerbook \\
    keith & mathew & 3d & gif & thanks & go & scsi & motherboard & apple & quadra \\
    okcforum & atheists & tiff & images & of & dos & controller & thanks & quadra & iisi \\
    atheism & livesey & image & format & cica & microsoft & vlb & ide & duo & centris \\
    livesey & of & image & pov  & dos & the & bios & isa & centris & mac \\
    \hline
    \multicolumn{2}{c}{\texttt{comp.windows.x}} & \multicolumn{2}{c}{\texttt{misc.forsale}} & \multicolumn{2}{c}{\texttt{rec.autos}} &
    \multicolumn{2}{c}{\texttt{rec.motorcycles}} & \multicolumn{2}{c}{\texttt{rec.sport.baseball}} \\
    \hline
    window & mit & sale & offer & car & dealer & dod & bmw & baseball & hitter\\ 
    motif & lcs & sale & forsale & cars & oil & bike & riding & braves & ball \\
    server & motif & for & the & engine & toyota & motorcycle & bikes & runs & year \\
    widget & xterm & sell & shipping & ford & eliot & ride & dod & phillies & players \\
    lcs & code & condition & offer & cars & cars & bike & bike & sox & players\\
    \hline
    \multicolumn{2}{c}{\texttt{rec.sport.hockey}} & \multicolumn{2}{c}{\texttt{sci.crypt}} &
    \multicolumn{2}{c}{\texttt{sci.electronics}} & \multicolumn{2}{c}{\texttt{sci.med}} & \multicolumn{2}{c}{\texttt{sci.space}} \\
    \hline
    hockey & bruins & encryption & crypto & circuit & radio & geb & msg & it & people \\
    nhl & pens & clipper & nsa & voltage & copy & medical & doctor & space & orbit \\
    game & detroit & chip & nsa & amp & battery & diet & disease & for & henry \\
    team & season & key & pgp & electronics & tv & cancer & geb & digex & moon \\
    leafs & espn & des & tapped & audio & power & photography & doctor & for & shuttle\\
    \hline
    \multicolumn{2}{c}{\texttt{soc.religion.christian}} & \multicolumn{2}{c}{\texttt{talk.politics.guns}} &
    \multicolumn{2}{c}{\texttt{talk.politics.mideast}} & \multicolumn{2}{c}{\texttt{talk.politics.misc}} &
    \multicolumn{2}{c}{\texttt{talk.religion.misc}}\\
    \hline
    god & sin & atf & fbi & israeli & israeli & cramer & government & sandvik & morality \\
    clh & bible & firearms & stratus & jews & armenians & optilink & drugs & koresh & jesus \\ 
    church & petch & guns & batf & israel & armenian & kaldis & president & sandvik & religion \\
    christian & mary & gun & stratus & arab & jake & clinton & br & bible & god \\
    heaven & church & handheld & waco & armenians & jewish & cramer & tax & christian & objective\\
    \hline
  \end{tabular}
  \caption{{\bf Most significant words in the newsgroup-specific factors for 20 newsgroups.} For each newsgroup, we include the top ten words in the newsgroup-specific components.}
  \label{tab:top10words}
\end{table}
\end{landscape}

\begin{table}
  \centering
  \begin{tabular}{c*{5}{c}}
    \hline
    Newsgroup classes & \multicolumn{2}{c}{Top ten shared words} & Newsgroup classes & \multicolumn{2}{c}{Top ten shared words} \\
    \hline
    \multirow{5}{*}{\thead {\texttt{comp.graphics}\\ \texttt{comp.os.ms-windows.misc}\\
        \texttt{comp.sys.ibm.pc.hardware}\\ \texttt{comp.sys.mac.hardware}\\ \texttt{comp.windows.x}}}
    & {windows} & {dos} & \multirow{5}{*}{\thead{\texttt{misc.forsale}}} & sale & shipping \\
    & {thanks} & {mac} & & sell & ca \\
    & {graphics} & {go} & & condition &wanted \\
    & {file} & {scsi} & & offer & thanks \\
    & {window} & {server}  &   & forsale & edu \\
    \hline
    \multirow{5}{*}{\thead{\texttt{rec.autos}\\ \texttt{rec.motorcycles} \\
        \texttt{rec.sport.baseball} \\ \texttt{rec.sport.hockey}}}
    & {dod} & {baseball} & \multirow{5}{*}{\thead{ \texttt{talk.politics.misc} \\
        \texttt{talk.politics.guns} \\
        \texttt{talk.politics.mideast}}} & {government} & {it} \\
    & {car} & {ride} &     & {israeli}   & {israel} \\
    & {bike} & {cars} &     & {jews} & {gun} \\
    & {motorcycle} & {bmw} &     & {atf} & {guns} \\
    & {game} & {team} &     & {firearms} & {batf} \\
    \hline
    \multirow{5}{*}{\thead {\texttt{sci.crypt} \\ \texttt{sci.electronics} \\
        \texttt{sci.med} \\ \texttt{sci.space}}}
    & {clipper} & {henry} & 
    \multirow{5}{*}{\thead{\texttt{talk.religion.misc} \\
        \texttt{alt.atheism} \\
        \texttt{soc.religion.christian}}} & {god} & {bible} \\
    & {encryption} & {orbit} &     & {bible} & {heaven} \\ 
    & {space} & {people} &     & {christian}  & {sandvik} \\
    & {chip} & {circuit} &     & {clh} & {faith} \\
    & {digex} & {voltage} &     & {jesus} & {church}\\
    \hline
  \end{tabular}
  \caption{{\bf Top ten words in the factors shared among specific subgroups of newsgroups.} In the shared recovered components corresponding to subsets of newsgroups, we show the ten most significant words in these shared components for six different subsets of newsgroups.}
  \label{tab:shared10words}
\end{table}

To assess prediction quality, we used the loadings and factors from
estimated from the training set to classify documents to one of 20
newsgroups in the held-out test set. To estimate the loadings in the
test set, we left-multiplied the test data matrix by the Moore-Penrose
pseudoinverse of factors estimated from training data. This gave a
rough estimate of the loading matrix for test data. Then test labels
were predicted using ten nearest neighbors using the loading rows
estimated for the labeled training documents. For the $200$ test
documents, BASS achieved approximately $58.3\%$ accuracy (Hamming
loss; Figure \ref{fig:newsgroup}B).
Because
some of the newsgroups were closely related to each other with respect
to topic, we partitioned the 20 newsgroups into six topics according
to subject matter. Then, the ten nearest neighbors were used to
predict the topic label of the held-out test data. In this experiment,
BASS achieved approximately $74.12\%$ accuracy (Hamming loss; Figure
\ref{fig:newsgroup}C; Table S3).

\begin{figure}[h!]
  \centering
  \includegraphics[width=0.95\textwidth]{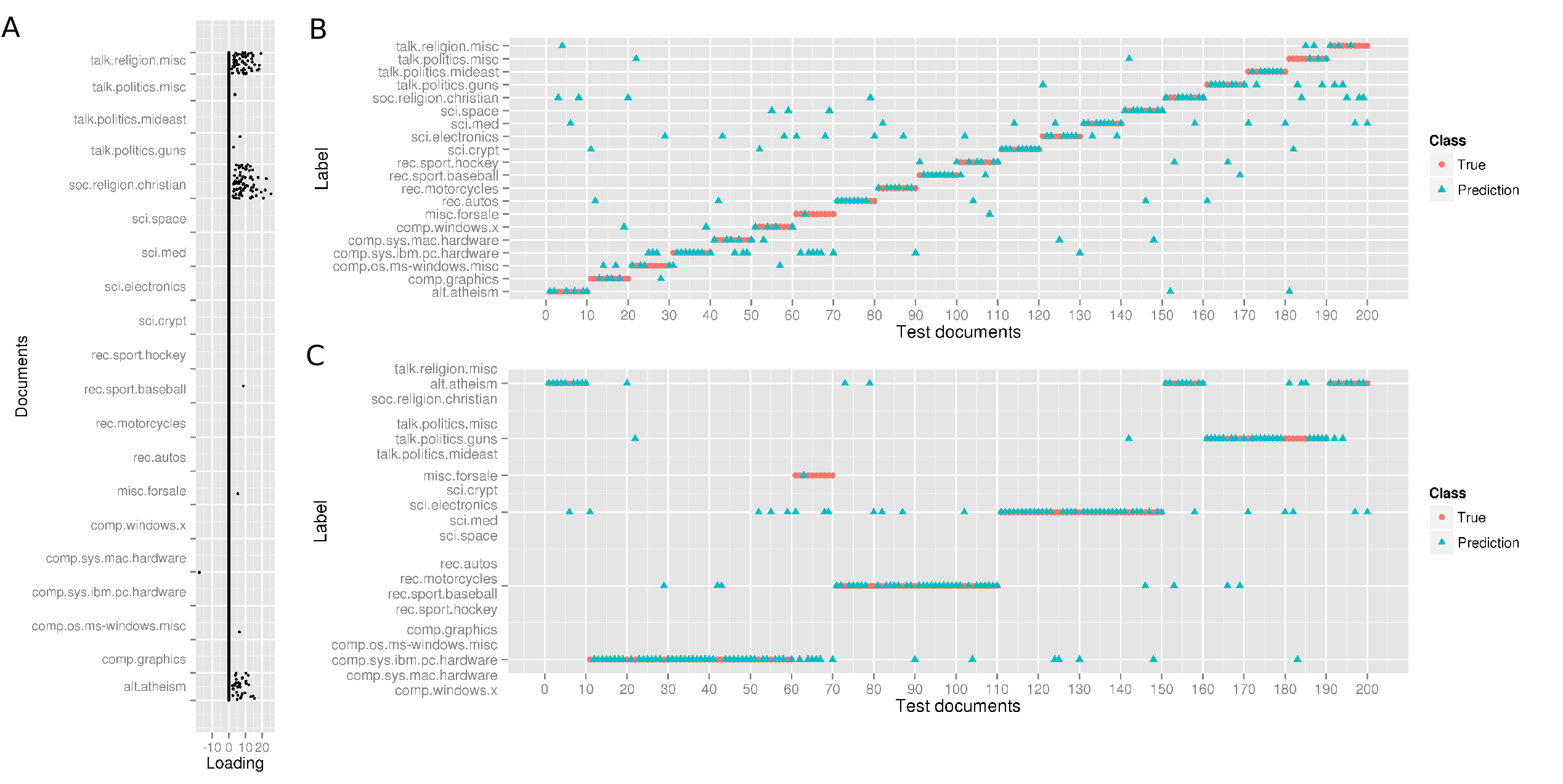}
  \caption{{\bf Newsgroup prediction on $200$ test documents.} Panel
    A: One factor loading selected as shared by three newsgroups
    (\texttt{talk.religion.misc}, \texttt{alt.atheism} and
    \texttt{soc.religion.christian}). Panel B: $20$ Newsgroups
    predictions on $100$ test documents using ten nearest neighbors
    from loadings estimated from the training data. Panel C: Document
    subgroup predictions based on six groups of similar newsgroups
    using ten nearest neighbors based on loadings estimated from the
    training data.}
  \label{fig:newsgroup}
\end{figure}

\section{Discussion}

There exists a rich set of methods to explore latent structure in
paired or multiple observations jointly
(e.g., \cite{parkhomenko_sparse_2009, witten_extensions_2009,
  zhao_co-module_2012} among others).  The multiple trajectories of
interpretation of these approaches as linear factor analysis models
includes the original inter-battery and multi-battery models
\citep{browne_maximum-likelihood_1979, browne_factor_1980}, the
probabilistic CCA model \citep{bach_probabilistic_2005}, the sparse
probabilistic projection \citep{archambeau_sparse_2008}, and, most
recently, the Bayesian CCA model \citep{klami_bayesian_2013} and GFA
models \citep{klami_group_2014}. Only recently has the idea of
column-wise shrinkage, or group-wise sparsity, been applied to develop
useful models for this problem. The advantage of column-wise shrinkage
is to decouple portions of the latent space from specific observations
and adaptively select the number of factors.

While the innovation of column-wise sparsity is primarily due to the
ideas developed in the Bayesian CCA model
\citep{virtanen_bayesian_2011}, additional layers of shrinkage were
required to create both column-wise and element-wise sparsity as is
essential in real data analyses.  The most recent attempt to develop
such combined effects is the sGFA model
\citep{khan_identification_2014} using a combination of an
element-wise ARD prior with spike-and-slab prior for column selection.
In our work here, we developed the necessary Bayesian prior and
methodology framework to realize these advantages for the analysis of
large data sets. In particular, we developed a structured sparse prior
using three hierarchical layers of the three parameter Beta ($\TPB$)
distribution. This carefully formulated prior combines both
column-wise and element-wise shrinkage with global shrinkage to adapt
the level of sparsity---both column-wise and element-wise---to the
underlying data, creating robustness to parameter settings that cannot
be achieved using a single-layer ARD prior.  The BASS model moreover
allows sparse and dense factor loadings, which proved essential for
real data scenarios that have this structure and has been pursued in
classical statistics~\citep{chandrasekaran_sparse_2009,
  candes_robust_2011, zhou_manifold_2011}. We showed in the
simulations that this level of regularization is essential for
problems in the $p \gg n$ data scenario, which motivated this
work.  With the assumption of full column rank of dense loadings and
one single observation, our model provides a Bayesian solution to the
sparse and low-rank decomposition problem.

Column-wise shrinkage in BASS is achieved using the
observation-specific global and column-specific $\TPB$ priors. With
current parameter settings, it is equivalent to the horseshoe prior
put on the entire column.  The horseshoe prior has been shown to
induce better shrinkage effects compared to the ARD prior, the Laplace
prior (Bayesian lasso), and other similar shrinkage priors while
remaining computationally tractable
\citep{carvalho_horseshoe_2010}. In addition, our local shrinkage
encourages element-wise sparsity. A two component mixture allows both
dense and sparse factors to be recovered for any subset of
observations. These shared factors have an interpretation as a
supervised low-rank projection when we consider one observation as
supervised labels (e.g., as in the Mulan Library data sets). To the
best of our knowledge, the BASS model is the first model in either the
Bayesian or classical statistical literature that is able to capture
low-rank and sparse decompositions among multiple observations.

We developed three algorithms that estimate the posterior distribution
of our model or MAP parameter values. We found that EM with random
initialization would occasionally get stuck in poor local optima. This
motivated the development of a fast and robust PX-EM algorithm by
introducing an auxiliary rotation
matrix~\citep{rockova2014fast}. Initializing EM with PX-EM enabled EM
to escape from poor initializations, as illustrated in simulations.
In addition, our PX-EM and EM algorithms have a better computational
complexity than the two competing approaches, GFA and sGFA, allowing
for real data application.


Extensions of multiple observation linear factor models to non-linear
or non-Gaussian models have been studied recently
\citep{salomatin_multifield_2009, damianou_manifold_2012,
  klami_group-sparse_2013,klami_polya-gamma_2014}. The ideas in this
paper of inducing structured sparsity in the loadings has parallels in
both of these settings. For example, we may consider structured
Gaussian process kernels in the non-linear setting, where structure
corresponds to known shared and observation-specific structure. A
number of issues remain, including robustness of the recovered sparse
factors across runs, scaling these methods to current studies in
genomics, neuroscience, or text analysis, allowing for missing data,
and developing approaches to include domain-specific structure across
samples or features.

\acks{}
The authors would like to thank David Dunson and Sanvesh Srivastava
for helpful discussions. The authors also appreciate the comments from
Arto Klami and three anonymous reviewers. BEE, CG, and SZ were funded
by NIH R00 HG006265 and NIH R01 MH101822.  SZ was also funded in part
by NSF DMS-1418261 and a Graduate Fellowship from Duke University. SM
was supported in part by NSF DMS-1418261, NSF DMS-1209155, NSF
IIS-1320357, and AFOSR under Grant FA9550-10-1-0436. All data are
publicly available: the gene expression data were acquired through GEO
GSE36868.  We acknowledge the PARC investigators and research team,
supported by NHLBI, for collection of data from the Cholesterol and
Pharmacogenetics clinical trial.


\newpage

\appendix

\section{Markov chain Monte Carlo (MCMC) Algorithm for Posterior
  Inference}
\label{app:MCMC}

We first derive the MCMC algorithm with Gibbs sampling steps for
BASS. We write the joint distribution of the full model as
\begin{align}
  p(\mb Y, \mb X, &\mb \Lambda, \mb \Theta, \mb \Delta, \mb \Phi, \mb
  T, \mb \eta, \mb \gamma, \mb Z, \mb \Sigma, \mb \pi) \notag \\
  &= p(\mb Y|\mb \Lambda, \mb X, \mb \Sigma) p(\mb X)  \notag \\
  & \qquad \times p(\mb \Lambda | \mb \Theta) p(\mb \Theta | \mb
  \Delta, \mb Z, \mb \Phi) p(\mb \Delta | \mb \Phi) p(\mb \Phi | \mb
  T) p(\mb T| \mb \eta) p(\mb \eta | \mb \gamma) \notag \\
  & \qquad \times p(\mb \Sigma) p(\mb Z|\mb \pi) p(\mb
  \pi), \notag
\end{align}
where $\mb \Theta = \{\sups{\theta}{w}_{jh}\}$, $\mb \Delta =
\{\sups{\delta}{w}_{jh}\}$, $\mb \Phi = \{\sups{\phi}{w}_h\}$, $\mb T
= \{\sups{\tau}{w}_h \}$, $\mb \eta = \{\sups{\eta}{w}\}$ and $\mb
\gamma = \{\sups{\gamma}{w}\}$ are the collections of
global-factor-local $\TPB$ prior parameters.

The full conditional distribution for latent factor $\mb x_i$ is
\begin{align}
  \mb x_i|- \sim \N_k \bigg( (\mb \Lambda^T \mb \Sigma^{-1} \mb
  \Lambda + \mb I)^{-1} \mb \Lambda^T \mb \Sigma^{-1} \mb y_i, (\mb
  \Lambda^T \mb \Sigma^{-1} \mb \Lambda + \mb I)^{-1} \bigg), \label{eq:xcond}
\end{align}
for $i = 1,\cdots,n$.

For $\mb \Lambda$, we derive the full conditional distributions of its
$p$ rows, $\mb \lambda_{j\cdot}$ for $j = 1\cdots p$,
\begin{align}
  \mb \lambda_{j\cdot}^T | - \sim \N_k \bigg((\sigma^{-2}_j \mb X
  \mb X^T + \mb D^{-1}_j ) ^{-1} \sigma^{-2}_j \mb X \mb
  y_{j\cdot}^T, (\sigma^{-2}_j \mb X \mb X^T + \mb D^{-1}_j
  )^{-1}\bigg), \notag
\end{align}
where 
\begin{align}
  \mb D^{-1}_j =
  \diag\bigg((\sups{\theta}{w_j}_{j1})^{I(\sups{z}{w_j}_1 = 1)}
  (\sups{\phi}{w_j}_1)^{I(\sups{z}{w_j}_1 = 0)}, \cdots,
  (\sups{\theta}{w_j}_{jk})^{I(\sups{z}{w_j}_k = 1)}
  (\sups{\phi}{w_j}_k)^{I(\sups{z}{w_j}_k = 0)}\bigg), \notag
\end{align}
and $w_j$ represents the observation that the $j^{th}$ row belongs to.

The full conditionals of $\sups{\theta}{w}_{jh}$,
$\sups{\delta}{w}_{jh}$ and $\sups{\phi}{w}_h$ with $\sups{z}{w}_h =
1$ are
\begin{align}
  \sups{\theta}{w}_{jh} | - &\sim \GIG\left(a - 1/2, 2
  \sups{\delta}{w}_{jh}, (\sups{\lambda}{w}_{jh})^2\right), \notag \\
  \sups{\delta}{w}_{jh} | - &\sim Ga\left(a + b, \sups{\phi}{w}_h +
  \sups{\theta}{w}_{jh}\right), \notag\\
  \sups{\phi}{w}_h | - &\sim Ga\left(p_w b + c,\sum_{j=1}^{p_w}
  \sups{\delta}{w}_{jh} + \sups{\tau}{w}_h\right), \notag
\end{align}
where $\GIG$ is the generalized inverse Gaussian distribution.

The full conditional of $\sups{\phi}{w}_h$ with $\sups{z}{w}_h=0$ is
\begin{align}
  \sups{\phi}{w}_h | - \sim \GIG\left(c - p_w/2, 2\sups{\tau}{w}_h,
  \sum_{j=1}^{p_w} (\sups{\lambda}{w}_{jh})^2\right). \notag
\end{align}
The full conditionals of the remaining parameters are
\begin{align}
  \sups{\tau}{w}_h | - &\sim Ga(c + d, \sups{\phi}{w}_h +
  \sups{\eta}{w}), \notag \\
  \sups{\eta}{w} | - &\sim Ga\left(kd + e, \sups{\gamma}{w} + \sum_{h=1}^k
  \sups{\tau}{w}_h\right), \notag \\
  \sups{\gamma}{w} | - &\sim Ga(e + f, \sups{\eta}{w} +
  \nu), \notag \\
  \sups{\pi}{w} | - &\sim Beta\left(1 + \sum_{h=1}^k \sups{z}{w}_h, 1+ k -
  \sum_{h=1}^k \sups{z}{w}_h\right). \notag
\end{align}
The full conditional of $\sups{z}{w}_h$ is
\begin{align}
  \Pr(\sups{z}{w}_h = 1|-) &\propto \sups{\pi}{w} \prod_{j=1}^{p_w}
  \N(\sups{\lambda}{w}_{jh}; 0, \sups{\theta}{w}_{jh})
  Ga(\sups{\theta}{w}_{jh}; a, \sups{\delta}{w}_{jh})
  Ga(\sups{\delta}{w}_{jh}; b, \sups{\phi}{w}_h), \notag\\
  \Pr(\sups{z}{w}_h = 0|-) &\propto (1-\sups{\pi}{w}) \prod_{j=1}^{p_w}
  \N(\sups{\lambda}{w}_{jh}; 0, \sups{\phi}{w}_h). \notag
\end{align}
We further integrate out $\sups{\delta}{w}_{jh}$ in $\Pr(\sups{z}{w}_h
= 1|-)$:
\begin{align}
  \Pr(\sups{z}{w}_h = 1|-) &\propto \sups{\pi}{w} \prod_{j=1}^{p_w}
  \int \N(\sups{\lambda}{w}_{jh}; 0, \sups{\theta}{w}_{jh})
  Ga(\sups{\theta}{w}_{jh}; a, \sups{\delta}{w}_{jh})
  Ga(\sups{\delta}{w}_{jh}; b, \sups{\phi}{w}_h) d
  \sups{\delta}{w}_{jh} \notag \\
  &= \sups{\pi}{w} \prod_{j=1}^{p_w} \N(\sups{\lambda}{w}_{jh}; 0,
  \sups{\theta}{w}_{jh}) \frac{\Gamma(a+b)}{\Gamma(a)
    \Gamma(b)}\frac{(\sups{\theta}{w}_{jh})^{a-1}
    (\sups{\theta}{w}_h)^b}{(\sups{\theta}{w}_{jh} +
    \sups{\phi}{w}_h)^{a+b}}. \notag
\end{align}
The full conditional of $\sigma^{-2}_j$ for $j = 1,\cdots,p$ is
\begin{align}
  \sigma^{-2}_j | - \sim Ga\bigg(n/2 + a_\sigma, 1/2(\mb y_{j\cdot} -
  \mb \lambda_{j\cdot} \mb X) (\mb y_{j\cdot} - \mb \lambda_{j\cdot}
  \mb X)^T + b_\sigma \bigg). \notag
\end{align}

\section{Variational Expectation Maximization (EM) Algorithm for MAP
  Estimates}
\label{app:EM}

{\bf Expectation Step}: Given model parameters, the distribution of
latent factor $\mb X$ was stated in Appendix A
(Equation~\ref{eq:xcond}). The expected sufficient statistics of
$\mb X$ can be derived
\begin{align}
  \la \mb x_{\cdot i} \ra &= (\mb \Lambda^T \mb \Sigma^{-1} \mb \Lambda
  + \mb I)^{-1} \mb \Lambda^T \mb \Sigma^{-1} \mb y_{\cdot
    i}, \label{eq:xess} \\
  \la \mb x_{\cdot i} \mb x_{\cdot i}^T \ra &= \la \mb x_{\cdot i} \ra
  \la \mb x_{\cdot i} \ra^T + (\mb \Lambda^T \mb \Sigma^{-1} \mb
  \Lambda + \mb I)^{-1}. \label{eq:xxess}
\end{align}
The expectation of the indicator variable $\sups{\rho}{w}_h = \la
\sups{z}{w}_h \ra $ is
\begin{align}
  \sups{\rho}{w}_h = \frac{\sups{\pi}{w} \prod_{j=1}^{p_w}
    \N(\sups{\lambda}{w}_{jh}; 0, \sups{\theta}{w}_{jh})
    Ga(\sups{\theta}{w}_{jh}; a, \sups{\delta}{w}_{jh})
    Ga(\sups{\delta}{w}_{jh}; b, \sups{\phi}{w}_h)} {(1-\sups{\pi}{w})
    \prod_{j=1}^{p_w} \N(\sups{\lambda}{w}_{jh}; 0, \sups{\phi}{w}_h)
    + \sups{\pi}{w} \prod_{j=1}^{p_w} \N(\sups{\lambda}{w}_{jh}; 0,
    \sups{\theta}{w}_{jh}) Ga(\sups{\theta}{w}_{jh}; a,
    \sups{\delta}{w}_{jh}) Ga(\sups{\delta}{w}_{jh}; b,
    \sups{\phi}{w}_h)}. \notag
\end{align}
{\bf Maximization Step}: The log posterior of $\mb \Lambda$ is written
as
\begin{align}
  \log (p(\mb \Lambda | -)) \propto \tr \bigg(\mb \Sigma^{-1} \mb
  \Lambda \mb S^{XY}\bigg) - \frac{1}{2} \tr\bigg(\mb \Lambda^T \mb
  \Sigma^{-1} \mb \Lambda \mb S^{XX} \bigg) - \frac{1}{2} \sum_{h=1}^k
  \mb \lambda_{\cdot h}^T \mb D_h \mb \lambda_{\cdot h}, \notag
\end{align}
where 
\begin{align}
  \mb D_h &= \diag\bigg(\frac{\sups{\rho}{1}_h}{\sups{\theta}{1}_{1h}}
  + \frac{1-\sups{\rho}{1}_h}{\sups{\phi}{1}_h},\cdots,
  \frac{\sups{\rho}{m}_h}{\sups{\theta}{m}_{p_m h}} +
  \frac{1-\sups{\rho}{m}_h}{\sups{\phi}{m}_h} \bigg), \notag \\
  \mb S^{XY} &= \sum_{i=1}^n \la \mb x_{\cdot i} \ra \mb y_{\cdot
    i}^T, \textrm{~and~} \mb S^{XX} = \sum_{i=1}^n \la \mb x_{\cdot i}
  \mb x_{\cdot i}^T \ra. \notag
\end{align}
We take the derivative with respect to the loading column $\mb
\lambda_{\cdot h}$ to get the MAP estimate. The derivative of first part in the
right hand side is
\begin{align}
  \frac{\partial \tr (\mb \Sigma^{-1} \mb \Lambda \mb
    S^{XY})}{\partial \mb \lambda_{\cdot h}} &= (\mb 1_k^h \otimes \mb
  I_p) \times \mbox{vec}[\mb \Sigma^{-1} \mb S^{YX}] =
  \mbox{vec}\bigg(\mb \Sigma^{-1} \mb S^{YX} \mb 1_k^h \bigg) \notag \\
  &= \mb \Sigma^{-1} \mb S^{YX} \mb 1_k^h, \notag
\end{align}
where $\mbox{vec}$ is the vectorization of a matrix, $ \mb 1_k^h \in
\mathbb{R}^{k \times 1}$ is a zero vector with a single $1$ in the
$h^{th}$ element, and $\mb S^{YX} = (\mb S^{XY})^T$. For the second
part
\begin{align}
  \frac{\partial \tr (\mb \Lambda^T \mb \Sigma^{-1} \mb \Lambda \mb
    S^{XX} )}{\partial \mb \lambda_{\cdot h}} &= 2 (\mb 1_k^h \otimes
  \mb I_p) \times \mbox{vec}[\mb \Sigma^{-1} \Lambda \mb S^{XX}]
  = 2 \times \mbox{vec}\bigg(\mb \Sigma^{-1} \mb \Lambda \mb S^{XX} \mb 1_k^h\bigg) \notag \\
  &= 2 \mb \Sigma^{-1} \mb \Lambda \mb S^{XX} \mb 1_k^h. \notag
\end{align}
For the third part, the derivative is $\mb D_h \mb \lambda_{\cdot
  h}$. The MAP estimate of $\mb \lambda_{\cdot h}$ is found by setting
the derivative to zero:
\begin{align}
  \hat{\mb \lambda}_{\cdot h} = [\mb S^{XX}_{hh} \mb I_p + \mb \Sigma
  \mb D_h]^{-1} \bigg(\mb S^{YX}_{\cdot h} - \sum_{h' \ne h} \mb
  \lambda_{\cdot h'} \mb S^{XX}_{h' h}\bigg), \notag
\end{align}
where $\mb S^{XX}_{ij}$ is the $(i, j)^{th}$ element of $\mb S^{XX}$,
and $\mb S^{YX}_{\cdot h}$ is the $h^{th}$ column of $\mb S^{YX}$. The
matrix inverse is for a diagonal matrix; thus $\hat{\mb
  \lambda}_{\cdot h}$ can be calculated efficiently. The MAP estimate
of the other model parameters can be obtained from their full
conditional distributions with the variables replaced by their
expectations. We list the parameter updates for those variables below
\begin{align}
  \sups{\hat{\theta}}{w}_{jh} &= \frac{2a - 3 + \sqrt{(2a - 3)^2 + 8
      (\sups{\lambda}{w}_{jh})^2 \sups{\delta}{w}_{jh} }}{4
    \sups{\delta}{w}_{jh}}, \notag \\
  \sups{\hat{\delta}}{w}_{jh} &= \frac{a+b}{\sups{\theta}{w}_{jh} +
    \sups{\phi}{w}_h }, \notag \\
  \sups{\hat{\phi}}{w}_h &= \frac{p'-1 + \sqrt{(p'-1)^2 + a' b'}}{a'}, \textrm{with} \notag \\
  & \qquad p' = \sups{\rho}{w}_h p_w b - (1-\sups{\rho}{w}_h) p_w/2 +
  c, \notag \\
  & \qquad a' = 2 (\sups{\rho}{w}_h \sum_{j = 1}^{p_w}
  \sups{\delta}{w}_{jh} + \sups{\tau}{w}_h), \notag \\
  & \qquad b' = (1-\sups{\rho}{w}_h)
  \sum_{j=1}^{p_w} (\sups{\lambda}{w}_{jh})^2 \notag \\
  \sups{\hat{\tau}}{w}_h &= \frac{c+d}{\sups{\phi}{w}_h +
    \sups{\eta}{w} }, \notag \\
  \sups{\hat{\eta}}{w} &= \frac{d k +
    e}{\sups{\gamma}{w} + \sum_{h=1}^k \sups{\tau}{w}_h}, \notag \\
  \sups{\hat{\gamma}}{w} &= \frac{e + f}{\sups{\eta}{w} + \nu}, \notag \\
  \sups{\hat{\pi}}{w} &= \frac{\sum_{h = 1}^k \sups{\rho}{w}_h }{k}, \notag \\
  \hat{\sigma}^{-2}_j &= \frac{n/2 + a_\sigma - 1}{1/2(\mb y_{j\cdot}
    - \mb \lambda_{j\cdot} \la \mb X \ra) (\mb y_{j\cdot} - \mb
    \lambda_{j\cdot} \la \mb X \ra )^T + b_\sigma}. \notag
\end{align}

\section{Parameter Expanded EM (PX-EM) Algorithm for Robust MAP Estimates}
\label{app:PX-EM}

We introduce a positive semidefinite matrix $\mb R$ in our original
model to obtain a parameter expanded version:
\begin{align}
  \mb y_i &= \mb \Lambda \mb R^{-1}_L \mb x_i + \mb \epsilon_i, \notag \\
  \mb x_i &\sim \N_k(\mb 0, \mb R), \notag \\
  \mb \epsilon_i &\sim \N_k(\mb 0, \mb \Sigma). \notag
\end{align}
Here, $\mb R_L$ is the lower triangular part of the Cholesky decomposition
of $\mb R$.  Marginally, the covariance matrix is still $\mb \Omega =
\mb \Lambda \mb \Lambda^T + \mb \Sigma$, as this additional parameter
keeps the likelihood invariant. This additional parameter reduces
the coupling effects between the updates of loading matrix and latent
factors \citep{liu_parameter_1998, dyk_art_2001} and serves to connect
different posterior modes with equal likelihood curves indexed by $\mb
R$ \citep{rockova2014fast}.

Let $\mb \Lambda^\star = \mb \Lambda \mb R^{-1}_L$ and $\mb \Xi^\star
= \{\mb \Lambda^\star, \mb \Theta, \mb \Delta, \mb \Phi, \mb T, \mb
\eta, \mb \gamma, \mb \pi, \mb \Sigma \}$. Then the parameters of our
expanded model are $\{ \mb \Xi^\star \cup \mb R \}$. We assign our
structured prior on $\mb \Lambda^\star$. Thus, the updates of $\mb
\Xi^\star$ are unchanged given the estimates of the first and second
moments of $\mb X$. The estimates of $\la \mb X \ra$ and $\la \mb X
\mb X^T \ra$ can be calculated using Equations~(\ref{eq:xess} and
\ref{eq:xxess}) in Appendix B after mapping the loading matrix back to
the original matrix: $\mb \Lambda = \mb \Lambda^\star \mb R_L$. It
remains to estimate $\mb R$.

Write the expected complete log likelihood in the expanded model as
\begin{align}
  Q(\mb \Xi^\star, \mb R | \subs{\mb \Xi}{s}) = \mathbb{E}_{\mb X, \mb
    Z| \subs{\mb \Xi}{s}, \mb Y, \mb R_0} \log\big(p(\mb \Xi^\star,
  \mb R, \mb X, \mb Z| \mb Y)\big). \notag
\end{align}
The only term involving $\mb R$ is $p(\mb X)$. Therefore, the $\mb R$
that maximizes this function is
\begin{align}
  \subs{\mb R}{s} = \argmax_{\mb R} Q(\mb \Xi^\star, \mb R | \subs{\mb
    \Xi}{s}) = \argmax_{\mb R} \bigg(\textrm{const}- \frac{n}{2} \log
  |\mb R| - \frac{1}{2} \tr \big(\mb R^{-1} \mb S^{XX}\big)\bigg).
  \notag
\end{align}
The solution is $\subs{\mb R}{s} = \frac{1}{n} \mb S^{XX}$.

The EM algorithm in this expanded parameter space generates the
sequence $\{ \subs{\mb \Xi^\star}{1} \cup \subs{\mb R}{1}, \subs{\mb
  \Xi^\star}{2} \cup \subs{\mb R}{2}, \cdots \}$. This sequence
corresponds to a sequence of parameter estimations in the original
space $\{\subs{\mb \Xi}{1}, \subs{\mb \Xi}{2}, \cdots\}$ where $\mb
\Lambda$ in the original space is equal to $\mb \Lambda^\star \mb R_L$
\citep{rockova2014fast}. We initialize $\subs{\mb R}{0} = \mb I_k$.

\vskip 0.2in
\bibliography{refs}

\begin{thebibliography}{95}
\providecommand{\natexlab}[1]{#1}
\providecommand{\url}[1]{\texttt{#1}}
\expandafter\ifx\csname urlstyle\endcsname\relax
  \providecommand{\doi}[1]{doi: #1}\else
  \providecommand{\doi}{doi: \begingroup \urlstyle{rm}\Url}\fi

\bibitem[Archambeau and Bach(2009)]{archambeau_sparse_2008}
C\'{e}dric Archambeau and Francis~R. Bach.
\newblock Sparse probabilistic projections.
\newblock In \emph{Advances in Neural Information Processing Systems 21}, pages
  73--80, 2009.

\bibitem[Armagan et~al.(2011)Armagan, Clyde, and
  Dunson]{armagan_generalized_2011}
Artin Armagan, Merlise Clyde, and David~B. Dunson.
\newblock Generalized beta mixtures of {G}aussians.
\newblock In \emph{Advances in Neural Information Processing Systems 24}, pages
  523--531, 2011.

\bibitem[Armagan et~al.(2013)Armagan, Dunson, and
  Lee]{armagan_generalized_Pareto_2011}
Artin Armagan, David~B. Dunson, and Jaeyong Lee.
\newblock Generalized double {P}areto shrinkage.
\newblock \emph{Statistica Sinica}, 23\penalty0 (1):\penalty0 119, 2013.

\bibitem[Bach and Jordan(2005)]{bach_probabilistic_2005}
Francis~R. Bach and Michael~I. Jordan.
\newblock A probabilistic interpretation of canonical correlation analysis.
\newblock \emph{Technical Report 688, Department of Statistics, University of
  California, Berkeley}, 2005.

\bibitem[Bhattacharya and Dunson(2011)]{bhattacharya_sparse_2011}
Anirban Bhattacharya and David~B. Dunson.
\newblock Sparse {B}ayesian infinite factor models.
\newblock \emph{Biometrika}, 98\penalty0 (2):\penalty0 291--306, 2011.

\bibitem[Bhattacharya et~al.(2014)Bhattacharya, Pati, Pillai, and
  Dunson]{bhattacharya_bayesian_2012}
Anirban Bhattacharya, Debdeep Pati, Natesh~S. Pillai, and David~B. Dunson.
\newblock Dirichlet-{L}aplace priors for optimal shrinkage.
\newblock \emph{Journal of the American Statistical Association}, Accepted for
  publication, 2014.

\bibitem[Brown et~al.(2013)Brown, Mangravite, and
  Engelhardt]{brown_integrative_2013}
Christopher~D. Brown, Lara~M. Mangravite, and Barbara~E. Engelhardt.
\newblock Integrative modeling of {eQTLs} and cis-regulatory elements suggests
  mechanisms underlying cell type specificity of {eQTLs}.
\newblock \emph{{PLoS} Genetics}, 9\penalty0 (8):\penalty0 e1003649, 2013.

\bibitem[Browne(1979)]{browne_maximum-likelihood_1979}
Michael~W. Browne.
\newblock The maximum-likelihood solution in inter-battery factor analysis.
\newblock \emph{British Journal of Mathematical and Statistical Psychology},
  32\penalty0 (1):\penalty0 75--86, 1979.

\bibitem[Browne(1980)]{browne_factor_1980}
Michael~W. Browne.
\newblock Factor analysis of multiple batteries by maximum likelihood.
\newblock \emph{British Journal of Mathematical and Statistical Psychology},
  33\penalty0 (2):\penalty0 184--199, 1980.

\bibitem[Cand{\`e}s et~al.(2011)Cand{\`e}s, Li, Ma, and
  Wright]{candes_robust_2011}
Emmanuel~J. Cand{\`e}s, Xiaodong Li, Yi~Ma, and John Wright.
\newblock Robust principal component analysis?
\newblock \emph{Journal of the ACM}, 58\penalty0 (3):\penalty0 11, 2011.

\bibitem[Carvalho et~al.(2008)Carvalho, Chang, Lucas, Nevins, Wang, and
  West]{carvalho_high-dimensional_2008}
Carlos~M. Carvalho, Jeffrey Chang, Joseph~E. Lucas, Joseph~R. Nevins, Quanli
  Wang, and Mike West.
\newblock High-dimensional sparse factor modeling: applications in gene
  expression genomics.
\newblock \emph{Journal of the American Statistical Association}, 103\penalty0
  (484), 2008.

\bibitem[Carvalho et~al.(2009)Carvalho, Polson, and
  Scott]{carvalho_handling_2009}
Carlos~M. Carvalho, Nicholas~G. Polson, and James~G. Scott.
\newblock Handling sparsity via the horseshoe.
\newblock In \emph{Proceedings of the Twelfth International Conference on
  Artificial Intelligence and Statistics}, volume~5, pages 73--80, 2009.

\bibitem[Carvalho et~al.(2010)Carvalho, Polson, and
  Scott]{carvalho_horseshoe_2010}
Carlos~M. Carvalho, Nicholas~G. Polson, and James~G. Scott.
\newblock The horseshoe estimator for sparse signals.
\newblock \emph{Biometrika}, 97\penalty0 (2):\penalty0 465--480, 2010.

\bibitem[Chandrasekaran et~al.(2009)Chandrasekaran, Sanghavi, Parrilo, and
  Willsky]{chandrasekaran_sparse_2009}
Venkat Chandrasekaran, Sujay Sanghavi, Pablo~A. Parrilo, and Alan~S. Willsky.
\newblock Sparse and low-rank matrix decompositions.
\newblock In \emph{47th Annual Allerton Conference on Communication, Control,
  and Computing}, pages 962--967, 2009.

\bibitem[Chandrasekaran et~al.(2011)Chandrasekaran, Sanghavi, Parrilo, and
  Willsky]{chandrasekaran_rank-sparsity_2011}
Venkat Chandrasekaran, Sujay Sanghavi, Pablo~A. Parrilo, and Alan~S. Willsky.
\newblock Rank-sparsity incoherence for matrix decomposition.
\newblock \emph{SIAM Journal on Optimization}, 21\penalty0 (2):\penalty0
  572--596, 2011.

\bibitem[Comon(1994)]{comon_independent_1994}
Pierre Comon.
\newblock Independent component analysis, {A} new concept?
\newblock \emph{Signal Processing}, 36\penalty0 (3):\penalty0 287--314, 1994.

\bibitem[Cunningham and Ghahramani(2015)]{Cunningham2014}
John~P Cunningham and Zoubin Ghahramani.
\newblock Linear dimensionality reduction: Survey, insights, and
  generalizations.
\newblock \emph{Journal of Machine Learning Research}, 16, 2015.

\bibitem[Damianou et~al.(2012)Damianou, Ek, Titsias, and
  Lawrence]{damianou_manifold_2012}
Andreas Damianou, Carl Ek, Michalis Titsias, and Neil Lawrence.
\newblock Manifold relevance determination.
\newblock In \emph{29th International Conference on Machine Learning}, pages
  145--152, 2012.

\bibitem[Dempster et~al.(1977)Dempster, Laird, and
  Rubin]{dempster_maximum_1977}
Arthur~P. Dempster, Nan~M. Laird, and Donald~B. Rubin.
\newblock Maximum likelihood from incomplete data via the {EM} algorithm.
\newblock \emph{Journal of the Royal Statistical Society. Series B},
  39\penalty0 (1):\penalty0 1--38, 1977.

\bibitem[Dyk and Meng(2001)]{dyk_art_2001}
David A.~van Dyk and Xiao-Li Meng.
\newblock The art of data augmentation.
\newblock \emph{Journal of Computational and Graphical Statistics}, 10\penalty0
  (1):\penalty0 1--50, 2001.

\bibitem[Edwards(2000)]{edwards_introduction_2000}
David Edwards.
\newblock \emph{Introduction to Graphical Modelling}.
\newblock Springer, New York, 2nd edition, June 2000.
\newblock ISBN 9780387950549.

\bibitem[Ek et~al.(2008)Ek, Rihan, Torr, Rogez, and
  Lawrence]{ek_ambiguity_2008}
Carl~Henrik Ek, Jon Rihan, Philip~H.S. Torr, Gr{\'e}gory Rogez, and Neil~D.
  Lawrence.
\newblock Ambiguity modeling in latent spaces.
\newblock In \emph{Machine Learning for Multimodal Interaction}, pages 62--73.
  Springer, 2008.

\bibitem[Engelhardt and Adams(2014)]{engelhardt_bayesian_2014}
Barbara~E. Engelhardt and Ryan~P. Adams.
\newblock Bayesian structured sparsity from {G}aussian fields.
\newblock \emph{{arXiv}:1407.2235}, 2014.

\bibitem[Engelhardt and Stephens(2010)]{engelhardt_analysis_2010}
Barbara~E. Engelhardt and Matthew Stephens.
\newblock Analysis of population structure: {A} unifying framework and novel
  methods based on sparse factor analysis.
\newblock \emph{{PLoS} Genetics}, 6\penalty0 (9):\penalty0 e1001117, 2010.

\bibitem[Gao et~al.(2013)Gao, Brown, and Engelhardt]{gao_latent_2013}
Chuan Gao, Christopher~D. Brown, and Barbara~E. Engelhardt.
\newblock A latent factor model with a mixture of sparse and dense factors to
  model gene expression data with confounding effects.
\newblock \emph{{arXiv:1310.4792}}, 2013.

\bibitem[Gao et~al.(2014)Gao, Zhao, McDowell, Brown, and Engelhardt]{Gao2014}
Chuan Gao, Shiwen Zhao, Ian~C. McDowell, Christopher~D. Brown, and Barbara~E.
  Engelhardt.
\newblock Differential gene co-expression networks via {B}ayesian biclustering
  models.
\newblock \emph{{arXiv:1411.1997}}, 2014.

\bibitem[Gonz{\'a}lez et~al.(2008)Gonz{\'a}lez, D{\'e}jean, Martin, and
  Baccini]{gonzalez_cca:_2008}
Ignacio Gonz{\'a}lez, S{\'e}bastien D{\'e}jean, Pascal~G.P. Martin, and Alain
  Baccini.
\newblock {CCA:} {A}n {R} package to extend canonical correlation analysis.
\newblock \emph{Journal of Statistical Software}, 23\penalty0 (12):\penalty0
  1--14, 2008.

\bibitem[Griffiths and Ghahramani(2011)]{griffiths_indian_2011}
Thomas~L. Griffiths and Zoubin Ghahramani.
\newblock The {I}ndian buffet process: {A}n introduction and review.
\newblock \emph{The Journal of Machine Learning Research}, 12:\penalty0
  1185--1224, 2011.

\bibitem[Hans(2009)]{Hans2009}
Chris Hans.
\newblock Bayesian lasso regression.
\newblock \emph{Biometrika}, 96\penalty0 (4):\penalty0 835--845, 2009.

\bibitem[Hoerl and Kennard(1970)]{hoerl_ridge_1970}
Arthur~E. Hoerl and Robert~W. Kennard.
\newblock Ridge regression: biased estimation for nonorthogonal problems.
\newblock \emph{Technometrics}, 12\penalty0 (1):\penalty0 55--67, 1970.

\bibitem[Hotelling(1933)]{hotelling_analysis_1933}
Harold Hotelling.
\newblock Analysis of a complex of statistical variables into principal
  components.
\newblock \emph{Journal of Educational Psychology}, 24\penalty0 (6):\penalty0
  417, 1933.

\bibitem[Hotelling(1936)]{hotelling_relations_1936}
Harold Hotelling.
\newblock Relations between two sets of variates.
\newblock \emph{Biometrika}, 28\penalty0 (3/4):\penalty0 321--377, 1936.

\bibitem[Huang et~al.(2011)Huang, Zhang, and Metaxas]{huang_learning_2011}
Junzhou Huang, Tong Zhang, and Dimitris Metaxas.
\newblock Learning with structured sparsity.
\newblock \emph{The Journal of Machine Learning Research}, 12:\penalty0
  3371--3412, 2011.

\bibitem[Jenatton et~al.(2010)Jenatton, Obozinski, and
  Bach]{jenatton_structured_2009}
Rodolphe Jenatton, Guillaume Obozinski, and Francis Bach.
\newblock Structured sparse principal component analysis.
\newblock In \emph{Proceedings of the Thirteenth International Conference on
  Artificial Intelligence and Statistics}, pages 366--373, 2010.

\bibitem[Jenatton et~al.(2011)Jenatton, Audibert, and
  Bach]{jenatton_structured_2011}
Rodolphe Jenatton, Jean-Yves Audibert, and Francis Bach.
\newblock Structured variable selection with sparsity-inducing norms.
\newblock \emph{The Journal of Machine Learning Research}, 12:\penalty0
  2777--2824, 2011.

\bibitem[Jia et~al.(2010)Jia, Salzmann, and Darrell]{jia_factorized_2010}
Yangqing Jia, Mathieu Salzmann, and Trevor Darrell.
\newblock Factorized latent spaces with structured sparsity.
\newblock In \emph{Advances in Neural Information Processing Systems 23}, pages
  982--990, 2010.

\bibitem[Joachims(1997)]{newsgroup_1997}
Thorsten Joachims.
\newblock A probabilistic analysis of the {R}occhio algorithm with {TFIDF} for
  text categorization.
\newblock In \emph{Proceedings of the 14th International Conference on Machine
  Learning}, pages 143--151, 1997.

\bibitem[Khan et~al.(2014)Khan, Virtanen, Kallioniemi, Wennerberg, Poso, and
  Kaski]{khan_identification_2014}
Suleiman~A. Khan, Seppo Virtanen, Olli~P. Kallioniemi, Krister Wennerberg,
  Antti Poso, and Samuel Kaski.
\newblock Identification of structural features in chemicals associated with
  cancer drug response: {A} systematic data-driven analysis.
\newblock \emph{Bioinformatics}, 30\penalty0 (17):\penalty0 i497--i504, 2014.

\bibitem[Klami(2014)]{klami_polya-gamma_2014}
Arto Klami.
\newblock {P}olya-gamma augmentations for factor models.
\newblock In \emph{The 6th Asian Conference on Machine Learning}, pages
  112--128, 2014.

\bibitem[Klami and Kaski(2008)]{klami_probabilistic_2008}
Arto Klami and Samuel Kaski.
\newblock Probabilistic approach to detecting dependencies between data sets.
\newblock \emph{Neurocomputing}, 72\penalty0 (1):\penalty0 39--46, 2008.

\bibitem[Klami et~al.(2013)Klami, Virtanen, and Kaski]{klami_bayesian_2013}
Arto Klami, Seppo Virtanen, and Samuel Kaski.
\newblock Bayesian canonical correlation analysis.
\newblock \emph{Journal of Machine Learning Research}, 14:\penalty0 965--1003,
  2013.

\bibitem[Klami et~al.(2014{\natexlab{a}})Klami, Bouchard, and
  Tripathi]{klami_group-sparse_2013}
Arto Klami, Guillaume Bouchard, and Abhishek Tripathi.
\newblock Group-sparse embeddings in collective matrix factorization.
\newblock In \emph{International Conference on Learning Representations},
  2014{\natexlab{a}}.

\bibitem[Klami et~al.(2014{\natexlab{b}})Klami, Virtanen, Lepp{\"a}aho, and
  Kaski]{klami_group_2014}
Arto Klami, Seppo Virtanen, Eemeli Lepp{\"a}aho, and Samuel Kaski.
\newblock Group factor analysis.
\newblock \emph{{IEEE} Transactions on Neural Networks and Learning Systems},
  Accepted for publication, 2014{\natexlab{b}}.

\bibitem[Knowles and Ghahramani(2011)]{knowles_nonparametric_2011}
David Knowles and Zoubin Ghahramani.
\newblock Nonparametric {B}ayesian sparse factor models with application to
  gene expression modeling.
\newblock \emph{The Annals of Applied Statistics}, 5\penalty0 ({2B}):\penalty0
  1534--1552, 2011.

\bibitem[Koller and Friedman(2009)]{koller_probabilistic_2009}
Daphne Koller and Nir Friedman.
\newblock \emph{Probabilistic Graphical Models: Principles and Techniques}.
\newblock The {MIT} Press, 1 edition, July 2009.

\bibitem[Kowalski(2009)]{kowalski_sparse_2009}
Matthieu Kowalski.
\newblock Sparse regression using mixed norms.
\newblock \emph{Applied and Computational Harmonic Analysis}, 27\penalty0
  (3):\penalty0 303--324, 2009.

\bibitem[Kowalski and Torr{\'e}sani(2009)]{kowalski_structured_2009}
Matthieu Kowalski and Bruno Torr{\'e}sani.
\newblock Structured sparsity: {F}rom mixed norms to structured shrinkage.
\newblock In \emph{Processing with Adaptive Sparse Structured Representations},
  2009.

\bibitem[Kyung et~al.(2010)Kyung, Gill, Ghosh, and
  Casella]{kyung_penalized_2010}
Minjung Kyung, Jeff Gill, Malay Ghosh, and George Casella.
\newblock Penalized regression, standard errors, and {B}ayesian lassos.
\newblock \emph{Bayesian Analysis}, 5\penalty0 (2):\penalty0 369--411, 2010.

\bibitem[Lawrence(2005)]{lawrence_probabilistic_2005}
Neil Lawrence.
\newblock Probabilistic non-linear principal component analysis with {G}aussian
  process latent variable models.
\newblock \emph{The Journal of Machine Learning Research}, 6:\penalty0
  1783--1816, 2005.

\bibitem[Leek et~al.(2010)Leek, Scharpf, Bravo, Simcha, Langmead, Johnson,
  Geman, Baggerly, and Irizarry]{Leek2010}
Jeffrey~T. Leek, Robert~B. Scharpf, Hector~Corrada Bravo, David Simcha,
  Benjamin Langmead, W.~Evan Johnson, Donald Geman, Keith Baggerly, and
  Rafael~A. Irizarry.
\newblock Tackling the widespread and critical impact of batch effects in
  high-throughput data.
\newblock \emph{Nature Reviews Genetics}, 11\penalty0 (10):\penalty0 733--739,
  2010.

\bibitem[Li(2002)]{li_genome-wide_2002}
Ker-Chau Li.
\newblock Genome-wide coexpression dynamics: {T}heory and application.
\newblock \emph{Proceedings of the National Academy of Sciences}, 99\penalty0
  (26):\penalty0 16875--16880, 2002.

\bibitem[Liu et~al.(1998)Liu, Rubin, and Wu]{liu_parameter_1998}
Chuanhai Liu, Donald~B. Rubin, and Ying~Nian Wu.
\newblock Parameter expansion to accelerate {EM}: {T}he {PX-EM} algorithm.
\newblock \emph{Biometrika}, 85\penalty0 (4):\penalty0 755--770, 1998.

\bibitem[Lucas et~al.(2010)Lucas, Kung, and Chi]{lucas_latent_2010}
Joseph~E. Lucas, Hsiu-Ni Kung, and Jen-Tsan~A. Chi.
\newblock Latent factor analysis to discover pathway-associated putative
  segmental aneuploidies in human cancers.
\newblock \emph{{PLoS} Computational Biology}, 6\penalty0 (9):\penalty0
  e1000920, 2010.

\bibitem[Ma et~al.(2011)Ma, Schadt, Kaplan, and Zhao]{ma_cosine_2011}
Haisu Ma, Eric~E. Schadt, Lee~M. Kaplan, and Hongyu Zhao.
\newblock {COSINE}: {Condition}-{specific} sub-{network} identification using a
  global optimization method.
\newblock \emph{Bioinformatics}, 27\penalty0 (9):\penalty0 1290--1298, 2011.

\bibitem[Mangravite et~al.(2013)Mangravite, Engelhardt, Medina, Smith, Brown,
  Chasman, Mecham, Howie, Shim, Naidoo, et~al.]{Mangravite2013}
Lara~M. Mangravite, Barbara~E. Engelhardt, Marisa~W. Medina, Joshua~D. Smith,
  Christopher~D. Brown, Daniel~I. Chasman, Brigham~H. Mecham, Bryan Howie,
  Heejung Shim, Devesh Naidoo, et~al.
\newblock A statin-dependent {QTL} for \emph{GATM} expression is associated
  with statin-induced myopathy.
\newblock \emph{Nature}, 502\penalty0 (7471):\penalty0 377--380, 2013.

\bibitem[McDonald(1970)]{mcdonald_three_1970}
Roderick~P. McDonald.
\newblock Three common factor models for groups of variables.
\newblock \emph{Psychometrika}, 35\penalty0 (1):\penalty0 111--128, 1970.

\bibitem[Mitchell and Beauchamp(1988)]{mitchell_bayesian_1988}
Toby~J. Mitchell and John~J. Beauchamp.
\newblock Bayesian variable selection in linear regression.
\newblock \emph{Journal of the American Statistical Association}, 83\penalty0
  (404):\penalty0 1023--1032, 1988.

\bibitem[Neal(1995)]{neal_bayesian_1995}
Radford~M. Neal.
\newblock \emph{Bayesian learning for neural networks}.
\newblock PhD thesis, University of Toronto, 1995.

\bibitem[O'Hagan(1979)]{o1979outlier}
Anthony O'Hagan.
\newblock On outlier rejection phenomena in {B}ayes inference.
\newblock \emph{Journal of the Royal Statistical Society. Series B},
  41\penalty0 (3):\penalty0 358--367, 1979.

\bibitem[Park and Casella(2008)]{park_bayesian_2008}
Trevor Park and George Casella.
\newblock The {B}ayesian lasso.
\newblock \emph{Journal of the American Statistical Association}, 103\penalty0
  (482):\penalty0 681--686, 2008.

\bibitem[Parkhomenko et~al.(2009)Parkhomenko, Tritchler, and
  Beyene]{parkhomenko_sparse_2009}
Elena Parkhomenko, David Tritchler, and Joseph Beyene.
\newblock Sparse canonical correlation analysis with application to genomic
  data integration.
\newblock \emph{Statistical Applications in Genetics and Molecular Biology},
  8\penalty0 (1):\penalty0 1--34, 2009.

\bibitem[Pedregosa et~al.(2011)Pedregosa, Varoquaux, Gramfort, Michel, Thirion,
  Grisel, Blondel, Prettenhofer, Weiss, Dubourg, et~al.]{scikit-learn}
Fabian Pedregosa, Ga{\"e}l Varoquaux, Alexandre Gramfort, Vincent Michel,
  Bertrand Thirion, Olivier Grisel, Mathieu Blondel, Peter Prettenhofer, Ron
  Weiss, Vincent Dubourg, et~al.
\newblock Scikit-learn: {M}achine learning in {P}ython.
\newblock \emph{The Journal of Machine Learning Research}, 12:\penalty0
  2825--2830, 2011.

\bibitem[Polson and Scott(2011)]{polson_shrink_2010}
Nicholas~G. Polson and James~G. Scott.
\newblock Shrink globally, act locally: {S}parse {B}ayesian regularization and
  prediction.
\newblock In \emph{Bayesian Statistics 9, eds. J.M. Bernardo et al.}, pages
  501--538. Oxford University Press, 2011.

\bibitem[Pournara and Wernisch(2007)]{pournara_factor_2007}
Iosifina Pournara and Lorenz Wernisch.
\newblock Factor analysis for gene regulatory networks and transcription factor
  activity profiles.
\newblock \emph{{BMC} Bioinformatics}, 8:\penalty0 61, 2007.

\bibitem[Pruteanu-Malinici et~al.(2011)Pruteanu-Malinici, Mace, and
  Ohler]{pruteanu-malinici_automatic_2011}
Iulian Pruteanu-Malinici, Daniel~L. Mace, and Uwe Ohler.
\newblock Automatic annotation of spatial expression patterns via {B}ayesian
  factor models.
\newblock \emph{{PLoS} Computational Biology}, 7\penalty0 (7):\penalty0
  e1002098, 2011.

\bibitem[Qu and Chen(2011)]{qu_sparse_2011}
Xinquan Qu and Xinlei Chen.
\newblock Sparse structured probabilistic projections for factorized latent
  spaces.
\newblock In \emph{Proceedings of the 20th {ACM} International Conference on
  {Information} and Knowledge Management}, pages 1389--1394, 2011.

\bibitem[Ray et~al.(2014)Ray, Zheng, Lucas, and Carin]{ray_bayesian_2014}
Priyadip Ray, Lingling Zheng, Joseph Lucas, and Lawrence Carin.
\newblock Bayesian joint analysis of heterogeneous genomics data.
\newblock \emph{Bioinformatics}, 30\penalty0 (10):\penalty0 1370--1376, 2014.

\bibitem[Rockov{\'a} and George(2014)]{rockova2014fast}
Veronika Rockov{\'a} and Edward~I. George.
\newblock Fast {B}ayesian factor analysis via automatic rotations to sparsity.
\newblock 2014.

\bibitem[Romberg et~al.(2001)Romberg, Choi, and
  Baraniuk]{romberg_bayesian_2001}
Justin~K. Romberg, Hyeokho Choi, and Richard~G. Baraniuk.
\newblock Bayesian tree-structured image modeling using wavelet-domain hidden
  markov models.
\newblock \emph{{IEEE} Transactions on Image Processing}, 10\penalty0
  (7):\penalty0 1056--1068, 2001.

\bibitem[Roweis(1998)]{roweis_em_1998}
Sam Roweis.
\newblock {EM} algorithms for {PCA} and {SPCA}.
\newblock In \emph{Advances in Neural Information Processing Systems 10}, pages
  626--632, 1998.

\bibitem[Salomatin et~al.(2009)Salomatin, Yang, and
  Lad]{salomatin_multifield_2009}
Konstantin Salomatin, Yiming Yang, and Abhimanyu Lad.
\newblock Multi-field correlated topic modeling.
\newblock In \emph{{SIAM} International Conference on Data Mining}, pages
  628--637, 2009.

\bibitem[Salzmann et~al.(2010)Salzmann, Ek, Urtasun, and
  Darrell]{salzmann_factorized_2010}
Mathieu Salzmann, Carl~H. Ek, Raquel Urtasun, and Trevor Darrell.
\newblock Factorized orthogonal latent spaces.
\newblock In \emph{Proceedings of the Thirteenth International Conference on
  Artificial Intelligence and Statistics}, pages 701--708, 2010.

\bibitem[Sch{\"a}fer and Strimmer(2005)]{schafer_empirical_2005}
Juliane Sch{\"a}fer and Korbinian Strimmer.
\newblock An empirical {B}ayes approach to inferring large-scale gene
  association networks.
\newblock \emph{Bioinformatics}, 21\penalty0 (6):\penalty0 754--764, 2005.

\bibitem[Shon et~al.(2005)Shon, Grochow, Hertzmann, and
  Rao]{shon_learning_2005}
Aaron Shon, Keith Grochow, Aaron Hertzmann, and Rajesh~P. Rao.
\newblock Learning shared latent structure for image synthesis and robotic
  imitation.
\newblock In \emph{Advances in Neural Information Processing Systems 18}, pages
  1233--1240, 2005.

\bibitem[Suvitaival et~al.(2014)Suvitaival, Parkkinen, Virtanen, and
  Kaski]{suvitaival_cross-organism_2014}
Tommi Suvitaival, Juuso~A. Parkkinen, Seppo Virtanen, and Samuel Kaski.
\newblock Cross-organism toxicogenomics with group factor analysis.
\newblock \emph{Systems Biomedicine}, 2:\penalty0 e29291, 2014.

\bibitem[Tibshirani(1996)]{tibshirani_regression_1996}
Robert Tibshirani.
\newblock Regression shrinkage and selection via the lasso.
\newblock \emph{Journal of the Royal Statistical Society. Series B},
  58\penalty0 (1):\penalty0 267--288, 1996.

\bibitem[Tipping(2001)]{tipping_sparse_2001}
Michael~E. Tipping.
\newblock Sparse {B}ayesian learning and the relevance vector machine.
\newblock \emph{The Journal of Machine Learning Research}, 1:\penalty0
  211--244, 2001.

\bibitem[Tipping and Bishop(1999{\natexlab{a}})]{tipping_mixtures_1999}
Michael~E. Tipping and Christopher~M. Bishop.
\newblock Mixtures of probabilistic principal component analyzers.
\newblock \emph{Neural Computation}, 11\penalty0 (2):\penalty0 443--482,
  1999{\natexlab{a}}.

\bibitem[Tipping and Bishop(1999{\natexlab{b}})]{tipping_probabilistic_1999}
Michael~E. Tipping and Christopher~M. Bishop.
\newblock Probabilistic principal component analysis.
\newblock \emph{Journal of the Royal Statistical Society: Series B},
  61\penalty0 (3):\penalty0 611--622, 1999{\natexlab{b}}.

\bibitem[Tsoumakas et~al.(2011)Tsoumakas, Spyromitros-Xioufis, Vilcek, and
  Vlahavas]{mulan}
Grigorios Tsoumakas, Eleftherios Spyromitros-Xioufis, Jozef Vilcek, and Ioannis
  Vlahavas.
\newblock Mulan: A java library for multi-label learning.
\newblock \emph{Journal of Machine Learning Research}, 12:\penalty0 2411--2414,
  2011.

\bibitem[Virtanen et~al.(2011)Virtanen, Klami, and
  Kaski]{virtanen_bayesian_2011}
Seppo Virtanen, Arto Klami, and Samuel Kaski.
\newblock Bayesian {CCA} via group sparsity.
\newblock In \emph{Proceedings of the 28th {International} {Conference} on
  {Machine} {Learning}}, pages 457--464, 2011.

\bibitem[Virtanen et~al.(2012)Virtanen, Klami, Khan, and
  Kaski]{virtanen_bayesian_2012}
Seppo Virtanen, Arto Klami, Suleiman~A. Khan, and Samuel Kaski.
\newblock Bayesian group factor analysis.
\newblock In \emph{Proceedings of the Fifteenth International Conference on
  Artificial Intelligence and Statistics}, volume~22, pages 1269--1277, 2012.

\bibitem[Wen and Yin(2013)]{wen_feasible_2013}
Zaiwen Wen and Wotao Yin.
\newblock A feasible method for optimization with orthogonality constraints.
\newblock \emph{Mathematical Programming}, 142\penalty0 (1-2):\penalty0
  397--434, 2013.

\bibitem[West(1987)]{west_scale_1987}
Mike West.
\newblock On scale mixtures of normal distributions.
\newblock \emph{Biometrika}, 74\penalty0 (3):\penalty0 646--648, 1987.

\bibitem[West(2003)]{west_bayesian_2003}
Mike West.
\newblock Bayesian factor regression models in the "large p, small n" paradigm.
\newblock In \emph{Bayesian Statistics 7, eds. J.M. Bernardo et al.}, pages
  723--732. Oxford University Press, 2003.

\bibitem[Witten et~al.(2013)Witten, Tibshirani, Gross, and
  Narasimhan]{RPackage_PMA}
Daniela Witten, Rob Tibshirani, Sam Gross, and Balasubramanian Narasimhan.
\newblock \emph{PMA: Penalized Multivariate Analysis}, 2013.
\newblock URL \url{http://CRAN.R-project.org/package=PMA}.
\newblock R package version 1.0.9.

\bibitem[Witten and Tibshirani(2009)]{witten_extensions_2009}
Daniela~M. Witten and Robert~J. Tibshirani.
\newblock Extensions of sparse canonical correlation analysis with applications
  to genomic data.
\newblock \emph{Statistical Applications in Genetics and Molecular Biology},
  8\penalty0 (1):\penalty0 1--27, 2009.

\bibitem[Witten et~al.(2009)Witten, Tibshirani, and
  Hastie]{witten_penalized_2009}
Daniela~M. Witten, Robert Tibshirani, and Trevor Hastie.
\newblock A penalized matrix decomposition, with applications to sparse
  principal components and canonical correlation analysis.
\newblock \emph{Biostatistics}, 10\penalty0 (3):\penalty0 515--534, 2009.

\bibitem[Yuan and Lin(2006)]{yuan_model_2006}
Ming Yuan and Yi~Lin.
\newblock Model selection and estimation in regression with grouped variables.
\newblock \emph{Journal of the Royal Statistical Society: Series B},
  68\penalty0 (1):\penalty0 49--67, 2006.

\bibitem[Zhao et~al.(2009)Zhao, Rocha, and Yu]{zhao_composite_2009}
Peng Zhao, Guilherme Rocha, and Bin Yu.
\newblock The composite absolute penalties family for grouped and hierarchical
  variable selection.
\newblock \emph{The Annals of Statistics}, 37\penalty0 ({6A}):\penalty0
  3468--3497, 2009.

\bibitem[Zhao and Li(2012)]{zhao_co-module_2012}
Shiwen Zhao and Shao Li.
\newblock A co-module approach for elucidating drug-disease associations and
  revealing their molecular basis.
\newblock \emph{Bioinformatics}, 28\penalty0 (7):\penalty0 955--961, 2012.

\bibitem[Zhou et~al.(2011)Zhou, Tao, and Wu]{zhou_manifold_2011}
Tianyi Zhou, Dacheng Tao, and Xindong Wu.
\newblock Manifold elastic net: {A} unified framework for sparse dimension
  reduction.
\newblock \emph{Data Mining and Knowledge Discovery}, 22\penalty0 (3):\penalty0
  340--371, 2011.

\bibitem[Zou and Hastie(2005)]{zou_regularization_2005}
Hui Zou and Trevor Hastie.
\newblock Regularization and variable selection via the elastic net.
\newblock \emph{Journal of the Royal Statistical Society. Series B},
  67\penalty0 (2):\penalty0 301--320, 2005.

\bibitem[Zou et~al.(2006)Zou, Hastie, and Tibshirani]{zou_sparse_2006}
Hui Zou, Trevor Hastie, and Robert Tibshirani.
\newblock Sparse principal component analysis.
\newblock \emph{Journal of Computational and Graphical Statistics}, 15\penalty0
  (2):\penalty0 265--286, 2006.

\bibitem[Zou et~al.(2013)Zou, Hsu, Parkes, and Adams]{zou2013}
James~Y Zou, Daniel~J Hsu, David~C Parkes, and Ryan~P Adams.
\newblock Contrastive learning using spectral methods.
\newblock In \emph{Advances in Neural Information Processing Systems}, pages
  2238--2246, 2013.

\end{thebibliography}

\end{document}